\documentclass{article}
\pdfoutput=1
\usepackage{jheppub}

\usepackage{braket}
\usepackage[table]{xcolor}
\usepackage{mathrsfs}
\usepackage[english]{babel}
\usepackage{graphicx}
\usepackage{caption}
\usepackage{subcaption}
\usepackage{amssymb}

\usepackage[makeroom]{cancel}

\def\nn{\nonumber}

\def\pd{\partial}

\def\cI{{\cal I}}
\def\cH{{\cal H}}

\def\cL{{\cal L}}

\def\cO{{\cal O}}

\def\cR{{\cal R}}
\def\cT{{\cal T}}

\def\cV{{\cal V}}
\def\cW{{\cal W}}

\def\cS{{\cal S}}

\def\bfp{{{\bf p}}}
\def\bfk{{{\bf k}}}
\def\bfq{{{\bf q}}}

\def\bfx{{{\bf x}}}
\def\bfy{{{\bf y}}}

\def\mfa{{\mathfrak{a}}}
\def\mfb{{\mathfrak{b}}}

\def\mfp{{\mathfrak{p}}}

\def\exd{{\hbox{d}}}

\def\bea{\begin{eqnarray}}
\def\eea{\end{eqnarray}}
\def\be{\begin{equation}}
\def\ee{\end{equation}}

\def\ssH{{\scriptscriptstyle H}}

\def\ssN{{\scriptscriptstyle N}}

\def\ssR{{\scriptscriptstyle R}}
\def\ssS{{\scriptscriptstyle S}}

\def\pref#1{(\ref{#1})}
\newcommand{\roughly}[1]{\mathrel{\raise.3ex\hbox{$#1$\kern-0.85em
\lower1ex\hbox{$\sim$}}}}

\newcommand{\gsim}{\roughly>}

\def\Tr{\mathrm{Tr}}

\def\TrB{\underset{}{\mathrm{Tr^{\,'}}}}

\def\vac{\mathrm{vac}}

\def\smath#1{\text{\scalebox{.85}{$#1$}}}
\def\sfrac#1#2{\smath{\frac{#1}{#2}}}

\numberwithin{equation}{section}
\title{Influence Through Mixing: Hotspots\\ as Benchmarks for Basic Black-Hole Behaviour}
 
\date{March 2021}

\author[a,b]{G. Kaplanek,}

\author[a,b]{C.P. Burgess}

\author[c]{and R. Holman}

\affiliation[a]{Department of Physics \& Astronomy, McMaster University, 1280 Main Street West, Hamilton ON, Canada.}
\affiliation[b]{Perimeter Institute for Theoretical Physics, 31 Caroline Street North, Waterloo ON, Canada.}
\affiliation[c]{Minerva Schools at KGI,
1145 Market Street, San Francisco, CA 94103, USA.}

\abstract{Effective theories are being developed for fields outside black holes, often with an unusual open-system feel due to the influence of large number of degrees of freedom that lie out of reach beyond the horizon. What is often difficult when interpreting such theories is the absence of comparisons to simpler systems that share these features. We propose here such a simple model, involving a single external scalar field that mixes in a limited region of space with a `hotspot' containing a large number of hot internal degrees of freedom. Since the model is at heart gaussian it can be solved explicitly, and we do so for the mode functions and correlation functions for the external field once the hotspot fields are traced out. We compare with calculations that work perturbatively in the mixing parameter, and by doing so can precisely identify its domain of validity. We also show how renormalization-group EFT methods can allow some perturbative contributions to be resummed beyond leading order, verifying the result using the exact expression. 
}

\begin{document}

\maketitle

\section{Introduction}
\label{sec:intro}

At long last the detection of gravitational waves \cite{LIGO} has made near-horizon black-hole physics an experimental science, and this is very likely to deepen our understanding of General Relativity (GR) and/or end its hundred-year reign as the paradigm of choice when describing gravity. With the advent of measurements --- eventually precision measurements --- it behooves theorists to raise their game when quantifying the kinds of physics one might hope to see in this new regime. And this they are doing; both by pushing the accuracy of GR gravitational-wave predictions, and by exploring more systematically the predictions of alternatives theories when gravitational fields are strong (for reviews see \cite{Yunes:2013dva, Blanchet:2013haa, Berti:2015itd, Barack:2018yly}). 

Effective field theories (EFTs) are usually important tools for this kind of work, because they allow predictions for physics on observable length scales that are robust to changes in the details of what goes on at smaller scales \cite{Weinberg:1978kz, Rothstein:2003mp, Burgess:2007pt, Levi:2018nxp, EFTBook}. This is useful both when these smaller scales are understood and when they are not. Although EFT methods have a long history, their use is even now still being developed for black hole applications \cite{Goldberger:2004jt, Goldberger:2005cd, Porto:2005ac, Kol:2007bc, Kol:2007rx, Gilmore:2008gq, Porto:2008jj, Damour:2009vw, Emparan:2009at, Damour:2009wj, Levi:2015msa, Allwright:2018rut, Cayuso:2017iqc, Cayuso:2020lca}; a development that has been slowed both by the relative novelty of EFT applications to gravity in general \cite{Donoghue:1994dn, Burgess:2003jk, Goldberger:2007hy, Porto:2016pyg, Donoghue:2017ovt, EFTBook} and by some of the novel aspects of black hole physics in particular, since these differ from more garden-variety applications of EFT techniques. 

One issue --- though not the only one \cite{Allwright:2018rut, Cayuso:2017iqc, Cayuso:2020lca} --- that complicates developing EFT methods for black-hole behaviour is their open and thermal nature, since the entanglement and decoherence that such physics can involve is not captured by traditional Wilsonian EFT tools. Such differences have led some to ask whether an effective description of extra-horizon physics might involve unusual features (such as nonlocality) or otherwise evade the arguments that usually preclude these phenomena from arising in a Wilsonian context \cite{Hawking:1976ra, Giddings:2006sj, Skenderis:2008qn, Almheiri:2012rt, Almheiri:2013hfa, Banks:1994ph, Mathur:2009hf}. 

What usually helps when developing EFT tools are concrete systems for which both UV and IR sectors are well-understood and within which the EFT description can be assessed by comparing to other methods. These kinds of comparisons are not yet available for black holes, and the search for effective descriptions of black-hole physics are the poorer for it. The purpose of this paper is to help fix this situation by providing a simple black-hole proxy that can help fill this void. On one hand the model should be simple enough to solve, but on the other hand share enough black hole properties to be informative about some of their putative EFT descriptions. 

The model we propose -- inspired by similar models in condensed matter systems \cite{FeynmanVernon, CaldeiraLeggett, Lin:2005uk} -- has a large number of degrees of freedom with a thermal character and no gap; to which an external field couples only in a small region of space; what we call here for brevity a `hotspot'. We model the thermal degrees of freedom as a collection of $N$ massless scalar fields -- $\chi^a$ with $a = 1, \cdots , N$ -- that are initially prepared in a thermal bath. These fields are meant to model the black hole's interior. We take these fields to `interact' with the external massless scalar field $\phi$, which is a proxy for the black hole's exterior. The word interact appears in quotes because $\phi$ only couples to the $\chi^a$ through a bilinear mixing term of the form
\be \label{MixDef}
   \cL_{\rm mix} = - g_a \, \chi^a \, \phi \,, 
\ee
and so the entire theory remains gaussian and can be solved in great detail. 

\begin{figure}[h]
\begin{center}
\includegraphics[width=60mm,height=60mm]{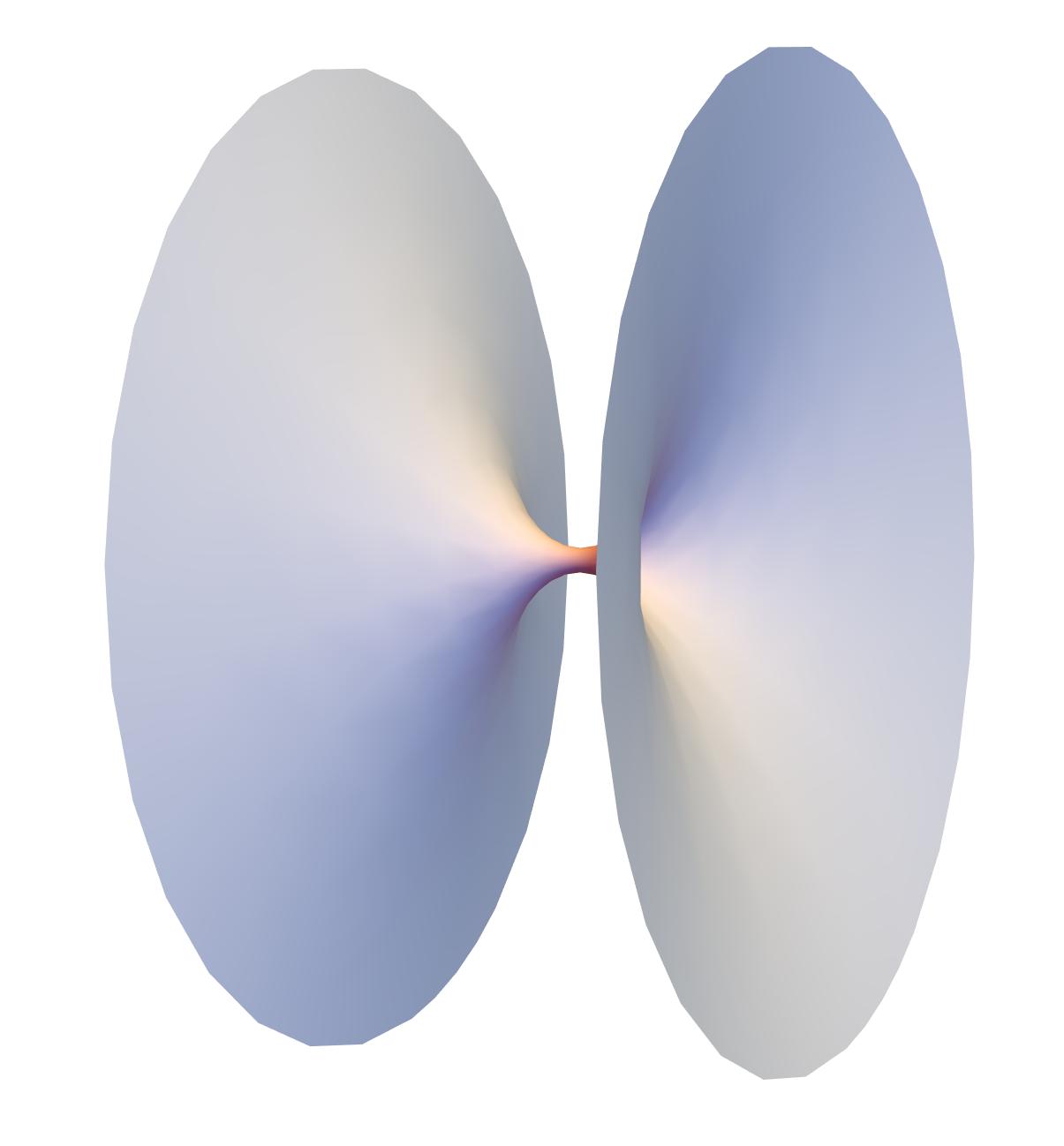} 
\caption{A cartoon of the two spatial branches, $\cR_+$ and $\cR_-$, in which the field $\phi$ and the $N$ fields $\chi^a$ repsectively live. The two types of fields only couple to one another in the localized throat region, which can be taken to be a small sphere of radius $\xi$ (or effectively a point in the limit that $\xi$ is much smaller than all other scales of interest).} \label{fig:FunnelFig} 
\end{center}
\end{figure}

So far this just describes a field mixing with a thermal bath. To make it more black-hole-like we imagine these two sectors only mix in a small localized region of space, and not interacting -- even gravitationally -- otherwise. In order to do this we imagine space at a given time to come with two spatial sheets, $\cR_+$ and $\cR_-$, with $\phi$ living only on $\cR_+$ and $\chi^a$ living on $\cR_-$. These two branches only intersect on a small spherical ball, $\cS_\xi$, of radius $\xi$, that plays the role of the black hole itself (see Fig.~\ref{fig:FunnelFig}).

In principle gravity can be included in this model, and does not generate couplings between the two sectors away from their overlap on $\cS_\xi$ (and this is why we take $\cR_\pm$ to be disjoint). We do not pursue this gravitational coupling further in this paper, focussing instead on how the field $\phi$ responds to the presence of the localized hotspot built from the thermal fields $\chi^a$. As a result our model does not capture the causal nature of the horizon or the exponential redshifts that arise in its vicinity for real black holes.

\begin{figure}[h]
\begin{center}
\includegraphics[width=100mm,height=50mm]{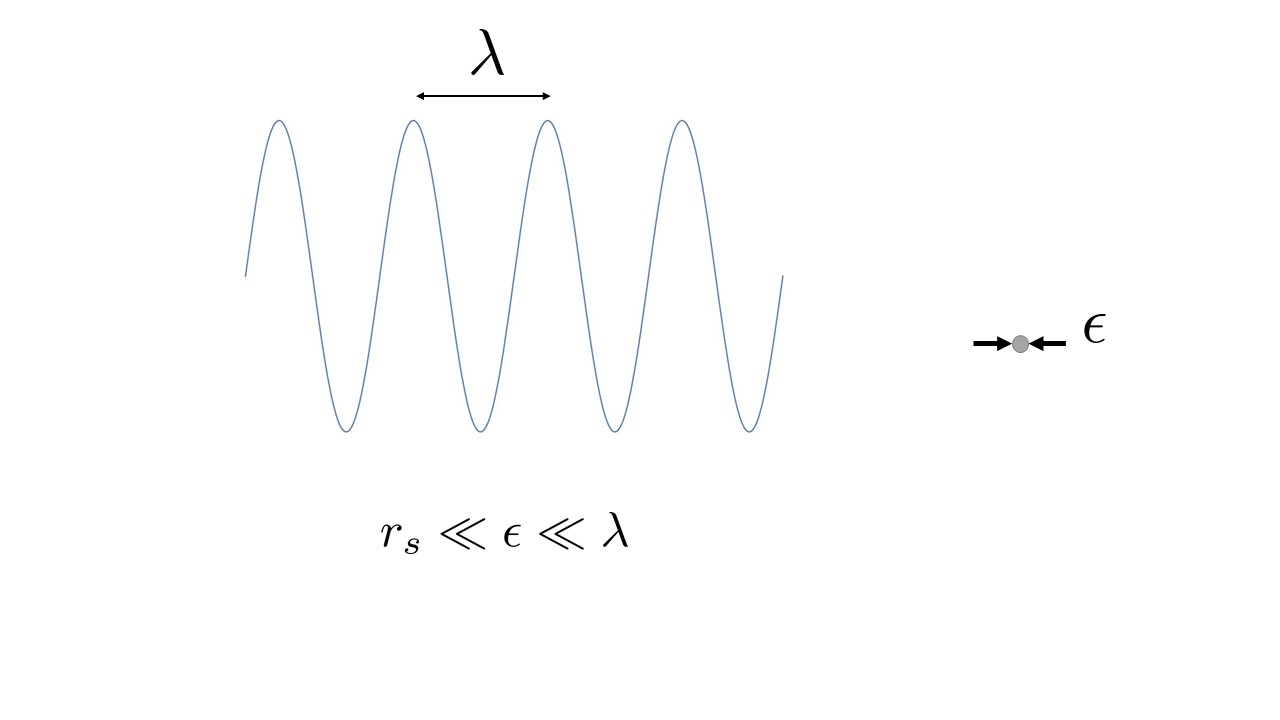} 
\vspace{-10mm}
\caption{An EFT regime appropriate for small black holes, for which the UV cutoff scale is much smaller than the length of any low-energy probe, $\epsilon \ll \lambda$, but much larger than the horizon scale $\epsilon \gg r_s$.} \label{fig:Scale1Fig} 
\end{center}
\end{figure}

Broadly speaking there are two types of black-hole EFTs that are usually pursued, and both can have counterparts in our hotspot model. The main variant is one that is appropriate to gravitational wave emission, and applies on length scales $\lambda \gg r_s$ that are much larger than the black hole's size (see Fig.~\ref{fig:Scale1Fig}). In this `world-line' or `point-particle' EFT the closest distance to the black hole that can be directly resolved corresponds to a cutoff that has size $\epsilon \gg r_s$ and so the black hole dynamics is described by its center-of-mass coordinate; it is regarded as a point mass moving along a trajectory in spacetime. The response of the black hole to applied `bulk' fields (and the back-reaction of the black hole back onto these fields) is described by an action defined as a functional of the bulk fields integrated along the black hole's one-dimensional world line. This type of EFT is obtained in the hotspot example by taking the radius $\xi$ of the interaction sphere $\cS_\xi$ to be much smaller than all other scales: $\xi \ll \epsilon \ll \lambda$.

The puzzle for this EFT is how it should capture the enormous number of degrees of freedom that are internal to the black hole, its perfect absorber properties and the Hawking radiation that comes with it. 
In \cite{Goldberger:2005cd, Galley:2009px, Galley:2012hx, Goldberger:2019sya} these are modelled by `integrating in' a large number of degrees of freedom, and in the hotspot model it is the $\chi^a$ fields that play this role. The drawback of this approach is the model-dependence that enters when choosing these extra degrees of freedom. Although the fluctuation-dissipation theorem implies that predictions in linear response do not depend on these details, it remains open the extent to which other predictions do, and if so whether the same might be true for low-energy black hole properties. Although the extra degrees of freedom can again be integrated out, they are not the traditional massive states of the usual Wilsonian treatments, and so can lead to actions with unusual properties including some forms of nonlocality. In companion papers \cite{Companion, Kaplanek:2021xxx} we use the hotspot model to explore some of these properties in an effort to ascertain the rules for such an EFT, and the extent to which locality and ordinary Wilsonian reasoning breaks down.

\begin{figure}[h]
\begin{center}
\includegraphics[width=100mm,height=50mm]{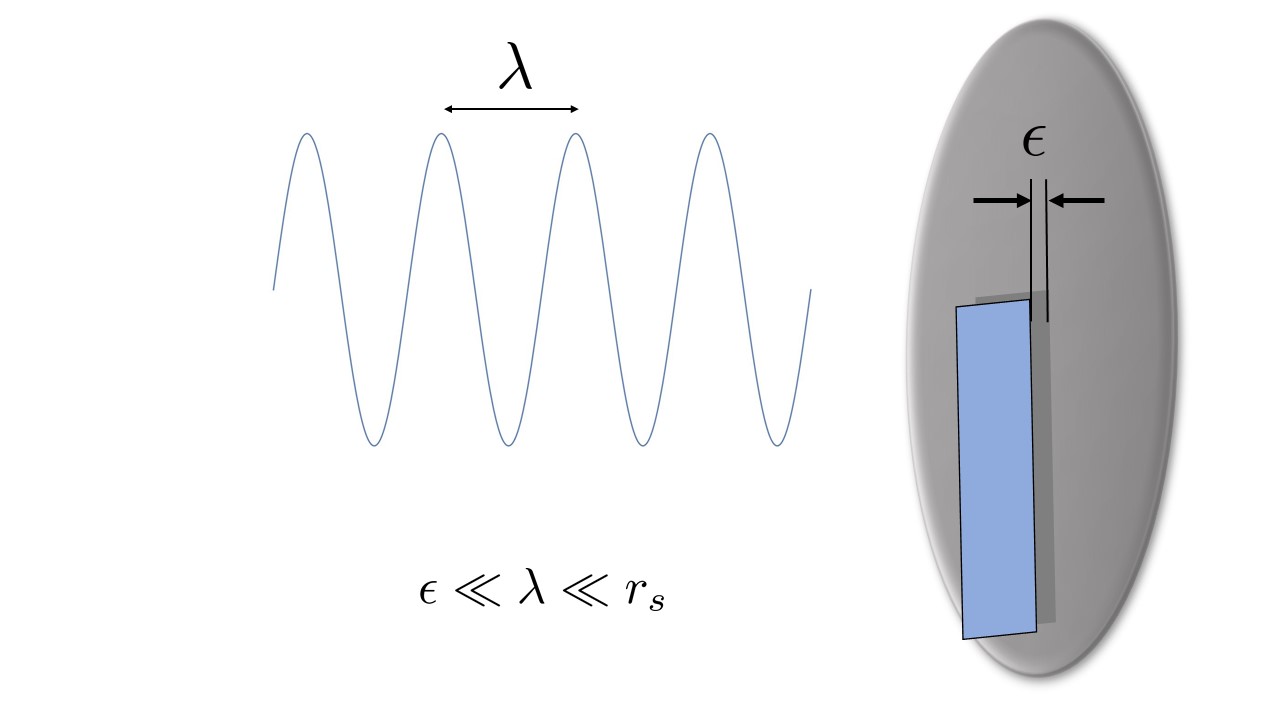} 
\caption{A `membrane' EFT regime for which hypothetical UV physics modifies near-horizon properties, for which the cutoff scale is much smaller than the length of any low-energy probe, which is in turn much smaller than the horizon scale $\epsilon \ll \lambda \ll r_s$.} \label{fig:Scale2Fig} 
\end{center}
\end{figure}

The second class of black hole models to which our hotspot setup can be relevant are those for which  probe scales, $\lambda$, and the UV cutoff length, $\epsilon$, are both much smaller than the horizon size, but where an effective description -- whether of conventional \cite{Price:1986yy, Thorne:1986iy, Damour:1978cg, Parikh:1997ma, Donnay:2019jiz} or more exotic \cite{Cardoso:2016rao, Abedi:2016hgu, Holdom:2016nek, Cardoso:2017cqb, Bueno:2017hyj, Mark:2017dnq, Conklin:2017lwb, Berti:2018vdi, Zhou:2016hsh} physics -- is envisaged to apply sufficiently near the event horizon (see Figure \ref{fig:Scale2Fig}). The beginnings of an EFT treatment of this kind of physics are developed in \cite{Burgess:2018pmm, Rummel:2019ads}, and involves an effective 3-dimensional action defined on a membrane that shrink-wraps the world-tube a distance $\epsilon$ from the black hole event horizon. EFT methods underline that the microscopic length $\epsilon$ is a regulator scale and so drops out of all physical predictions (as regulators always do), and this makes the EFT framework particularly useful for understanding the physical significance\footnote{In particular, the relevant physical scale involves both couplings and the intrinsic UV length scale, and so for weak coupling is often smaller than are the physical length scales of any micro-physics that may be involved \cite{Burgess:2018pmm, Rummel:2019ads}.} of the length-scales involved in these types of models. This type of EFT can be studied within the hotspot framework by allowing the radius $\xi$ of the interaction sphere $\cS_\xi$ remain larger than the cutoff scale: $\epsilon \ll \lambda \ll \xi$. We do not pursue this variant further in this paper.

The remainder of this paper sets up the hotspot framework and derives the equations that govern how $\phi$ responds to the hotspot (topics that dominate the discussion of \S\ref{sec:setup}). Along the way we also make some preliminary explorations of its physical implications (with more to follow in \cite{Companion, Kaplanek:2021xxx}). We find in particular the following noteworthy properties.
\begin{itemize}
\item
The field equations satisfied by the Heisenberg-picture field $\phi$ are solved explicitly under the assumption that the hotspot couplings $g_a$ of \pref{MixDef} turn on suddenly at time $t = 0$ and remain constant thereafter. The result is first computed perturbatively in the hotspot coupling $g_a$ in \S\ref{sec:LargeNHeis}, and then as an exact expression in \S\ref{sec:exactcorr}. Using the mode expansion of \pref{linearexp1} and \pref{linearexp2} our perturbative solution for the mode functions appearing in $\phi$ is given in \pref{S0order} and \pref{sa0order}, while the exact result is given in \pref{Ssol} and \pref{sasol}. The quantity $\tilde g$ appearing in these expressions is defined by $\tilde g^2 := \sum_a g_a^2 = N g^2$. 

Using the Heisenberg picture allows us to work in position space where we can follow the passage of the initial transient wave (generated by the turn-on of the couplings) as well as watch how the $\phi$ field settles down at later times in the on-going presence of the hotspot coupling. Computing both exact and perturbative results allows us to identify precisely which small dimensionless parameter controls the perturbative expansion.
\item
The Heisenberg-picture evolution is used to compute the Wightman function $\cW(t,\bfx; t', \bfx') = \langle \phi(t,\bfx) \, \phi(t',\bfx') \rangle$ for the external field, assuming the $\phi$ field starts in its vacuum at $t=0^-$ and the $\chi^a$ fields are prepared in a thermal state. The perturbative result is given in \pref{pertcorr} while the exact expression is in \pref{fullW}, \pref{curlySanswer} and \pref{curlyEanswer}. These results are computed for arbitrary spacetime separations for the fields, but we also obtain specific formulae for the regime where $t > |\bfx|$ and $t' > |\bfx'|$, but $t-t'$ is otherwise arbitrary. 

This result has a thermal character (in the sense that its temperature-dependent part satisfies a detailed-balance relation -- the Kubo-Martin-Schwinger (or KMS) condition \cite{Kubo,Martin} -- though does so in a way that depends on the distance from the hotspot. 
\item
\S\ref{sec:BC} detours to explore the consequences of supplementing the basic hotspot interaction of \pref{MixDef} with a self-coupling, also localized at the hotspot, having the form
\be \label{lambdaDef}
   \cL_{\lambda} = -\frac{\lambda}2  \,  \phi^2 \,.
\ee
Including this coupling is not simply an intellectual exercise because its presence is often required to renormalize divergences that arise because fields like $\phi$ diverge at the hotspot position once couplings are turned on there. As is well-known from other contexts \cite{Goldberger:2001tn, deRham:2007mcp, Burgess:2008yx, Bayntun:2009im, PPEFT, PPEFT2, PPEFT3, PPEFTDis, PPEFTHe, PPEFTH1, PPEFTH2} having fields divergence at the position of a source like this is fairly generic --- the simplest example being the Coulomb potential diverging at the position of a source charge. From an EFT perspective the presence of couplings like $\lambda$ is often compulsory, because the requirement that UV divergences drop out of physical observables causes the couplings to run and $\lambda = 0$ need not be a fixed point of this renormalization-group (RG) flow. 

\S\ref{sec:BC} computes the renormalization-group evolution implied for the coupling $\lambda$ in the hotspot model, along the way showing how this can be used to resum contributions to all orders in $\lambda(\epsilon)/4\pi \epsilon$ (where $\epsilon$ is a near-hotspot regularization scale) along the lines explored in \cite{PPEFT, PPEFT2}. The $\lambda$-dependence of the Wightman function is also given in the general expressions quoted above, and comparison with the exact result --- {\it c.f.}~eqs.~\pref{fullW}, \pref{curlySanswer} and \pref{curlyEanswer} --- verifies how the RG resummation captures the $\lambda$-dependence of the full expression. 
\end{itemize}
Finally, \S\ref{sec:Conclusions} briefly summarizes some of our conclusions and discusses some directions for future work. Many of the calculational details are given in a collection of appendices.

\section{Modelling the hotspot}
\label{sec:setup}

This section sets up the benchmark model whose properties we study. We do so using the language of open systems, with degrees of freedom divided up into an observable system and an `environment' --- a proxy for the black hole interior --- whose properties are never measured.

\subsection{Hotspot definition}
\label{ssec:HotspotDef}

For the observable sector we choose a single real scalar field, $\phi(x)$, and take it to live in a spatial region, $\cR_+$, of infinite extent. The environment is given by $N$ real scalar fields, $\chi^{a}$ with $a=1,\cdots,N$, that reside in a different spatial region $\cR_-$. While one or both of $\cR_+,\cR_-$ could in principle be curved, we take them to be flat for simplicity. We also take all of these fields to be massless. 

We suppose that the fields interact with one another locally and only do so on a relatively small codimension-1 2-sphere, $\cS_\xi$, of radius $\xi$ which is the only place where $\cR_+$ and $\cR_-$ actually touch one another (see Fig.~\ref{fig:FunnelFig}). In practice this means that both $\cR_+$ and $\cR_-$ have a small sphere excised from the origin (for all time) and the surface of this sphere is identified in the two spaces. 

Our interest for much of this paper is in scales much larger than $\xi$ and so consider the idealization of taking $\xi \to 0$, in which case $\cS_\xi$ reduces to a single point of contact between $\cR_+$ and $\cR_-$, which we take to be the origin $\bfx = \mathbf{0}$ of both $\cR_\pm$. In this limit the couplings of $\phi$ to $\chi^a$ are captured by an effective action localized at $\bfx = 0$. 

\subsubsection{Action and Hamiltonian}

The action that defines the model is therefore taken to be $S = S_+ + S_- + S_{\rm int}$ where the kinematics of $\phi$ and $\chi^a$ are described by
\be \label{Naction}
S_+ = - \frac{1}{2} \int_{\cR_+} \exd^{4}x \; \partial_{\mu} \phi \, \partial^{\mu} \phi \quad\hbox{and}\quad
S_- =  -\frac12 \int_{\cR_-} \exd^4x\; \delta_{ab} \, \partial_{\mu} \chi^{a} \partial^{\mu} \chi^{b}  \,.
\ee
Our later interest is usually in the case where the $\chi^{a}$ couplings do not break the $O(N)$ symmetry of their kinetic terms. 

The lowest-dimension interaction (mixing, really) that involves $\phi$ on the interaction surface is given by
\be \label{Sint3}
S_{\mathrm{int}} = - \int_{\cS_\xi^t} \exd^4 x \; \left[ G_a\,  \chi^{a}  \phi + \frac{G_\phi}2 \, \phi^2 \right]  \,,
\ee
in which the integration is over the world-tube, $\cS_\xi^t$ swept out by the surface $\cS_\xi$ over time. The Einstein summation convention applies, so there is an implied sum over $a$. The couplings $G_a$ and $G_\phi$ here have dimension mass: $[G_a] = [G_\phi] = +1$. 

In the limit $\xi \to 0$ the 2-sphere $\cS_\xi$ degenerates to a point and this interaction becomes
\be \label{Sint}
S_{\rm int} \simeq   - \int   \exd t \;\left[  g_a \, \chi^{a}(t,\mathbf{0}) \,  \phi(t,\mathbf{0}) + \frac{\lambda}2 \, \phi^2(t,\mathbf{0}) \right] \,,  
\ee
where the integration is over the proper time of the interaction point $\bfx = 0$ in both $\cR_+$ and $\cR_-$. The couplings appearing here are $g_a = 4\pi \xi^2 G_a$ and $\lambda = 4\pi \xi^2 G_\phi$ and have dimensions $[g_a] = [\lambda] = -1$. Although the coupling $\lambda$ might seem unnecessary, in later sections we see how it can be generated by the presence of the couplings $g_a$.

In what follows we allow the couplings $g_a$ and $\lambda$ to depend on time, and in particular will use this time dependence to turn on suddenly the interaction between the fields at $t = 0$. Doing so allows us both to study transient effects associated with the couplings turning on as well as late-time effects after the transients have passed.

The quantization of this model follows closely the treatment of a field coupled to a central qubit given in \cite{Lin:2005uk}. The canonical momenta for this problem are 
\bea
\mfp  := \partial_t  {\phi}   \quad \quad   \mathrm{and}\quad  \quad \Pi_a  := \delta_{ab} \, \partial_t  {\chi}^b  \ , \label{canmom}
\eea
and quantization proceeds by demanding these satisfy the equal-time commutation relations 
\be \label{CCR1}
\Bigl[  {\phi} (t,\bfx) ,  {\mfp} (t,\bfy) \Bigr] = i \delta^{3}(\bfx - \bfy) \quad \hbox{and} \quad
\Bigl[  { \chi} ^{a}(t,\bfx) ,  {\Pi}_b(t, \bfy) \Bigr] = i \delta^{a}_b \delta^{3}(\bfx - \bfy) \,.  
\ee

The free Hamiltonian is ${H}_0 :=  {\cH}_+ \otimes \cI_- + \cI_+ \otimes  {\cH}_-$, where $\cH_\pm$ and $\cI_\pm$ are the Hamiltonian and identity operators acting separately within the $\phi$- and $\chi$-sectors of the Hilbert space. Explicitly
\be
 {\cH}_{+}  := \frac12  \int_{\cR_+} \exd^3\bfx \; \Bigl[  {\mfp}^2 + \big( \boldsymbol{\nabla}  {\phi} \big)^2 \Bigr]    \quad\hbox{and}\quad
 {\cH}_{-}    :=   \frac12 \int_{\cR_-} \exd^{3}x \;\Bigl[ \delta^{ab} {\Pi}_a{\Pi}_b + \delta_{ab}   \boldsymbol{\nabla} {\chi}^{a} \cdot   \boldsymbol{\nabla} {\chi}^{b} \Bigr] \ .
\ee
The interaction Hamiltonian (in the limit of a point-like interaction surface) is similarly
\be 
 {H}_{\mathrm{int}}    =    g_a \, {\phi} (t,\mathbf{0}) \otimes  {\chi}^a (t, \mathbf{0}) + \frac{\lambda}2 \, \phi^2(t, \mathbf{0}) \otimes \cI_- \ .
\ee

\subsubsection{Initial conditions and the sudden approximation}

For later calculations we assume the state of the total system at $t = 0$ to be of the form
\be\label{initialstate} 
\rho(0) = \rho_+ \otimes \rho_- \,,
\ee
for separate density matrices $\rho_\pm$ in the two sectors. In general, interactions introduce correlations and so do not preserve this factorized form, and it is for this reason that we imagine the couplings between $\phi$ and $\chi^a$ to be initially absent, being turned on suddenly with 
\be
    g_a(t) = \Theta(t) \, g_a \,,
\ee
where $\Theta(t)$ is the Heaviside step function. This allows us to prepare initially uncorrelated states and then observe how the joint system reacts to the onset of coupling. 

In practice we choose the $\phi$ sector initially to be in its vacuum,
\be
\rho_+ = \ket{\vac} \bra{\vac}  \label{initialstate+}
\ee
where $\ket{\vac}$ is the standard Minkowski vacuum defined by $\mfa_\bfp \ket{\vac} = 0$. With eventual comparison to black holes in mind we take the $\chi^a$ sector to be in a thermal state,
\be \label{envthermstate}
\rho_- = \varrho_{\beta} := \frac{e^{ - \beta \cH_{-} } }{ \TrB[ e^{ - \beta \cH_{-}} ] }  \,,
\ee
with inverse temperature $\beta > 0$. The prime on the trace indicates that it is only taken over the $\chi$ sector. 

\subsection{Time evolution in different pictures}
\label{subsec:schropic}

Our goal is to solve for the time-evolution of the $\phi$-sector of the system and because $S_{\rm int}$ is bilinear in $\phi$ and $\chi^a$ the system's evolution can be evaluated in quite some generality. An exact solution is in particular given in \S\ref{sec:exactcorr}, after first detouring in \S\ref{sec:LargeNHeis} to describe an approximate solution that is evaluated perturbatively in the following combination of hotspot couplings
\be \label{gtildedef}
  \tilde g^2  := \delta^{ab} g_a g_b = N g^2 \,,
\ee
where the second equality specializes to the case where all couplings are equal. 

Although not required when solving the model, a large-$N$ limit can be defined wherein the coupling $\tilde g$ is held fixed (and need not itself be particularly small) as $N \to \infty$. This limit is briefly discussed in \S\ref{ssec:LargeNThermal}, where it is shown that the behaviour of the $\chi^a$ fields becomes particularly simple since they become oblivious to the presence of the $\phi$ field. The large-$N$ limit is not used elsewhere in this paper, besides in \S\ref{ssec:LargeNThermal}. 

\subsubsection{Interaction picture}

For perturbative evaluation we first diagonalize the free Hamiltonian. This is done in the usual way, by writing (with time-dependence as appropriate for the interaction picture)
\be \label{phiexpINT} 
 {\phi}(x) = \int \frac{\exd^{3}p}{ \sqrt{ (2\pi)^3 2 E_p } } \big[ e^{  i  p \cdot  x}  {\mfa}_{\bfp} + e^{- i  p \cdot  x}  {\mfa}^{\ast}_{\bfp} \big]  \quad \hbox{and} \quad
 {\chi}^{a}(x) = \int \frac{\exd^{3}p}{ \sqrt{ (2\pi)^2 2 E_p } } \big[ e^{ ip \cdot  x}  {\mfb}^{a}_{\bfp} + e^{-i p \cdot  x}  {\mfb}^{a\ast}_{\bfp} \big]  
\ee
where $p \cdot x := p_\mu x^\mu = - E_p t + \bfp \cdot \bfx$ with $E_{p} \ = \ |\mathbf{p}|$, and the canonical commutation relations imply the usual creation- and annihilation-operator algebra: $[{\mfa}_{\bfp}, {\mfa}_{\bfq} ] =[{\mfb}_{\bfp}^{a}, {\mfb}_{\bfq}^{b} ] = 0$ (together with their adjoints) and $[{\mfa}_{\bfp}, {\mfa}_{\bfq}^{\ast} ] = \delta^{3}(\bfp - \bfq)$ while $[{\mfb}_{\bfp}^{a}, {\mfb}_{\bfq}^{b\ast} ] = \delta^{ab} \delta^{3}(\bfp - \bfq)$. This diagonalizes ${\cH}_{\pm}$:
\bea
 {\cH}_{+}   =   \int \exd^{3}p  \; \frac{ E_p}{2} \bigg[  {\mfa}_{\bfp}  {\mfa}_{\bfp}^{\ast} +  {\mfa}_{\bfp}^{\ast}  {\mfa}_{\bfp} \bigg] 
\quad \hbox{and} \quad
 {\cH}_{-}  =  \int \exd^{3}p  \; \frac{ E_p}{2} \delta_{ab} \bigg[  {\mfb}^{a}_{\bfp}  {\mfb}_{\bfp}^{b\ast} +  {\mfb}_{\bfp}^{a \ast}  {\mfb}^{b}_{\bfp} \bigg] \ .
\eea

The interaction-picture interaction Hamiltonian in the pointlike limit ($\xi \to 0$) similarly becomes
\bea
 {H}_{\mathrm{int}}(t) &  = &    g_a   {\phi}(t, \mathbf{0}) \otimes  {\chi}^a(t, \mathbf{0}) + \frac{\lambda}{2} \, \phi^2(t,\mathbf{0}) \otimes \cI_{-} \nn\\
& = &  \int \frac{\exd^{3}p\; \exd^{3}q}{2(2\pi)^3  \sqrt{E_{p}E_{q}}} \; \Bigl[  g_{a}  \big(  {\mfa}_{\bfp} e^{- i E_p t}  + {\mfa}^{\ast}_{\bfp} e^{+ i E_p t}  \big) \otimes \big(  {\mfb}^{a}_{\bfq} e^{- i E_q t}  + {\mfb}^{a\ast}_{\bfq} e^{+ i E_q t}  \big) \\
&&\qquad\qquad\qquad\qquad\qquad + \frac{\lambda}{2} \big(  {\mfa}_{\bfp} e^{- i E_p t}  + {\mfa}^{\ast}_{\bfp} e^{+ i E_p t}  \big)   \big(  {\mfa} _{\bfq} e^{- i E_q t}  + {\mfa}^{ \ast}_{\bfq} e^{+ i E_q t}  \big) \otimes \cI_{-} \Bigr] \,. \nn
\eea
Matrix elements of this can be used in standard fashion to compute the evolution of the system's state.

\subsubsection{Heisenberg picture}
\label{subsec:heispic}

Later sections solve explicitly for time evolution, and do so by solving how the fields evolve in Heisenberg picture, including the effects of the couplings in $H_{\rm int}$. To this end it is worth briefly setting up the Heisenberg picture quantities and in particular exposing differences from the interaction-picture description given above.

Keeping in mind that we later entertain time-dependent couplings, $g_a(t)$, the full time-evolution operator ${U}(t,t')$ can be defined as the solution to $\partial_t {U}(t,t_0) = - i  {H} (t)  {U}(t,t_0)$ that satisfies $U(t=t_0) = \cI$. This leads to the usual time-ordered form
\be 
 {U}(t,t_0)   =   \cT\exp\left( - i \int_{t_0}^t \exd s\  {H} (s) \right) \ . \label{fullU}
\ee
It is this transformation that is used to construct time-dependent Heisenberg-picture operators, $A_\ssH(t)$, from Schr\"odinger-picture operators, ${A}_{\ssS}$, using: 
\be\label{AHvsAS}
 A_{\ssH}(t) = U^{\ast}(t,0) A_{\ssS} U(t,0) \,. 
\ee
We assume here that the two pictures agree at $t = 0$. 

A virtue of transforming to the Heisenberg picture that the state does not evolve at all. In Heisenberg picture it is the field operators that carry the burden of any time evolution when computing correlation functions or transition amplitudes. This means that the factorized form \pref{initialstate} for $\rho$ can also be used at later times, ensuring that $\chi^a$-sector expectation values can always be taken using the thermal state \pref{envthermstate}. 

Eq.~\pref{AHvsAS} implies in particular that the Heisenberg picture field operators are given by
\be  \label{Hpicfield}
 {\phi}_{\ssH}(t,\bfx)   :=   {U}^{\ast}(t,0) \big[  {\phi}_{\ssS}(\bfx) \otimes  \cI_- \big]  {U}(t,0) \quad \hbox{and} \quad
 {\chi}^{a}_{\ssH}(t,\bfx)   :=  {U}^{\ast}(t,0) \big[ \cI_+ \otimes  {\chi}^{a}_{\ssS}(\bfx) \big]  {U}(t,0) \,,
\ee
and similarly for their conjugate momenta. An important conceptual point about this definition is that the presence of the interaction term in $H$ implies that the Heisenberg field operators do not only act separately on the two sectors of the Hilbert space. In particular, expansion of $\phi_\ssH(t,\bfx)$ in terms of creation and annihilation operators involve {\it both} $\mfa_\bfp$ and $\mfb^a_\bfp$, as does the expansion of the $\chi_\ssH^a(t,\bfx)$.

In later sections the time-evolution of the fields $\phi_\ssH$ and $\chi^a_\ssH$ is determined by explicitly integrating their Heisenberg-picture field equations. These express the differential version of \pref{AHvsAS}, 
\be \label{HeisEOM}
\partial_t A_{\ssH}(t) = - i U^{\ast}(t,0) [ A_{\ssS} , H_{\ssS}(t) ] U(t,0) = - i [ A_{\ssH}(t) , H_{\ssH}(t) ] \ .
\ee
To work out the implications of \pref{HeisEOM} for the field operators explicitly we first record the following Schr\"odinger-picture commutators with the full Hamiltonian 
\be 
-i \Bigl[  \phi _{\ssS}(\bfx) , H_{\ssS}(t) \Bigr] = \mfp_{\ssS}(\bfx)  \,, \qquad
-i \Bigl[  {\chi}^a_{\ssS}(\bfx), H^{\ssS}(t) \Bigr] =  {\Pi}^{a}_{\ssS}(\bfx) \,,
\ee
\be 
-i \Bigl[  \Pi^{a}_{\ssS}(\bfx) , H_{\ssS}(t) \Bigr] =  \nabla^2 \chi_{\ssS}^{a}(\bfx) - g_a   \delta^{3}(\bfx) \; \phi _{\ssS}(\mathbf{0})   \nn
\ee
and
\be 
-i \Bigl[ \mfp_{\ssS}(\bfx)  , H_{\ssS}(t) \Bigr] = \nabla^2 \phi _{\ssS}(\bfx) - \delta^{3}(\bfx) \bigg(   g_a \;   \chi_{\ssS}^{a}(\mathbf{0}) + \lambda \phi _{\ssS}(\mathbf{0})    \bigg) \,.
\ee
Using these in \pref{HeisEOM} yields the equations of motion\footnote{Use of \pref{HeisEOM} assumes no further time-dependence arises through a time-dependence of couplings after they are initially turned on, which amounts to assuming the `sudden' approximation when turning on couplings at $t = 0$.}
\be
( - \partial_t^2 + \nabla^2 ) \phi _{\ssH}(t,\bfx) = \delta^{3}(\bfx) \bigg[ \lambda \phi _{\ssH}(t,\mathbf{0}) +  g_a \chi_{\ssH}^a(t,\mathbf{0}) \bigg] \label{heis1}  
\ee
and
\be
\delta_{ab} ( - \partial_t^2 + \nabla^2 ) \chi^{b}_{\ssH}(t,\bfx) = \delta^{3}(\bfx) \; g_a \phi _{\ssH}(t,\mathbf{0}) \label{heis2} \,.
\ee

These equations can be solved because they are linear in all of the fields, a consequence of $H_{\rm int}$ describing more of a mixing between $\phi$ and $\chi^a$ than an honest-to-God interaction. It is convenient to do so by first expanding the fields in terms of mode functions and then using the field equations to set up a coupled series of linear differential equations. That is, writing 
\be
   \phi_\ssH(t,\bfx) = \phi(t,\bfx) + \hat\phi(t,\bfx) \quad\hbox{and} \quad
   \chi^a_\ssH(t,\bfx) = \chi^a(t,\bfx) + \hat \chi^a(t,\bfx) \,,
\ee
with $\phi$ and $\chi^a$ being the interaction-picture fields given by \pref{phiexpINT}, then the deviations from the interaction picture are
\be \label{linearexp1}
\hat\phi(t,\bfx) = \int \frac{\exd^{3}p}{\sqrt{(2\pi)^3 2 E_p } } \left\{ \Bigl[   S_{\bfp}(t,\bfx)  \mfa_{\bfp} +  S_{\bfp}^{\ast}(t,\bfx)   \mfa^{\ast}_{\bfp} \Bigr] \; \otimes \cI_-  + \; \cI_+ \otimes \; \delta_{ab}\Bigl[ s^{a}_{\bfp}(t,\bfx) \; \mfb^{b}_{\bfp} + s_{\bfp}^{a\ast}(t,\bfx) \; \mfb^{b\ast}_{\bfp} \Bigr] \right\} 
\ee
and  
\be \label{linearexp2}
\hat \chi^{a}(t,\bfx) = \int \frac{\exd^{3}p}{\sqrt{(2\pi)^3 2 E_p } } \left\{ \Bigl[ R^{a}_{\bfp}(t,\bfx) \; \mfa_{\bfp} + R^{a}_{\bfp}(t,\bfx)^{\ast} \; \mfa^{\ast}_{\bfp} \Bigr]  \; \otimes \cI_-   +   \cI_+ \otimes \; \delta_{bc} \Bigl[  r^{ab}_{\bfp}(t,\bfx) \mfb^{c}_{\bfp} +  r_{\bfp}^{ab\ast}(t,\bfx)  \mfb^{c\ast}_{\bfp} \Bigr] \right\}
\ee
where the to-be-determined mode functions $\{ S_\bfp, s_{\bfp}^a, R^{a}_{\bfp}, r^{ab}_{\bfp} \}$ vanish in the absence of $H_{\rm int}$.

Inserting \pref{linearexp1} and \pref{linearexp2} into the Heisenberg equations of motion (\ref{heis1}) and (\ref{heis2}) leads to the following set of coupled equations for the mode functions $\{ S_\bfp, s_{\bfp}^a, R^{a}_{\bfp}, r^{ab}_{\bfp} \}$:
\bea \label{ModeEOMs}
\left( - \partial_t^2 + \nabla^2 \right) S_{\bfp}(t,\bfx) & = &  \bigg[ \lambda \Bigl( e^{-i E_p t} + S_{\bfp}(t,\mathbf{0}) \Bigr) + g_b R_{\bfp}^{b}(t,\mathbf{0})  \bigg] \delta^3(\bfx)  \nn\\
\left( - \partial_t^2 + \nabla^2 \right) s^{a}_{\bfp}(t,\bfx) & = & \bigg[ \lambda \, s^{a}_{\bfp}(t,\mathbf{0}) + \delta^{ab} g_{b} \, e^{-i E_p t} +  g_b \, r^{ba}_{\bfp}(t,\mathbf{0})  \bigg]  \delta^3(\bfx) \\
\delta_{ab} \left( - \partial_t^2 + \nabla^2 \right) R^{b}_{\bfp}(t,\bfx) & = & g_{a} \Bigl[ e^{-i E_p t} + S_{\bfp}(t,\mathbf{0}) \Bigr]   \delta^3(\bfx) \nn\\
\delta_{ac}\left( - \partial_t^2 + \nabla^2 \right) r^{cb}_{\bfp}(t,\bfx) & = &  g_{a} s^b_{\bfp}(t,\mathbf{0}) \,  \delta^3(\bfx) \,.\nn
\eea

\subsection{Integrating out $\chi^a$}

We wish to understand how the $\phi$ field responds to the presence of the hotspot, and we do so under the assumption that no measurements directly involve the fields $\chi^a$. Because no $\chi^a$ measurements are made the $\chi^a$ mode functions can be solved as functions of the $\phi$ mode functions to obtain a reduced set of equation to solve.

To see how this works explicitly consider preparing the $\phi$ field in its vacuum and then suddenly turn on hotspot couplings at $t = 0$. This should generate a flurry of transient behaviour before the $\phi$ field settles down at late times into a new adiabatic vacuum whose properties we wish to compute. To this end write $g_a(t) = g_a \, \Theta(t)$ and $\lambda(t) = \lambda \, \Theta(t)$, and so the time-dependence of eqs.~\pref{ModeEOMs} can be made more explicit:
\bea \label{ModeEOMsSudden1}
\left( - \partial_t^2 + \nabla^2 \right) S_{\bfp}(t,\bfx) & = & \Theta(t) \bigg[ \lambda \big[ e^{-i E_p t} + S_{\bfp}(t,\mathbf{0}) \big] + g_a  R_{\bfp}^{a}(t,\mathbf{0})  \bigg] \delta^3(\bfx) \\
\delta_{ab} \left( - \partial_t^2 + \nabla^2 \right) R^{b}_{\bfp}(t,\bfx) & = & \Theta(t) \, g_a \Bigl[ e^{-i E_p t} + S_{\bfp}(t,\mathbf{0}) \Bigr]   \delta^3(\bfx) \nn
\eea
and
\bea \label{ModeEOMsSudden2}
\left( - \partial_t^2 + \nabla^2 \right) s^{a}_{\bfp}(t,\bfx) & = &\Theta(t) \bigg[ \lambda \,s^{a}_{\bfp}(t,\mathbf{0}) + \delta^{ab} g_b\, e^{-i E_p t} +  g_b  \, r^{ba}_{\bfp}(t,\mathbf{0})  \bigg]  \delta^3(\bfx) \\
\left( - \partial_t^2 + \nabla^2 \right) r^{ab}_{\bfp}(t,\bfx) & = & \Theta(t)\, \delta^{ac} g_c s^b_{\bfp}(t,\mathbf{0}) \,  \delta^3(\bfx) \,. \nn
\eea
These are to be solved subject to the initial conditions
\bea \label{ICs}
  &&S_{\bfp}(0,\bfx) = \pd_t S_{\bfp}(0,\bfx) =  s^{a}_{\bfp}(0,\bfx)  =  \pd_t s^{a}_{\bfp}(0,\bfx) = 0\nn\\
 \hbox{and} \quad &&R^{a}_{\bfp}(0,\bfx) = \pd_t R^{a}_{\bfp}(0,\bfx) =  r^{ab}_{\bfp}(0,\bfx)  =  \pd_t r^{ab}_{\bfp}(0,\bfx)  =  0  \,.
\eea

\subsubsection{Solving the $\chi^a$ equations}

The mode functions associated with $\chi^a$ can be eliminated from the coupled equations \pref{ModeEOMsSudden1} and \pref{ModeEOMsSudden2} with initial conditions \pref{ICs} by using the retarded propagator
\be
G_{\ssR}(t,\bfx ; t', \bfy) = \frac{\Theta(t - t')}{4 \pi |\bfx - \bfy|}\, \delta\Bigl[ (t - t') - |\bfx - \bfy| \Bigr] \ , \label{GRprop}
\ee
that satisfies the equation of motion 
\be
\left( - \partial_t^2 + \nabla^2 \right) G_{\ssR}(t,\bfx ; t', \bfy) = - \delta(t - t') \delta^3(\bfx - \bfy) \ . \label{GReom}
\ee
In terms of this the formal solutions for $R^{a}_{\bfp}$ and $r^{ab}_{\bfp}$ (the mode functions appearing in $\chi^a$) are
\bea \label{IntRa}
R_{\bfp}^a(t,\bfx) & = & - \delta^{ab} g_b \int_0^\infty \exd s\; G_{\ssR}(t,\bfx ; \tau, \mathbf{0})   \Bigl[ e^{-i E_p \tau} + S_{\bfp}(\tau,\mathbf{0}) \Bigr]  \nn\\
& = & - \delta^{ab} g_b  \frac{\Theta(t - |\bfx|)}{4 \pi |\bfx|} \Bigl[ e^{-i E_p (t - |\bfx|)} + S_{\bfp}(t - |\bfx|,\mathbf{0}) \Bigr]
\eea
and
\be \label{Inrab}
r^{ab}_{\bfp}(t, \bfx) = - \delta^{ac} g_c \frac{\Theta(t - |\bfx|)}{4 \pi |\bfx|} s^{b}_{\bfp}(t-|\bfx|,\mathbf{0}) \,.
\ee
These solutions have support only in the forward lightcone of the event where the couplings turn on, and there give the mode functions at a distance $r = |\bfx|$ from the hotspot in terms of their values at the hotspot position, but as a function of the retarded time $t_r := t - r$ and with an amplitude that is suppressed by a power of $1/r$. 

Using these solutions to eliminate $R^a_{\bfp}$ and $r^{ab}_{\bfp}$ from \pref{ModeEOMsSudden1} and \pref{ModeEOMsSudden2} leaves a coupled set of equations involving only the mode functions appearing in $\phi$:
\bea \label{Seliminated}
\left( - \partial_t^2 + \nabla^2 \right) S_{\bfp}(t,\bfx) & = & \Theta(t) \bigg( \lambda \big[ e^{-i E_p t} + S_{\bfp}(t,\mathbf{0}) \big] \\
& \ & \qquad \qquad \qquad  - \frac{\tilde{g}^2 \Theta(t - |\bfy|)}{4 \pi |\bfy|} \Bigl[ e^{-i E_p (t - |\bfy|)} + S_{\bfp}(t - |\bfy|,\mathbf{0}) \Bigr]  \bigg|_{|\bfy| = 0} \bigg) \; \delta^3(\bfx) \nn
\eea
as well as 
\bea \label{saeliminated}
\left( - \partial_t^2 + \nabla^2 \right) s^{a}_{\bfp}(t,\bfx) & = &\Theta(t) \bigg[ \lambda \,s^{a}_{\bfp}(t,\mathbf{0}) + \frac{\tilde{g}}{\sqrt{N}} \, e^{-i E_p t} \\
& \ & \qquad \qquad \qquad - \frac{\tilde{g}^2 \Theta(t - |\bfy|)}{4 \pi |\bfy|} s^a_{\bfp}(t - |\bfy|,\mathbf{0}) \bigg|_{|\bfy|=0} \bigg) \; \delta^3(\bfx) \nn
\eea
where we specialize to the case where all of the $g_a$'s have the same size, and use \pref{gtildedef} to write $g_a = \tilde g/\sqrt N$ for all $a$. The factor of $\sqrt N$ is extracted here for convenience because it cancels an explicit factor of $N$ that comes from the summation over the index `$a$' in \pref{Seliminated}.

Eqs.~\pref{Seliminated} and \pref{saeliminated}  reveal a characteristic `Coloumb' singularity as $|\bfy| \to 0$, which at face value appears to threaten any program to solve \pref{Seliminated} and \pref{saeliminated} iteratively as a series in $\tilde{g}^2$ and $\lambda$. In what follows, this divergence at $|\bfy|=0$ is regularized by instead evaluating $\bfy$ at the microscopically small scale $|\bfy| = \epsilon$. This divergence problem is a general issue that arises when exploring effective field theories describing compact sources, where the domain of validity of the low-energy/long-wavelength theory does not allow sufficient spatial resolution to resolve the source's structure; it is generic that external fields diverge at the position of a compact source. 

But the example of the Coulomb field for a small charge distribution also suggests that evaluating the $1/r$ divergence at $r = 0$ is really an artefact of trying to extrapolate to zero an external solution that is not actually appropriate in the microscopic theory within which the source's structure can be resolved. A general EFT treatment of these issues is possible \cite{PPEFT, PPEFT2, PPEFT3, PPEFTDis, EFTBook} (and tested in detail calculating nuclear finite-size effects in atoms \cite{PPEFTHe, PPEFTH1, PPEFTH2}), and shows how all such divergences get renormalized by the effective couplings in the action -- such as the coupling $\lambda$ of hotspot action \pref{Sint} --- that describes the source's low-energy properties (as we also see in detail below). 

\subsubsection{Renormalization of $\lambda$ and $\epsilon$-regularization}

Regulating the field equations on the microscopic surface $|\bfy| = \epsilon$ allows \pref{Seliminated} and \pref{saeliminated} to be rewritten 
\be \label{Seliminated2}
\left( - \partial_t^2 + \nabla^2 \right) S_{\bfp}(t,\bfx) =  \bigg( \Theta(t) \lambda \big[ e^{-i E_p t} + S_{\bfp}(t,\mathbf{0}) \big] - \frac{\tilde{g}^2 \Theta(t - \epsilon)}{4 \pi \epsilon} \Bigl[ e^{-i E_p (t - \epsilon)} + S_{\bfp}(t - \epsilon,\mathbf{0}) \Bigr]  \bigg) \; \delta^3(\bfx) 
\ee
and
\be \label{saeliminated2}
\left( - \partial_t^2 + \nabla^2 \right) s^{a}_{\bfp}(t,\bfx)=  \bigg(\Theta(t) \big[ \lambda \,s^{a}_{\bfp}(t,\mathbf{0}) + \frac{\tilde{g}}{\sqrt{N}} \, e^{-i E_p t} \big] - \frac{\tilde{g}^2 \Theta(t - \epsilon)}{4 \pi \epsilon} s^a_{\bfp}(t - \epsilon,\mathbf{0}) \bigg) \; \delta^3(\bfx)
\ee
where we use $\Theta(t) \Theta(t - \epsilon) = \Theta(t - \epsilon)$ since $\epsilon > 0$. 

These equations can also be formally integrated using the retarded propagator \pref{GRprop} to give 
\bea \label{Seliminated3}
S_{\bfp}(t,\bfx) &=&  - \frac{\lambda \Theta(t - |\bfx|) }{ 4 \pi |\bfx| } \Bigl( e^{-i E_p (t - |\bfx|)} + S_{\bfp}(t - |\bfx|,\mathbf{0}) \Bigr) \nn\\
&& \qquad\qquad\qquad\qquad\qquad + \frac{\tilde{g}^2 \Theta(t - |\bfx| - \epsilon)}{16 \pi^2 |\bfx| \epsilon} \Bigl[ e^{-i E_p (t - |\bfx| - \epsilon)} + S_{\bfp}(t -|\bfx| - \epsilon,\mathbf{0}) \Bigr]  \nn\\
& \simeq & - \bigg( \frac{\Theta(t - |\bfx|) }{ 4 \pi |\bfx| } \bigg[ \lambda - \frac{\tilde{g}^2}{4 \pi \epsilon}  \bigg] + \frac{\tilde{g}^2 \delta(t - |\bfx|)}{16 \pi^2 |\bfx| } \bigg) \big[ e^{-i E_p (t - |\bfx|)} + S_{\bfp}(t - |\bfx|,\mathbf{0}) \big] \\
&& \qquad \qquad \qquad \qquad\qquad \qquad - \frac{\tilde{g}^2 \Theta(t - |\bfx|)}{16 \pi^2 |\bfx|} \Bigl[ (- i E_p ) e^{-i E_p (t - |\bfx|)} + \partial_t S_{\bfp}(t -|\bfx| ,\mathbf{0}) \Bigr] \nn
\eea
and
\bea \label{saeliminated3}
s^{a}_{\bfp}(t,\bfx) &=&  - \frac{\Theta(t - |\bfx|)}{4 \pi |\bfx| } \big[ \lambda \, s^{a}_{\bfp}(t - |\bfx|,\mathbf{0}) + \frac{\tilde{g}}{\sqrt{N}} \, e^{-i E_p (t - |\bfx|)} \big] +\frac{\tilde{g}^2 \Theta(t - |\bfx| - \epsilon)}{16 \pi^2 |\bfx| \epsilon} s^a_{\bfp}(t - |\bfx| - \epsilon,\mathbf{0}) \nn\\
 & \simeq &  - \bigg( \frac{\Theta(t - |\bfx|)}{4 \pi |\bfx| } \bigg[ \lambda - \frac{\tilde{g}^2}{4 \pi \epsilon} \bigg] - \frac{\tilde{g}^2 \delta(t -|\bfx|)}{16\pi^2 |\bfx|} \bigg) \, s^{a}_{\bfp}(t - |\bfx|,\mathbf{0}) \\
&& \qquad \qquad \qquad \qquad \qquad \qquad  -  \frac{\tilde{g} \Theta(t - |\bfx|)}{4 \pi \sqrt{N} |\bfx| } \, e^{-i E_p (t - |\bfx|)} - \frac{\tilde{g}^2 \Theta(t - |\bfx|)}{16 \pi^2 |\bfx|} \partial_t s^a_{\bfp}(t - |\bfx| ,\mathbf{0}) \nn
\eea
where the approximate equalities exploit the fact that $\epsilon$ is a microscopic quantity to expand each of the last terms in powers of $\epsilon$, and dropping terms that are $\cO(\epsilon)$. Note that this expansion in $\epsilon$ implicitly assumes that $E_{p} \epsilon \ll 1$ for all modes when expanding the exponential function. 

Although the $1/\epsilon$ term diverges, this divergence can be absorbed by redefining  
\be \label{lambdaren}
\lambda_{\ssR} := \lambda - \frac{\tilde{g}^2}{4 \pi \epsilon} \ ,
\ee
showing that the divergence renormalizes the $\phi$ self-coupling $\lambda$. Dropping the subscript `$R$' for notational simplicity, eqs.~\pref{Seliminated3} and \pref{saeliminated3} become\footnote{In arriving at \pref{Seliminated5} and \pref{saeliminated5} we simplify the terms which come multiplied by $\delta(t - |\bfx|)$ by using the initial conditions to eliminate $S_{\bfp}(0,\mathbf{0})=s^a_{\bfp}(0,\mathbf{0})=0$ in the final result.}
\bea \label{Seliminated5}
S_{\bfp}(t,\bfx) & = & - \frac{\lambda \Theta(t - |\bfx|) }{ 4 \pi |\bfx| } \Bigl( e^{-i E_p (t - |\bfx|)} + S_{\bfp}(t - |\bfx|,\mathbf{0}) \Bigr) - \frac{\tilde{g}^2 \delta(t - |\bfx|)}{16 \pi^2 |\bfx| } \\
& \ & \qquad \qquad \qquad \qquad  - \frac{\tilde{g}^2 \Theta(t - |\bfx|)}{16 \pi^2 |\bfx|} \Bigl[ (- i E_p ) e^{-i E_p (t - |\bfx|)} + \partial_t S_{\bfp}(t -|\bfx| ,\mathbf{0}) \Bigr] \nn
\eea
and
\be \label{saeliminated5}
s^{a}_{\bfp}(t,\bfx) = - \frac{\tilde{g} \,\Theta(t - |\bfx|)}{4 \pi \sqrt{N} |\bfx| } \, e^{-i E_p (t - |\bfx|)} - \frac{\lambda \Theta(t - |\bfx|)}{4 \pi |\bfx| } s^{a}_{\bfp}(t - |\bfx|,\mathbf{0}) - \frac{\tilde{g}^2 \Theta(t - |\bfx|)}{16 \pi^2 |\bfx|} \partial_t s^a_{\bfp}(t - |\bfx| ,\mathbf{0})  \ .
\ee
These are the equations that are to be solved in the next sections to determine the mode functions for $\phi$, and from these also determine its response to the hotspot.

\section{Perturbative response}
\label{sec:LargeNHeis}

This section provides one of the points of comparison for the exact results of \S\ref{sec:exactcorr}. Here we solve eqs.~\pref{Seliminated5} and \pref{saeliminated5} iteratively in $\tilde g$ and $\lambda$, and use the lowest order solutions to determine perturbatively how the fields evolve in time.  

\subsection{Mode functions}

\begin{figure}
\centering
\includegraphics[height=60mm]{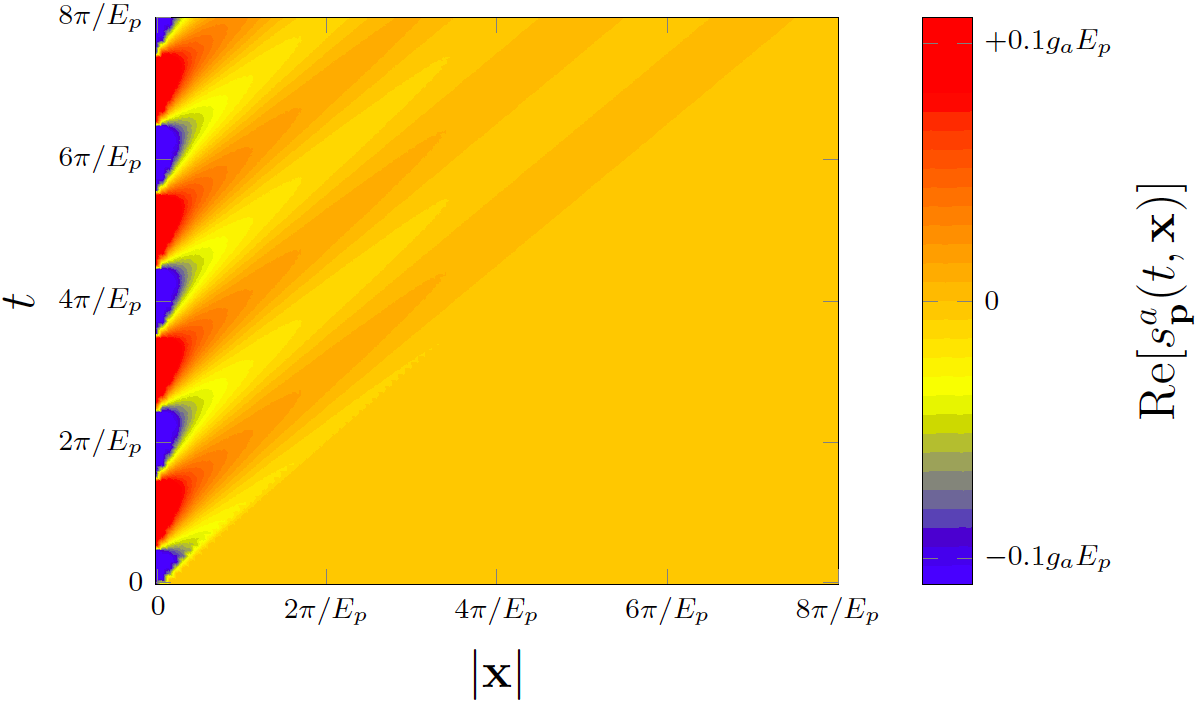}
\caption{Re$[s_\bfp^a(t,\bfx)]$ from \pref{sa0order} {\it vs} $t$ and $|\bfx|$, showing the wave-front emanating from $t=\bfx=0$, the growth for small $|\bfx|$ and the oscillatory behaviour with wavelength set by $E_p$. (Colour online.)}
\label{figureRespa}
\end{figure}

The iterative solution to \pref{Seliminated5} and \pref{saeliminated5} gives the perturbative result
\be \label{S0order}
S_{\bfp}(t,\bfx) \simeq  - \bigg( \lambda - \frac{ i \tilde{g}^2 E_p}{4\pi } \bigg) \frac{\Theta(t - |\bfx|)}{4 \pi |\bfx|} e^{- i E_{p}(t- |\bfx|)} - \frac{\tilde{g}^2 \delta(t - |\bfx|)}{16 \pi^2 |\bfx| } \qquad (\mathrm{perturbative})
\ee
and
\be \label{sa0order}
s^{a}_{\bfp}(t,\bfx) \simeq - \frac{\tilde{g} \,\Theta(t - |\bfx|)}{4 \pi \sqrt{N} |\bfx| } \, e^{-i E_p (t - |\bfx|)}  \qquad (\mathrm{perturbative})
\ee
to leading nontrivial order in $\lambda$ and $\tilde{g}$. The real part of the perturbative solution \pref{sa0order} is shown in Fig.~\ref{figureRespa}, which shows how the result is nonzero only after the passage of the wave-front that radiates out from the turn-on event at $t = \bfx = 0$.

Using these mode functions in the expansion for $\phi$, the leading-order perturbative limit of the Heisenberg-picture fields truncated at order $\tilde{g}^2$ can be written
\bea \label{phiHsolpert}
\phi_{\ssH}(t,\bfx) & \simeq & \bigg( \phi(t,\bfx) - \frac{\lambda \Theta(t - |\bfx|)}{4 \pi |\bfx|} \,\phi(t-|\bfx|,\mathbf{0}) - \frac{\tilde{g}^2\Theta(t - |\bfx|)}{16 \pi^2 |\bfx|} \,\mfp(t - |\bfx|,\mathbf{0})  - \frac{\tilde g^2 \delta(t - |\bfx|)}{16 \pi^2 |\bfx|} \,\phi(0,\mathbf{0}) \bigg) \otimes \cI_{-} \nn \\
&& \qquad \qquad \qquad\qquad \qquad \qquad  - \frac{\tilde{g} \,\Theta(t - |\bfx|)}{4 \pi \sqrt{N} |\bfx| }\sum_{a=1}^N \cI_{+} \otimes \chi^a(t-\bfx,\mathbf{0})
\eea
where (as above) $\phi$ and $\chi$ are the interaction-picture fields given in \pref{phiexpINT} and $\mathfrak{p} = \partial_t \phi$ is the canonical momentum defined in \pref{canmom}. The Heaviside step functions show how $\phi_{\ssH}(t,\bfx)$ does not respond to the turn-on of the hotspot couplings at $t = 0$ until after the transient wave reaches the particular point $\bfx$, after which mode interference occurs. The sum over $a$ is written explicitly in \pref{phiHsolpert} to underline the necessity of keeping this term, even in the large-$N$ limit despite the factor of $1/\sqrt{N}$.

\subsection{Two-point $\phi$ correlator}

The physical implications of the field evolution just calculated gets communicated to observables through field correlators, and because the model considered here is gaussian the two-point function carries all of this information. For observers situated in $\cR_+$ only the correlators of the field $\phi$ can be accessed, and so we therefore next compute the two-point correlator,
\be \label{CorrelationStart}
W_\beta(t,\bfx; t',\bfx') := \Tr\Bigl[ \phi _{\ssH}(t,\bfx) \phi _{\ssH}(t',\bfx') \rho_0 \Bigr] = \frac{1}{Z_\beta} \mathrm{Tr}\Bigl[ \phi _{\ssH}(t,\bfx) \phi _{\ssH}(t',\bfx') \big( \ket{\mathrm{vac}} \bra{\mathrm{vac}} \otimes  e^{ - \beta \cH_{-}} \big) \Bigr]  \,,
\ee
where $\rho_0$ denotes the system's state, for which we use the state given in \pref{initialstate}, \pref{initialstate+} and \pref{envthermstate}. $Z_\beta := \TrB\left[ e^{- \beta \cH_{-}} \right]$ is the partition function for the $N$ thermal $\chi^a$ fields. 

Using the perturbative solution for $\phi_{\ssH}$ given in \pref{phiHsolpert} allows the leading-order in $\tilde{g}^2$ and $\lambda$ contribution to be written in terms of the free correlation functions,
\bea
&\ & W_\beta(t,\bfx; t',\bfx') \simeq \langle \vac | \phi(t,\bfx) \phi(t',\bfx') | \vac \rangle \nn\\
&\ & \qquad \qquad  - \frac{\lambda \Theta(t-|\bfx|)}{4 \pi |\bfx|} \langle \vac | \phi(t-|\bfx|,\mathbf{0}) \phi(t',\bfx') | \vac \rangle - \frac{\lambda \Theta(t'-|\bfx'|)}{4 \pi |\bfx'|} \langle \vac | \phi(t,\bfx) \phi(t'-|\bfx'|,\mathbf{0}) | \vac \rangle \nn \\
&\ & \qquad \qquad + \frac{\tilde{g}^2 \Theta(t-|\bfx|) \Theta(t' - |\bfx'|)}{16 \pi^2 |\bfx| |\bfx'| } \, \TrB\left[ \chi^a(t-|\bfx|,\mathbf{0}) \chi^b(t'-|\bfx'|,\mathbf{0})  \varrho_{\beta} \right]  \\
&\ & \qquad \qquad - \frac{\tilde{g}^2 \Theta(t-|\bfx|)}{16 \pi^2 |\bfx|} \langle \vac | \mfp(t-|\bfx|,\mathbf{0}) \phi(t',\bfx') | \vac \rangle - \frac{\tilde{g}^2 \Theta(t'-|\bfx'|)}{16 \pi^2 |\bfx'|} \langle \vac | \phi(t,\bfx) \mfp(t'-|\bfx'|,\mathbf{0}) | \vac \rangle \nn \\
&\ & \qquad \qquad - \frac{\tilde{g}^2 \delta(t-|\bfx|)}{16 \pi^2 |\bfx|} \langle \vac | \phi(0,\mathbf{0}) \phi(t',\bfx') | \vac \rangle - \frac{\tilde{g}^2 \delta(t'-|\bfx'|)}{16 \pi^2 |\bfx'|} \langle \vac | \phi(t,\bfx) \phi(0,\mathbf{0}) | \vac \rangle \,.\nn
\eea
This can be simplified using the following explicit forms for the free correlators
\be
\langle \vac | \phi(t,\bfx) \phi(t',\bfx') | \vac \rangle = \frac{1}{4\pi^2 \big[ - (t - t' - i \delta)^2 + |\bfx - \bfx'|^2 \big]} \,,
\ee
\be
\langle \vac | \mfp(t,\bfx) \phi(t',\bfx') | \vac \rangle = \frac{t - t'}{2\pi^2 \big[ - (t - t' - i \delta)^2 + |\bfx - \bfx'|^2 \big]^2} \,,
\ee
and
\be
\langle \vac | \phi(t,\bfx)  \mfp(t',\bfx')  | \vac \rangle = - \langle \vac | \mfp(t,\bfx) \phi(t',\bfx') | \vac \rangle = \frac{- t + t'}{2\pi^2 \big[ - (t - t' - i \delta)^2 + |\bfx - \bfx'|^2 \big]^2} \,,
\ee
leading to 
\bea
&\ & W_\beta(t,\bfx; t',\bfx') \simeq \frac{1}{4\pi^2 \big[ - (t - t' - i \delta)^2 + |\bfx - \bfx'|^2 \big]}\nn \\ 
&\ & \qquad + \frac{\lambda}{16\pi^3} \bigg( \frac{\Theta(t-|\bfx|)}{|\bfx|} \frac{1}{(t-t'-|\bfx| - i \delta)^2 - |\bfx'|^2} + \frac{\Theta(t'-|\bfx'|)}{|\bfx'|} \frac{1}{(t-t'+|\bfx'| - i \delta)^2 - |\bfx|^2} \bigg) \nn \\
& \ & \qquad - \frac{\tilde{g}^2 \Theta(t-|\bfx|) \Theta(t' - |\bfx'|)}{64 \pi^2 \beta^2 |\bfx| |\bfx'| \sinh^2 \left[ \frac{\pi}{\beta}( t - |\bfx| - t' + |\bfx'| - i \delta ) \right]} \\
&\ & \qquad + \frac{\tilde{g}^2}{32 \pi^4} \bigg( - \frac{\Theta(t-|\bfx|)}{|\bfx|} \frac{t-t'-|\bfx|}{\big[ (t-t'-|\bfx| - i \delta)^2 - |\bfx'|^2 \big]^2} + \frac{\Theta(t'-|\bfx'|)}{|\bfx'|} \frac{t-t'+|\bfx'|}{\big[ (t-t'+|\bfx'| - i \delta)^2 - |\bfx|^2 \big]^2} \bigg) \nn \\
&\ & \qquad + \frac{\tilde{g}^2}{64\pi^4} \bigg( \frac{\delta(t - |\bfx|)}{|\bfx| \big[ - (t' + i \delta)^2 - |\bfx'|^2 \big]} + \frac{\delta(t' - |\bfx'|)}{|\bfx'| \big[ - (t - i \delta)^2 - |\bfx|^2 \big]} \bigg) \,, \nn
\eea
where the inverse temperature $\beta = 1/T$ arises from the thermal average in the $\chi$ sector.

Of particular interest is the form of this result after the passage of the transients, with both $t$ and $t'$ chosen to lie in the future light cone of the switch-on event ({\it i.e.} $t>|\bfx|$ and $t' > |\bfx'|$). In this region the above expression becomes
\bea  \label{pertcorr}
W_\beta(t,\bfx; t',\bfx') & \simeq & \frac{1}{4\pi^2 \big[ - (t - t' - i \delta)^2 + |\bfx - \bfx'|^2 \big]} + \frac{\lambda}{16\pi^3 |\bfx| |\bfx'|} \bigg[ \frac{|\bfx| + |\bfx'|}{(t - t' - i \delta)^2 - (|\bfx + |\bfx'|)^2 } \bigg] \nn\\
& \ & \quad - \frac{ \tilde{g}^2 }{64 \pi^2 \beta^2 |\bfx| |\bfx'| \sinh^2 \left[ \frac{\pi}{\beta}( t - |\bfx| - t' + |\bfx'| - i \delta ) \right]}  \\
&\ & \qquad + \frac{\tilde{g}^2}{32 \pi^4} \bigg( - \frac{1}{|\bfx|} \frac{t-t'-|\bfx|}{\big[ (t-t'-|\bfx| - i \delta)^2 - |\bfx'|^2 \big]^2} + \frac{1}{|\bfx'|} \frac{t-t'+|\bfx'|}{\big[ (t-t'+|\bfx'| - i \delta)^2 - |\bfx|^2 \big]^2} \bigg) \nn
\eea
As expected, in this limit the $\cO(\lambda)$ and $\cO(\tilde g^2)$ terms break translation invariance, though time-translation invariance is restored once the transients due to the coupling turn-on have passed. Rotations about the position of the hotspot remain a symmetry. Apart from a global $1/r$ fall-off the thermal $\cO(\tilde g^2)$ term depends only on the retarded times $t_r = t - |\bfx|$ and $t'_r = t' - |\bfy|$, with correlations that die exponentially once $t_r - t_r' \gg \beta$. By contrast, the temperature-independent $\cO(\tilde{g}^2)$ term --- and the $\cO(\lambda)$ contributions --- preserve the power-law fall-off for large $t - t'$, but modify its amplitude in a way that becomes less important further from the hotspot.

\begin{figure}
\centering
\begin{subfigure}{.5\textwidth}
  \centering
  \includegraphics[width=.9\linewidth]{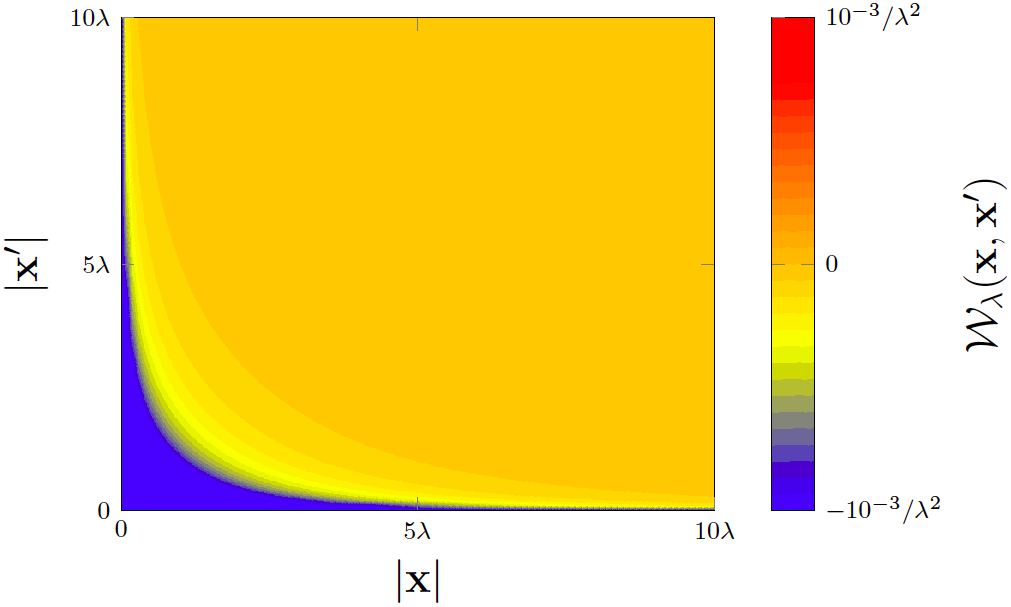}
  \caption{$\lambda$-dependent term}
  \label{fig:sub1}
\end{subfigure}%
\begin{subfigure}{.5\textwidth}
  \centering
  \includegraphics[width=.9\linewidth]{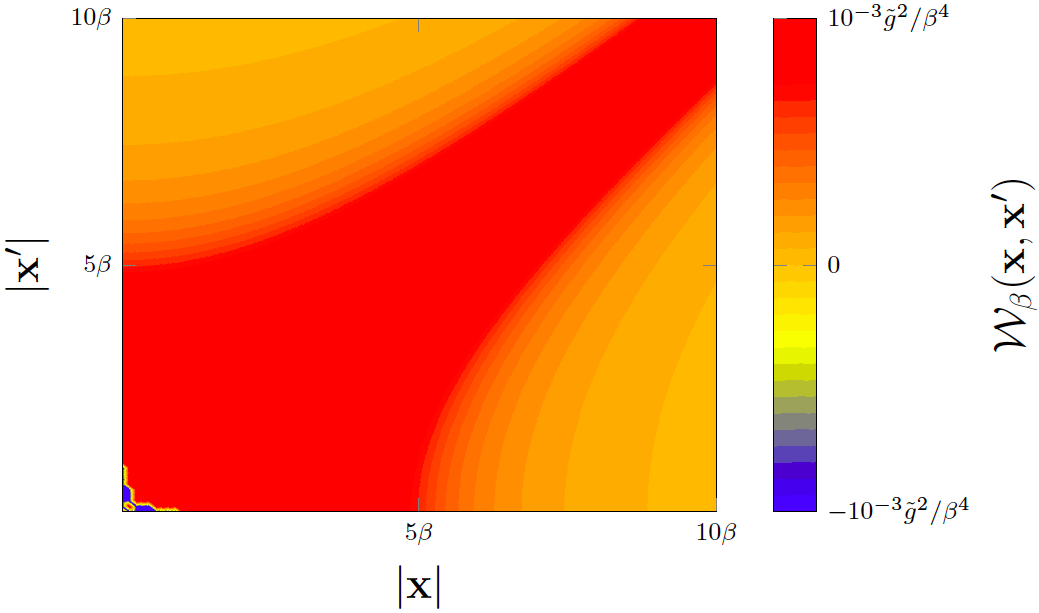}
  \caption{$\tilde g^2$-dependent term}
  \label{fig:sub2}
\end{subfigure}
\caption{The equal-time $t=t'$ limits of the $\lambda$-dependent term and $\tilde g^2$-dependent term of the Wightman function given in eq.~\pref{pertcorrequal}. (Colour online.) }
\label{figure:Wglambda}
\end{figure}

For some applications it is the equal-time correlator evaluated with $t=t'$ that is of interest (at late times $t> |\bfx|, |\bfx'|$). In this case the above simplifies to\footnote{Note that this formula has no $i \delta$'s left in it --- the reason for this is that any poles located at $|\bfx|+ |\bfx'|$ can safely have $\delta \to 0^{+}$ taken. For the remaining poles at $|\bfx| - |\bfx'|$, we use the identity $\frac{1}{(z \pm i \delta)^2} = \frac{1}{z^2} \pm i \pi \delta'(z)$ and notice that a cancellation occurs.}
\bea \label{pertcorrequal}
W_\beta(t,\bfx; t,\bfx') & \simeq & \frac{1}{4\pi^2 |\bfx - \bfx'|^2} - \frac{\lambda}{16\pi^3 |\bfx| |\bfx'| ( |\bfx| + |\bfx'| )}  \\
& \ & \qquad \qquad - \frac{\tilde{g}^2}{64 \pi^2 \beta^2 |\bfx| |\bfx'| \sinh^2 \left[ \frac{\pi}{\beta}( |\bfx| - |\bfx'| ) \right]} + \frac{\tilde{g}^2}{16 \pi^4 ( |\bfx|^2 - |\bfx'|^2 )^2} \nn
\eea
Notice \pref{pertcorrequal} is real-valued, as should be the case for unitary time-evolution.  The $\lambda$- and the $\tilde{g}^2$-dependent terms of \pref{pertcorrequal} are plotted in Fig.~\ref{figure:Wglambda}

\subsection{Two-point $\chi^a$ correlator in the large-$N$ limit}
\label{ssec:LargeNThermal}

Although the $\langle \phi \phi \rangle$ correlator does not simplify in the large-$N$ limit, the same is not true for $\langle \chi \chi \rangle$ correlations. This can be seen by inserting the perturbative formulae \pref{S0order} and \pref{sa0order} for $S_\bfp$ and $s^a_\bfp$ into the implicit solutions \pref{IntRa} and \pref{Inrab} for the $\chi^a$-mode functions $R_{\bfp}^a$ and $r_{\bfp}^{ab}$, which gives
\bea \label{Rapert}
R_{\bfp}^a(t,\bfx) & \simeq & - \frac{\tilde{g}}{\sqrt{N}} \cdot  \frac{\Theta(t - |\bfx|)}{4 \pi |\bfx|} e^{-i E_p (t - |\bfx|)} + \ldots
\eea
and
\be \label{rabpert}
r^{ab}_{\bfp}(t, \bfx) = \frac{\tilde{g}^2}{N} \cdot \frac{\Theta(t - |\bfx| - \epsilon)}{16 \pi^2 |\bfx| \epsilon} e^{ - i E_p (t - |\bfx| - \epsilon)} \, \ldots 
\ee
to leading order in $\tilde{g}^2$ and $\lambda$. The solution \pref{rabpert} contains a $1/\epsilon$ divergence, which can be absorbed into the coupling for a self-interaction proportional to $\cI_{+} \otimes \delta_{ab} \chi^{a} \chi^b$ (although for brevity we do not do so here). 

Eqs.~\pref{Rapert} and \pref{rabpert} show that the mixing of $\chi^a$ with $\phi$ is suppressed by powers of $1/N$, and so become negligible in the large-$N$ limit. The same suppression does not occur in the $\langle \phi \phi \rangle$ correlator because the explicit $1/N$ suppression is compensated by the sum over $a$ and $b$ in the combination $g_a g_b \langle \chi^a \,\chi^b\rangle$. It follows that these correlators satisfy
\bea
\mathrm{Tr}[ \chi_{\ssH}^{a}(t,\bfx) \chi_{\ssH}^b(t',\bfx') \rho_0 ] = \TrB\left[ \chi^{a}(t,\bfx) \chi^b(t',\bfx')  \varrho_{\beta} \right] + \cO(1/N)
\eea
and so in the limit $N \gg 1$ are simply the thermal correlation functions for free fields, as if the $\phi$ field did not exist.

For completeness we quote here the explicit form for this free thermal correlator, with details of the calculation given in \S\ref{sec:freethermaltrace}. The result evaluated at spacetime points $x = (t, \mathbf{x})$ and $x' = (t', \mathbf{x}')$ is\footnote{Note the given $i \delta$-prescription is only valid for real time arguments. Given that the $N$ environment fields are assumed to be prepared in a thermal state, this correlation function must obey the Kubo-Martin-Schwinger (KMS) condition $\langle \chi^a(t - i\beta, \bfx) \chi^{b}(0, \mathbf{0}) \rangle_{\beta} = \langle \chi^a(t, \bfx) \chi^{a}(0, \mathbf{0}) \rangle_{\beta}^{\ast}$, which assumes a shift in imaginary time --- for the correct $i \delta$-prescription in this case, see (\ref{thermcorrKMS}) in Appendix \ref{sec:freethermaltrace} (which agrees with the above prescription for real time arguements).}
\bea \label{thermalcorrelatorab}
\langle \chi^a(x) \chi^b(x') \rangle_{\beta}  &: =& \TrB\left[ \chi^{a}(t,\bfx) \chi^b(t',\bfx')  \varrho_{\beta} \right] \\
& = & \frac{\delta^{ab}}{8 \pi \beta |\mathbf{x} - \mathbf{x}'|}\left\{ \coth\left[  \dfrac{\pi}{\beta} \left( t - t' + |\mathbf{x} - \mathbf{x}'| - i \delta \right) \right] - \coth\left[ \dfrac{\pi}{\beta} \left( t - t' - |\mathbf{x} - \mathbf{x}'| - i \delta \right) \right] \right\} \,, \notag
\eea
in agreement with standard formulae \cite{Thermal}. In this expression the limit $\delta \to 0^{+}$ is to be taken at the end of the calculation. Notice that eq.~\pref{thermalcorrelatorab} obeys the required reality property (for real scalars)
\be
\langle \chi^a(y) \chi^b(x) \rangle_{\beta}  =   \Bigl[ \langle \chi^b(x) \chi^a(y) \rangle_{\beta} \Bigr]^{\ast}  \,,
\ee
and at zero temperature ($\beta \to \infty$) goes over to
\be
\langle \chi^a(x) \chi^b(x') \rangle_{\beta}  \to
\frac{\delta^{ab}}{4 \pi^2[ -( t - t' - i \delta)^2 + |\mathbf{x} - \mathbf{x}' |^2]} \,, 
\ee
as it should. 

\section{RG Improvement and resumming the $\lambda$ expansion}
\label{sec:BC}

This section studies the dependence of hotspot physics on the self-coupling $\lambda$, in particular exploring how limiting it is to treat its implications perturbatively. Although the validity of expansions in $\lambda$ might seem to be a tangential issue if one's focus is on the thermal coupling $\tilde g$, it really is not. As discussed earlier, the response of a field like $\phi$ to the hotspot typically diverges at the hotspot position --- {\it c.f.} for example equations \pref{Seliminated3} and \pref{saeliminated3} --- and these divergences are ultimately handled by being renormalized into couplings like $\lambda$, as in \pref{lambdaren}. As a consequence of this renormalization couplings like $\lambda$ run in the renormalization-group sense, and (as we show here, following \cite{PPEFT}) this can make it inconsistent to set them to zero at all scales. 

This section derives precisely how the coupling $\lambda$ runs, and along the way shows that the dimensionless expansion parameter that justifies treating $\lambda$ perturbatively turns out to be $\lambda/4\pi \epsilon$, where $\epsilon$ is the very small regularization length scale used to regulate the divergences (as in eqs. \pref{Seliminated3} and \pref{saeliminated3}). Physically, both $\lambda$ and $\epsilon$ might reasonably be expected to be of order the size $\xi$ of the compact hotspot; a length scale that has been assumed to be much smaller than the other scales of physical interest.  If perturbative calculations actually require $\lambda \ll 4\pi \epsilon$ then they may not be that useful, since $\lambda$ would have to be much smaller even than this already very microscopic scale. The renormalization-group arguments presented here show how perturbative predictions can be extended to the regime $\lambda \gsim 4\pi \epsilon$, providing results that can be compared to the exact calculations to follow in \S\ref{sec:exactcorr}.

\subsection{Effective interactions and boundary condition}
\label{ssec:PPEFT}

To better understand the effects of $\lambda$ beyond perturbation theory this section temporarily turns off the coupling $\tilde g$ in order to eliminate unnecessary distractions. Non-perturbative information is then extracted by leaving $\lambda$ nonzero for all time and exploring more systematically how it modifies the dynamics of the $\phi$ field. A natural framework for this is the language of point-particle (or world-line) EFTs, since these systematically incorporate the effects of small objects on their surroundings, organized in powers of $k a$ (where $a$ is the object's size and $k$ is the momentum of a typical probe). In practice we therefore work completely in the $\cR_+$ sector, following closely the logic of \cite{PPEFT, EFTBook}, with the bulk field interacting only with the contact interaction
\be
H_{\mathrm{int}}(t) =   \frac{\lambda}{2} \int \exd^{3}x \;  \phi ^2(t,\mathbf{x}) \, \delta^3(\bfx) = \frac{\lambda}{2} \; \phi ^2(t,\mathbf{0}) \ ,
\ee
with $\lambda$ independent of time.

The implications of $\lambda$ are incorporated by identifying the mode functions that are appropriate in the presence of this interaction. Since the Heisenberg equation of motion in this case -- {\it c.f.} equation \pref{heis1} -- is
\be \label{heiseternal}
(-\partial_t^2 + \nabla^2 ) \phi (t,\bfx) = \lambda \, \phi (t,\mathbf{0}) \,  \delta^3(\bfx) \ ,
\ee
this is also the equation satisfied by each mode function, $u_{\omega \ell m}(t,\bfx)$, in an expansion ({\it c.f.}~equation \pref{phiexpINT}) like
\be
\phi (t,\bfx) = \sum_{\ell = 0 }^{\infty} \sum_{m = - \ell}^{+\ell} \int \exd \omega \; \bigg[ u_{\omega \ell m}(t,\bfx) \mfa_{\omega \ell m } + u^{\ast}_{\omega \ell m}(t,\bfx) \mfa^{\ast}_{\omega \ell m } \bigg] \ .
\ee
Once the $\lambda$-dependence of these mode functions is identified by solving \pref{heiseternal}, the implications for the Wightman function are obtained from formulae like
\be
\bra{\vac} \phi (t,\bfx)\phi (s,\bfy) \ket{\vac} = \sum_{\ell = 0 }^{\infty} \sum_{m = - \ell}^{+\ell} \int \exd \omega\; u_{\omega \ell m}(t,\bfx)  u^{\ast}_{\omega \ell m}(s,\bfy)  \,.
\ee
Here $\ket{\vac}$ satisfies $\mfa_{\omega \ell m} \ket{\vac} = 0$, and $\mfa_{\omega\ell m}$ satisfies the standard commutation relations $[\mfa_{\omega \ell m} , \mfa^*_{\tilde{\omega} \tilde{\ell} \tilde{m}} ] = \delta (\omega - \tilde{\omega} )\delta_{\ell \tilde{\ell}} \delta_{m\tilde{m}}$, and the mode functions $u_{\omega \ell m}$ are assumed to be properly normalized.

The main observation is that the dependence of $u_{\omega \ell m}$ on $\lambda$ can be inferred by integrating its equation of motion
\be \label{modeEOM}
(-\partial_t^2 + \nabla^2 ) u_{\omega \ell m}(t,\bfx) = \lambda \,  u_{\omega \ell m}(t,\bfx)  \, \delta^3(\bfx)\ ,
\ee
over a tiny sphere $B_{\epsilon}   :=   \left\{ \; \mathbf{x} \in \mathbb{R}^{3} \; \big| \; |\bfx| \leq \epsilon \; \right\}$ of radius $\epsilon>0$ centred around the origin. Following standard steps \cite{PPEFT, PPEFT2, PPEFTDis} this integration leads to a $\lambda$-dependent boundary condition near the hotspot, of the form
\be  \label{BC}
4 \pi \epsilon^2 \left( \frac{ \partial u_{\omega \ell m}(t,\bfx)}{\partial r} \right)_{r = \epsilon} = \lambda \; u_{\omega \ell m}(t,\bfx) \Bigr|_{r=\epsilon}  \,.
\ee
That is, for $r > \epsilon$ mode functions simply satisfy the Klein-Gordon equation 
\be
( - \partial_{t}^2 + \nabla_{\bfx}^2 ) u_{\bfp}(t,\bfx) = 0 \ ,
\ee
and only learn about the coupling $\lambda$ through its appearance in the boundary condition \pref{BC}. 

Concretely, expanding the solution in terms of spherical harmonics, 
\be
u_{\omega \ell m}(t,\bfx) = e^{- i \omega t } R_{\omega\ell}(r) Y_{\ell m}(\theta, \phi) ,
\ee
the radial solutions are spherical Bessel functions
\be \label{Rsolagain}
R_{\omega\ell}(r) = C_{\ell }(\omega) j_{\ell}(\omega r) + D_{\ell }(\omega) y_{\ell}(\omega r) \ , 
\ee
where $C_{\ell }(\omega)$ and $D_{\ell }(\omega)$ are integration constants, whose ratio is determined by the boundary condition \pref{BC} and so is $\lambda$-dependent. Explicitly, the boundary condition \pref{BC} implies
\be \label{BCbeforeratio}
4 \pi \epsilon^2 \; \partial_r R_{\omega\ell}( \epsilon) = \lambda \, R_{\omega\ell }( \epsilon) 
\ee

Substituting the solution \pref{Rsolagain} into \pref{BCbeforeratio}, and using the Bessel function identity
\be
\partial_r f_{\ell}(\omega r) = \frac{\ell}{r}\,  f_{\ell}(\omega r) - \omega f_{\ell + 1}(\omega r)   
\ee
(that holds for both $f_\ell = j_\ell$ and $f_\ell = y_\ell$), shows that the boundary condition \pref{BCbeforeratio} becomes
\be \label{BClambdavsDC}
  \frac{\lambda}{4\pi \epsilon} = \Bigl( {r \,\partial_r} \ln R_{\omega\ell} \Bigr)_{r = \epsilon} = \frac{\ell j_\ell(\omega \epsilon) - \omega \epsilon j_{\ell+1}(\omega\epsilon) + (D_\ell/C_\ell)[\ell y_\ell(\omega \epsilon) - \omega \epsilon y_{\ell+1}(\omega \epsilon)]}{j_\ell(\omega\epsilon) + (D_\ell/C_\ell) \, y_\ell(\omega \epsilon)} \,,
\ee
and this, once solved, leads to the following solution for the $\lambda$-dependence of $D_\ell/C_\ell$
\be \label{BCDoverC}
\frac{D_{\ell }(\omega)}{C_{\ell }(\omega)} 
= - \; \frac{ [( \lambda/  4 \pi \epsilon ) - \ell] j_{\ell}(\omega \epsilon) +   \omega \epsilon  j_{\ell + 1}(\omega \epsilon)  }{ [( \lambda/  4 \pi \epsilon ) - \ell] y_{\ell}(\omega \epsilon) +   \omega \epsilon  y_{\ell + 1}(\omega \epsilon) } \,.
\ee

These expressions simplify in the limit of practical interest, where $\omega \epsilon \ll 1$. In this limit we may use the expansions
\be
j_{\ell}(z) = \frac{\sqrt{\pi} \; z^{\ell} }{2^{\ell+1}  \Gamma(\ell + \frac{3}{2} ) } 
+ \cO(z^{\ell+2}) \quad \hbox{and} \quad
y_{\ell}(z) = -  \frac{2^{\ell}\, \Gamma( \frac{1}{2} + \ell )}{ \sqrt{\pi} \; z^{ \ell +1}  } 
+ \cO(z^{-\ell + 1}) \,,
\ee
to find 
that \pref{BCDoverC} becomes
\be \label{BCDoverC2}
 \frac{D_{\ell }(\omega)}{C_{\ell }(\omega)}   \simeq \frac{\pi}{\Gamma(\ell+\frac32) \, \Gamma(\ell + \frac12)} \left[  \frac{( \lambda/  4 \pi \epsilon ) - \ell  
}{ ( \lambda/  4 \pi \epsilon ) + \ell  + 1 }  \right] \; \left( \frac{\omega \epsilon}{2} \right)^{2\ell+1} \,.
\ee
The coefficient here can be simplified using
\be
\frac{\pi}{\Gamma(\ell+\tfrac{1}{2})\Gamma(\ell +\tfrac{3}{2})} =\frac{2^{4 \ell + 2} [ \ell !]^2 }{2 (2\ell + 1) [ (2\ell)! ]^2} \,.
\ee

The inverse of \pref{BCDoverC2} --- or equivalently, the small $\omega\epsilon$ limit of \pref{BClambdavsDC} --- similarly becomes
\be \label{BClambdavsDC2}
  \frac{\lambda}{4\pi \epsilon}\simeq \frac{\ell (\omega\epsilon/2)^{2\ell+1} + (\ell+1) X_\ell(\omega) }{  (\omega\epsilon/2)^{2\ell+1} -X_\ell(\omega)}
\ee
where
\be \label{BClambdavsDC22}
 X_\ell(\omega) := \frac{1}{\pi} \, \Gamma \left(\ell + \sfrac32 \right) \, \Gamma \left(\ell + \sfrac12 \right) \, \frac{D_\ell(\omega)}{C_\ell(\omega)} \,.
\ee
Whether the numerator and denominator of thse last expressions can be further expanded depends on how the quantities $(\lambda/4\pi \epsilon) - \ell$ and $D_\ell/C_\ell$ behave when $\omega \epsilon \ll 1$. 

\subsection{Renormalization group and the interpretation of $\epsilon$}

There are two ways to read the above boundary conditions. The naive way is as given in \pref{BCDoverC} or \pref{BCDoverC2}: they give $D_\ell(\omega)/C_\ell(\omega)$ as an explicit function of $\ell$ and the two dimensionless variables $\lambda/4\pi \epsilon$ and $\omega \epsilon$. What is bothersome about this interpretation is that it makes $D_\ell/C_\ell$ depend not only on the coupling $\lambda$, but also on the arbitrary regularization scale $\epsilon$. 

But if $D_\ell/C_\ell$ depends on $\epsilon$ then so also will the physical observables that are built from it. Normally regularization dependence in a calculation drops out of physical quantities because it gets renormalized into a redefinition of the couplings. Or, equivalently, it is cancelled by an implicit regularization dependence that is hidden within couplings like $\lambda$.

\subsubsection{Running of $\lambda$}

This observation suggests a different way to interpret the above boundary condition \cite{PPEFT}. This alternative reading insists physical quantities cannot depend on arbitrary regularization scales, and because of this neither can $D_\ell/C_\ell$. In this case expressions like \pref{BClambdavsDC} or \pref{BClambdavsDC2} should be reinterpreted as making explicit how $\lambda = \lambda(\epsilon)$ must depend on $\epsilon$ in order to ensure that $D_\ell/C_\ell$ remains $\epsilon$-independent. That is to say, in this interpretation \pref{BClambdavsDC2} is an RG equation for the coupling $\lambda(\epsilon)$. 

\begin{figure}[h!]
\centering
\includegraphics[width=75mm,height=50mm]{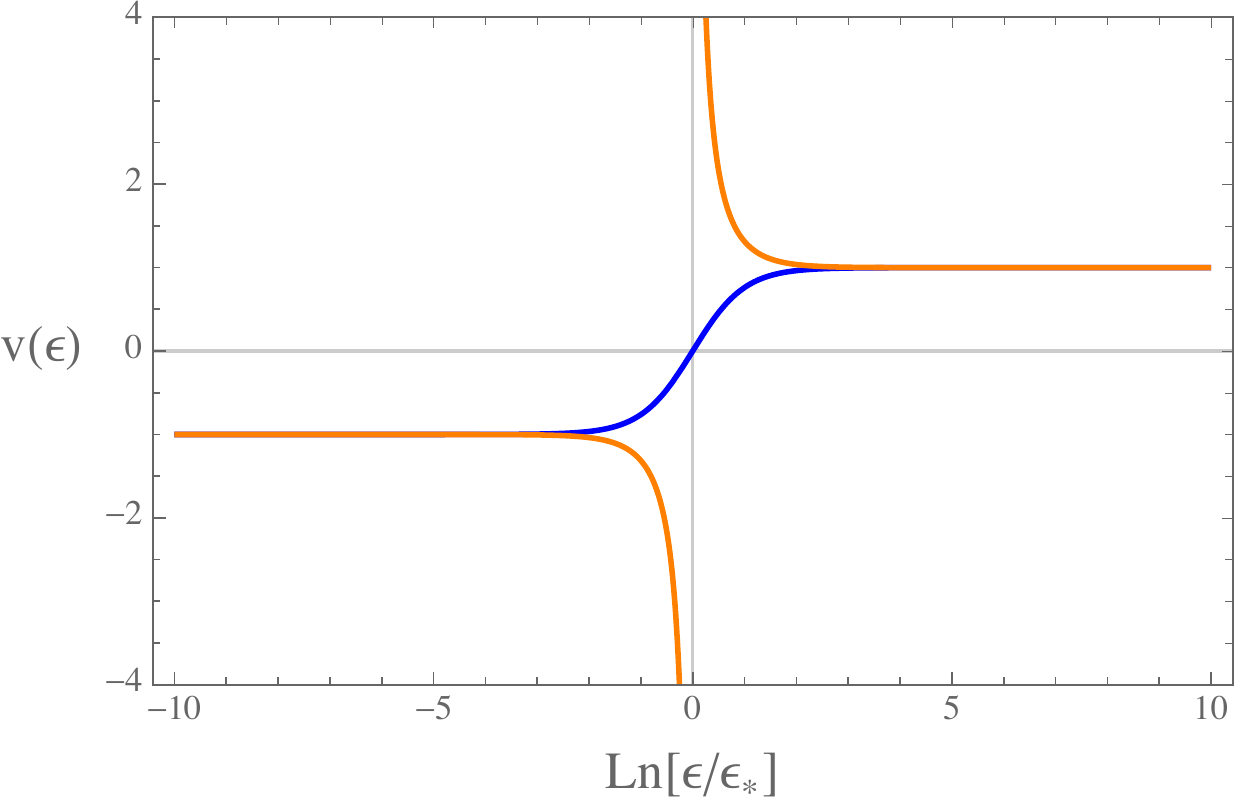}
\caption{The two categories of RG flow described by  the evolution equation \pref{universalDE}. This figure plots the universal variable $v(\epsilon)$ against $\ln(\epsilon/\epsilon_\star)$. An example is shown with both $|v| > 1$ and $v  < 1$, and the plot shows the sign of $|v| -1$ is invariant because $v = \pm 1$ are fixed points. (Figure taken from \cite{PPEFTH2}.)}
\label{figureFlow}
\end{figure}

To see what the evolution implied by \pref{BClambdavsDC2} means more explicitly, it is worth expressing it in differential form. As explored in detail in Appendix \ref{AppendixF}, this can be put into a universal form by defining the new variable $v(\epsilon)$ using
\be \label{lambdavsvdefs}
  \frac{\lambda}{2\pi\epsilon} =  \left( 2\ell+1 \right) v - 1   \,,
\ee
for which differentiation of \pref{BClambdavsDC2} becomes
\be \label{universalDE}
  \epsilon \, \frac{\exd v}{\exd \epsilon} = \left( \ell + \sfrac12 \right) (1-v^2) \,.
\ee
As is easily verified, the solution of \pref{universalDE} subject to the initial condition $v(\epsilon_0) = v_0$ is given by
\be \label{RGsolution}
  v(\epsilon) = \frac{ (v_0 + 1)(\epsilon/\epsilon_0)^{2\ell+1} + (v_0 - 1)}{(v_0 + 1)(\epsilon/\epsilon_0)^{2\ell+1} - (v_0 - 1)} 
\ee
and this agrees with \pref{BClambdavsDC2} once \pref{lambdavsvdefs} is used, with integration constant $v_0$ determining the combination $D_\ell/C_\ell$. These generically describe evolution from $v = -1$ to $v = +1$ as $\epsilon$ ranges from 0 to $\infty$. A plot of two representative solutions to \pref{RGsolution} is given in Fig.~\ref{figureFlow}. 

Of course there is nothing wrong with simply regarding the boundary condition as specifying $D_\ell/C_\ell$ once a coupling $\lambda_0$ is specified using a specific choice of regularization scale $\epsilon_0$. What the RG interpretation tells us is that once this choice is made, we are completely free to use any other regularization scale, $\epsilon_1$ instead, 
provided that we also change the value of the coupling to $\lambda_1$ where both pairs $(\epsilon_0, \lambda_0)$ and $(\epsilon_1, \lambda_1)$ lie on the same RG trajectory $\lambda(\epsilon)$ defined by \pref{BClambdavsDC2} (or, equivalently by \pref{lambdavsvdefs} and \pref{RGsolution}). It is only when the coupling and regulator are changed in this correlated way that physical quantities remain unchanged.

\subsubsection{RG-invariant characterization of coupling strength} 

Because physical observables depend only on the coupling {\it trajectories} it is more informative to specify the strength of the coupling by labelling the coupling trajectories using a more convenient RG-invariant parameterization, rather than simply by specifying its value $\lambda_0 = \lambda(\epsilon_0)$ for a specific (but arbitrary) regularizations scale $\epsilon_0$. This section follows \cite{PPEFT} and identifies a particular choice of RG-invariant parameterization that is convenient because (unlike the value $\lambda_0$, say) the parameters are simply related to physical quantities.

To this end the first observation is that the evolution equation \pref{universalDE} has two fixed points, $v = \pm 1$, along which $v$ is $\epsilon$-independent. Trajectories that {\it do} evolve therefore cannot cross $v = \pm 1$ and so fall into two distinct categories, distinguished by
\be
\eta_{\star}   :=  \hbox{sign}( v^2 - 1) \,.
\ee
$\eta_\star$ is an RG-invariant quantity inasmuch as the sign of $v^2-1$ does not depend on $\epsilon$ for any $v(\epsilon)$ satisfying \pref{universalDE}. 

Any specific curve can be uniquely characterized in an RG-invariant way by specifying both $\eta_{\star}$ and the new variable $\epsilon_{\star}$, defined as the place where the curve passes through zero (if $\eta_\star = -1$) or where it diverges (if $\eta_\star = +1$). Using this definition the general solution \pref{RGsolution} simplifies to
\be \label{RGsolutionINV}
  v(\epsilon) = \frac{ (\epsilon/\epsilon_\star)^{2\ell+1} + \eta_\star}{(\epsilon/\epsilon_\star)^{2\ell + 1} - \eta_\star} 
\ee
and this shows that the pair $(\eta_{\star},\epsilon_{\star})$ are related to any specific choice of initial condition $(\epsilon_0, v_0)$ by $\eta_{\star} = \hbox{sign}(v_0^2-1)$ and
\be \label{epsilonstarvsv0}
\left( \frac{\epsilon_{\star}}{\epsilon_0} \right)^{2\ell +1} = \eta_\star \left( \frac{v_0 - 1}{v_0 + 1} \right) = \left| \frac{v_0 - 1}{v_0 + 1} \right| \,.
\ee

What makes these variables convenient is that $\epsilon_\star$ provides an invariant length scale that is shared by all representatives $(\epsilon,v)$ or $(\epsilon,\lambda)$ along a particular RG trajectory. It is consequently this length scale --- and {\it not} $\epsilon_0$ or $\lambda_0$, say ---  that is physical and so whose size characterizes the values of physical observables. This is shown in detail in \cite{PPEFT, PPEFT2, PPEFT3, PPEFTDis, PPEFTH1, PPEFTH2}, where cross sections and energy shifts in many examples are evaluated and found to be simply related to $\epsilon_\star$. 

To see explicitly why this is so, we write $\lambda(\epsilon)$ in terms of $\epsilon_\star$ by combining \pref{lambdavsvdefs} and \pref{RGsolutionINV} to get  
\be\label{lambdavsepsexplicit}
\frac{\lambda}{2 \pi \epsilon} + 1 =  (2 \ell + 1) v(\epsilon) = (2 \ell + 1) \, \frac{\left( {\epsilon}/{\epsilon_\star} \right)^{2\ell + 1} + \eta_{\star} }{ \left( {\epsilon}/{\epsilon_\star} \right)^{2\ell + 1} - \eta_{\star} }  \ ,
\ee
and for later purposes also record its inverse ({\it c.f.}~eqs.~\pref{lambdavsvdefs} and \pref{epsilonstarvsv0})
\be \label{epsilonstarvslambda0}
\left( \frac{\epsilon_{\star}}{\epsilon} \right)^{2\ell +1} = \left| \frac{v - 1}{v + 1} \right|= \left| \frac{[\lambda/(4\pi \epsilon)] - \ell}{[\lambda/(4\pi \epsilon)] + \ell + 1} \right| \,.
\ee
Eq.~\pref{lambdavsepsexplicit} can be used in \pref{BCDoverC2} to determine the integration constant ratio $D_\ell/C_\ell$ from which physical quantities are ultimately determined. This exercise gives
\be \label{DCvsEpsStar}
  \frac{D_\ell(\omega)}{C_\ell(\omega)} = \frac{\pi \eta_\star}{\Gamma(\ell+\frac32) \, \Gamma(\ell+\frac12)} \left( \frac{\omega \epsilon_\star}{2} \right)^{2\ell+1} \,,
\ee
verifying that the explicit dependence on $\epsilon$ and $\lambda$ combines into the invariant combinations $\eta_\star$ and $\epsilon_\star$. In particular, it is the dimensionless quantity $\omega \epsilon_\star$ that controls the size of any physical response, and \pref{DCvsEpsStar} shows quantitatively in this language how small angular momenta $\ell$ are preferred when $\omega \epsilon_\star \ll 1$.

In practical examples $\epsilon_\star$ is set by the size of the underlying object (in this case the hotspot) times the appropriate coupling that controls the interactions through which it is probed. For example, when a similar analysis is applied to describing the effects of finite nuclear size on the energy levels in pionic atoms, one finds $\epsilon_\star \sim R_\ssN$ is of order the nuclear radius \cite{PPEFT2}. But the same analysis when describing nuclear-size effects on Hydrogen energy levels finds $\epsilon_\star \sim \alpha R_\ssN$, with $\alpha$ the fine-structure constant \cite{PPEFT}. 

By comparison, if boundary conditions must be imposed for an effective theory outside the nucleus then $\epsilon > R_\ssN$.  Concrete examples like these show that a small source probed by a weakly coupled field tends to produce $\epsilon_\star \ll \epsilon$, if $\epsilon$ is regarded to be typical linear size of the compact object. 

\subsubsection{Resumming all orders in $\lambda/4\pi\epsilon$}
\label{ssec:Resummation}

It is instructive to explore the connection between $\lambda(\epsilon)$ and $\epsilon_\star$ explicitly in the weak-coupling limit, by expanding \pref{lambdavsepsexplicit} in powers of $\epsilon_\star/\epsilon$. In this limit \pref{RGsolutionINV} simplifies to $v(\epsilon) \simeq 1 + 2 \eta_{\star} \left( {\epsilon_{\star}}/{\epsilon} \right)^{2\ell + 1} + \cdots$, and so
\be
   \frac{\lambda}{2\pi \epsilon} \simeq 2\ell + 2 \eta_\star (2\ell + 1) \left( \frac{\epsilon_\star}{\epsilon} \right)^{2\ell + 1} \,.
\ee
For the $\ell = 0$ mode in particular $\lambda$ becomes $\epsilon$-independent in the perturbative limit, with
\be \label{lambdaleading}
 \lambda \simeq  4 \pi \eta_{\star} \epsilon_{\star} \qquad (\hbox{for} \; \ell =0 \; \hbox{and} \; \epsilon_\star \ll \epsilon) \ .
\ee
This shows that for $s$-wave processes, expressions for physical quantities as functions of $\omega \epsilon_\star$ (for mode frequency $\omega$) can be turned into corresponding expressions as functions of $\lambda\omega$, by using \pref{lambdaleading}. Provided powers of $\epsilon_\star/\epsilon$ are negligible these expressions need have no dependence on regulators like $\epsilon$, making any discussion of RG evolution completely unnecessary. 

But what happens if $\epsilon$ is now decreased and $\lambda(\epsilon)$ adjusted along a particular RG flow to a point where $\epsilon_\star/\epsilon$ is no longer negligible and $\lambda/4\pi \epsilon$ is no longer small? The answer for the physical observable as a function of $\omega\epsilon_\star$ does not change at all, because physics depends only on which RG trajectory one lives, and not on the particular point one sits along this trajectory. All that changes as $\epsilon$ and $\lambda$ are varied is that expression \pref{lambdaleading} can no longer be used to trade $\epsilon_\star$ for $\lambda$; instead one must go back to the full result \pref{lambdavsepsexplicit} when doing so. 

This observation provides a way to resum all orders in $\lambda/4\pi \epsilon$ while holding quantities like $\omega \epsilon_\star$ fixed. Suppose one computes an observable as a function of two dimensionless quantities $O = O[\omega \epsilon, \lambda/4\pi\epsilon]$, and does so perturbatively in $\lambda/4\pi \epsilon$. The result can be turned into an expression $O = O[\omega\epsilon, \epsilon_\star/\epsilon]$ by trading $\lambda$ for $\epsilon_\star$ using \pref{lambdaleading}. But we know that $\epsilon$ is just a calculational artefact that is not actually in the physical result, which must therefore really only be a function of the one variable $\omega\epsilon_\star$. 

The result for the same observable elsewhere on the RG trajectory, where $\lambda/4\pi \epsilon$ and $\epsilon_\star/\epsilon$ are {\it not} small, is given by the same expression $O = O[\omega\epsilon_\star]$ since it does not depend at all on $\epsilon$. Expressing this result in terms of $O = O[\omega\epsilon, \lambda/4\pi \epsilon]$ using \pref{lambdavsepsexplicit} then gives an explicit resummation of the observable to all orders in $\lambda(\epsilon)/4\pi\epsilon$. 

\subsection{Resummation of the two-point function}
\label{sec:norm}

The above reasoning can be applied to the perturbative calculation of $\phi$-field response given above, allowing results that are derived to lowest nontrivial order in $\lambda/4\pi\epsilon$ to be promoted into expressions that work to all orders in this variable. 

\subsubsection{$s$-wave resummation}

We derive the resummed results here, and then check (in this section) that they capture a full mode sum using the boundary condition \pref{BC}. In later sections we also verify that the results found for large $\lambda/4\pi \epsilon$ in this way also agree with the exact expressions derived in \S\ref{sec:exactcorr}.

The starting point is the perturbative expression \pref{pertcorr}, that is given again here in the special case $\tilde g = 0$:
\be  \label{pertcorr2}
W_\beta(t,\bfx; t',\bfx')   \simeq   \frac{1}{4\pi^2 \big[ - (t - t' - i \delta)^2 + |\bfx - \bfx'|^2 \big]} + \frac{\lambda}{16\pi^3 |\bfx| |\bfx'|} \bigg[ \frac{|\bfx| + |\bfx'|}{(t - t' - i \delta)^2 - (|\bfx + |\bfx'|)^2 } \bigg] \,.
\ee
Anticipating the dominance of the $\ell = 0$ mode (as is appropriate in applications for which $\epsilon_\star \ll |\bfx|, |\bfx'|$ say, see next section), we may trade $\lambda$ in this expression for $\epsilon_\star$ using \pref{lambdaleading} to find 
\be  \label{pertcorr3}
W_\beta(t,\bfx; t',\bfx')   \simeq   \frac{1}{4\pi^2 \big[ - (t - t' - i \delta)^2 + |\bfx - \bfx'|^2 \big]} + \frac{\eta_\star \epsilon_\star}{4\pi^2 |\bfx| |\bfx'|} \bigg[ \frac{|\bfx| + |\bfx'|}{(t - t' - i \delta)^2 - (|\bfx + |\bfx'|)^2 } \bigg] \,.
\ee
But an expression with broader validity than \pref{pertcorr2} can be obtained from \pref{pertcorr3} by using in this result the more general $\ell = 0$ relation giving $\epsilon_\star$ in terms of $\lambda$ given in \pref{epsilonstarvslambda0}, leading to
\bea \label{WightmanlambdastarResum}
\braket{\phi (t,\bfx)\phi(t',\bfx')} & \simeq &\frac{1}{4\pi^2} \left[ \frac{1}{- (t-t' - i\eta)^2 + |\bfx - \bfx'|^2 } \right]\\
& \ & \qquad\qquad\qquad  + \frac{1}{16\pi^3|\bfx| |\bfx'|}  \left| \frac{\lambda}{(\lambda/4\pi \epsilon) + 1} \right|\bigg[  \frac{|\bfx| + |\bfx'|}{(t - t' - i \delta)^2 - ( |\bfx| + |\bfx'|)^2 } \bigg] \,. \notag
\eea
This clearly agrees with \pref{pertcorr2} for small $\lambda/4\pi \epsilon$, but its validity is now extended to include the regime $\lambda \gsim 4\pi \epsilon$ provided only that $\ell = 0$ modes dominate when computing the hotspot influence. The conditions under which this is true are explored more fully in the next section, which verifies \pref{WightmanlambdastarResum} starting directly from a mode-sum using modes that satisfy the boundary condition \pref{BC}.

\subsubsection{Mode-sum calculation}
\label{sec:norm}

We next recompute \pref{WightmanlambdastarResum} by evaluating the $\phi$-field correlator as an exact function of $\lambda$, by calculating the sum over mode-functions whose $\lambda$-dependence is acquired through the boundary condition \pref{BC}. As described above, this boundary condition fixes the ratio of integration constants $D_\ell/C_\ell$ to be given in terms of $\epsilon_\star$ as in \pref{DCvsEpsStar}, repeated here for conenience:
\be \label{DCvsEpsStar2}
  \frac{D_\ell(\omega)}{C_\ell(\omega)} = \frac{\pi \eta_\star}{\Gamma(\ell+\frac32) \, \Gamma(\ell+\frac12)} \left( \frac{\omega \epsilon_\star}{2} \right)^{2\ell+1} \,.
\ee

\subsubsection*{Mode normalization}

The integration constants $C_\ell$ and $D_\ell$ are determined separately by combining \pref{DCvsEpsStar2} with mode-function normalization, which requires
\be
\langle u_{\omega\ell m}, u_{\tilde{\omega} \tilde{\ell} \tilde{m} } \rangle = \delta(\omega - \tilde{\omega}) \delta_{\ell \tilde{\ell}} \delta_{m \tilde{m}} \,, \quad 
\langle u_{\omega\ell m}, u^{\ast}_{\tilde{\omega} \tilde{\ell} \tilde{m} } \rangle = 0 \,, \quad
\langle u^{\ast}_{\omega\ell m}, u^{\ast}_{\tilde{\omega} \tilde{\ell} \tilde{m} } \rangle = - \delta(\omega - \tilde{\omega}) \delta_{\ell \tilde{\ell}} \delta_{m \tilde{m}} 
\ee
where the angle brackets denote the Klein-Gordon inner product 
\be \label{KGinnerproduct}
\langle F, G \rangle := i \int \exd^{3}x\; \bigg( F(t,\bfx) \partial_t {G}^{\ast}(t,\bfx) - \partial_t {F}(t,\bfx) G^{\ast}(t,\bfx) \bigg) \,.
\ee
As is easily verified, this inner product is time-independent when evaluated for any solutions $F,G$ to the Klein-Gordon equation, and this remains true even in the presence of the modified boundary condition \pref{BC}, whenever the effective coupling $\lambda$ is real. To see why notice that this boundary condition implies the radial flux density of Klein-Gordon probability at $r = \epsilon$ is 
\be
   J_r(\epsilon) \propto  \Bigl[ F(t,\bfx) \partial_r {G}^{\ast}(t,\bfx) - \partial_r {F}(t,\bfx) G^{\ast}(t,\bfx)  \Bigr]_{r=\epsilon} = (\lambda^* - \lambda) \Bigl[ F(t,\bfx)  {G}^{\ast}(t,\bfx)  \Bigr]_{r=\epsilon} 
\ee
and so vanishes for real $\lambda$. 

The normalization integrals are computed in Appendix \ref{sec:norm}, leading to the following $\epsilon_\star$-dependent results for $C_\ell$ and $D_\ell$ separately
\be \label{CellForm}
C_{\ell m}(\omega) = \sqrt{\frac{\omega}{\pi}} \left\{ 1 +\left[ \frac{\pi}{\Gamma(\ell+\tfrac{1}{2}) \Gamma(\ell +\tfrac{3}{2})} \bigg( \frac{\omega \epsilon_{\star}}{2} \bigg)^{2 \ell + 1}\right]^2  \right\}^{-1/2} \,,
\ee
and
\bea \label{DellForm}
D_{\ell m}(\omega) & = & \frac{\pi\eta_{\star}}{\Gamma(\ell+\tfrac{1}{2})\Gamma(\ell +\tfrac{3}{2})}  \left( \frac{\omega \epsilon_{\star}}{2} \right)^{2 \ell + 1}   \sqrt{\frac{\omega}{\pi}} \left\{ 1 +\left[ \frac{\pi}{\Gamma(\ell+\tfrac{1}{2}) \Gamma(\ell +\tfrac{3}{2})} \bigg( \frac{\omega \epsilon_{\star}}{2} \bigg)^{2 \ell + 1}\right]^2  \right\}^{-1/2}  \,.
\eea
With these choices the mode functions and field operators satisfy the required boundary condition at $r = \epsilon$, and this completely determines their dependence on $\epsilon_\star$. 

The main approximation made in deriving \pref{CellForm} and \pref{DellForm} is to assume that $\omega\epsilon$ is small enough to allow the replacement of $j_\ell(\omega\epsilon)$ and $y_\ell(\omega\epsilon)$ with their leading asymptotic forms; that is by using \pref{BCDoverC2} instead of \pref{BCDoverC}. Since the Bessel functions are explicitly known this approximation can be improved to any desired order in $\omega \epsilon$, by upgrading condition \pref{DCvsEpsStar2} using a more accurate representation of the Bessel functions.

\subsubsection*{Mode Sum}

We are now in a position to compute the Wightman function in terms of a mode sum, using the above $\epsilon_\star$-dependent form for the modes,
\be \label{ModeForm}
u_{\omega\ell m}(t,\bfx) = C_{\ell m }(\omega) e^{- i \omega t} \left[  j_{\ell}(\omega r) + \frac{\pi\eta_{\star}}{\Gamma(\ell+\tfrac{1}{2})\Gamma(\ell +\tfrac{3}{2})}  \left( \frac{\omega \epsilon_{\star}}{2} \right)^{2 \ell + 1} y_{\ell}(\omega r) \right] Y_{\ell m} (\theta,\phi ) \ ,
\ee
that properly matching at the boundary condition at $r = \epsilon$ (dropping subdominant terms in $\omega\epsilon$). As argued above, all explicit dependence on $\epsilon$ drops out of this exression once evaluated at $|\bfx| = \epsilon$, cancelling between any explicit dependence and the $\epsilon$-dependence implicit without the coupling $\lambda$, leaving a dependence only on the RG-invariant quantity $\epsilon_\star$. In particular, eqs.~\pref{CellForm} and \pref{ModeForm} do {\it not} assume validity of the weak-coupling limit $\epsilon_\star \ll \epsilon$ (or equivalently $\lambda/4\pi \epsilon$ need not be much smaller than unity).

The Wightman function is given in terms of these modes by
\bea
\braket{\phi (t,\bfx)\phi(t',\bfx')} & = & \sum_{\ell =0}^{\infty} \sum_{m=-\ell}^{+\ell} \int_0^\infty \exd \omega\; u_{\omega \ell m}(t,\bfx) u^{\ast}_{\omega \ell m}(t',\bfx') \\
& = & \sum_{\ell =0}^{\infty} \sum_{m=-\ell}^{+\ell} \int_0^\infty \exd \omega\; e^{- i \omega (t - t')} |C_{\ell m }(\omega)|^2 \left[ j_{\ell}(\omega |\bfx|) + \sfrac{\pi \eta_{\star}}{\Gamma(\ell+\tfrac{1}{2})\Gamma(\ell +\tfrac{3}{2})} \left( \frac{\omega \epsilon_{\star}}{2} \right)^{2 \ell + 1} y_{\ell}(\omega |\bfx| ) \right] \nn \\
& \ & \qquad  \times \left[ j_{\ell}(\omega |\bfx'|) + \sfrac{\pi \eta_{\star}}{\Gamma(\ell+\tfrac{1}{2})\Gamma(\ell +\tfrac{3}{2})} \left( \frac{\omega \epsilon_{\star}}{2} \right)^{2 \ell + 1} y_{\ell}(\omega |\bfx'| ) \right] Y_{\ell m} (\theta,\phi) Y^{\ast}_{\ell m} (\theta',\phi') \ .\notag
\eea
Performing the sum over $m$ (see Appendix \ref{App:ModeSum} for details) leads to the intermediate expression
\bea
\braket{\phi (t,\bfx)\phi(t',\bfx')} & = & \sum_{\ell =0}^{\infty} \frac{2\ell + 1}{4\pi} \int_0^\infty \exd \omega\; e^{- i \omega (t - t')} |C_{\ell 0 }(\omega)|^2 \left[ j_{\ell}(\omega |\bfx|) + \sfrac{\pi \eta_{\star}}{\Gamma(\ell+\tfrac{1}{2})\Gamma(\ell +\tfrac{3}{2})} \left( \frac{\omega \epsilon_{\star}}{2} \right)^{2 \ell + 1} y_{\ell}(\omega |\bfx| ) \right] \qquad \qquad \\
& \ & \qquad \qquad \qquad \times \left[ j_{\ell}(\omega |\bfx'|) + \sfrac{\pi \eta_{\star}}{\Gamma(\ell+\tfrac{1}{2})\Gamma(\ell +\tfrac{3}{2})} \left( \frac{\omega \epsilon_{\star}}{2} \right)^{2 \ell + 1} y_{\ell}(\omega |\bfx'| ) \right] P_{\ell}( \cos \theta)\ . \notag
\eea
The terms in this sum involving two factors of $j_\ell$ reproduce the standard vacuum Minkowski Wightman function in the absence of the hotspot source. 

\subsubsection*{Leading order in $\omega \epsilon_\star$}

To make further progress we assume $\omega \epsilon_\star \ll 1$, which is a natural limit for modes with energies much smaller than the UV scale $1/\epsilon_\star$. In this case because $\omega\epsilon_\star$ appears raised to the power $2\ell + 1$ it follows that the leading regime comes purely from the $s$-wave partial wave with $\ell = 0$.  Using
\be
j_{0}( \omega |\bfx| ) = \frac{\sin(x)}{x} \quad \quad \mathrm{and} \quad \quad y_{0}(x) = - \frac{\cos(x)}{x} \ ,
\ee
the contribution up to leading (linear) nontrivial order in $\omega\epsilon_\star$ can be simplified to
\bea
\braket{\phi (t,\bfx)\phi(t',\bfx')} & \simeq &\frac{1}{4\pi^2} \int_0^\infty \exd \omega\; e^{- i \omega (t - t')}  \left\{\omega \sum_{\ell =0}^{\infty}(2\ell + 1)  j_{\ell}(\omega |\bfx|) j_{\ell}(\omega |\bfx'|) P_{\ell}( \cos \theta) \right. \nn\\
&& \qquad\qquad\left.  - \frac{ (4 \pi \eta_{\star} \epsilon_{\star} )}{16\pi^3|\bfx| |\bfx'|} \;  \bigg[ \sin(\omega |\bfx|) \cos(\omega |\bfx'|) + \cos(\omega |\bfx|) \sin(\omega |\bfx'|) \bigg] + \cO(\omega^2 \epsilon_{\star}^2) \right\}  \ .\notag \\
  & \simeq &\frac{1}{4\pi^2|\bfx - \bfx'|} \int_0^\infty \exd \omega\; e^{- i \omega (t - t')} \left\{ \sin \left(\omega  |\bfx - \bfx'| \right) \phantom{\frac12} \right. \\
&& \qquad\qquad\qquad\qquad\qquad\qquad\qquad\qquad \left.  - \frac{ (4 \pi \eta_{\star} \epsilon_{\star} )}{16\pi^3|\bfx| |\bfx'|}  \, \sin\Bigl[ \omega  \big( |\bfx| + |\bfx'| \big)\Bigr] + \cO(\omega^2 \epsilon_{\star}^2) \right\} \ . \notag
\eea
Evaluating the remaining integrals then gives
\bea \label{Wightmanlambdastar}
\braket{\phi (t,\bfx)\phi(t',\bfx')} & \simeq &\frac{1}{4\pi^2} \left[ \frac{1}{- (t-t' - i\delta)^2 + |\bfx - \bfx'|^2 } \right]\\
& \ & \qquad\qquad\qquad  + \frac{ (4 \pi \eta_{\star} \epsilon_{\star} )}{16\pi^3|\bfx| |\bfx'|} \bigg[  \frac{|\bfx| + |\bfx'|}{(t - t' - i \delta)^2 - ( |\bfx| + |\bfx'|)^2 } \bigg] \ , \notag
\eea
where $\delta = 0^+$ is the usual positive infinitesimal that is taken to zero at the end of the calculation. 

Notice that \pref{Wightmanlambdastar} precisely agrees with the result found in \pref{pertcorr3}, and so guarantees that the result \pref{WightmanlambdastarResum} is found once the combination $\eta_\star \epsilon_\star$ is traded for $\lambda(\epsilon)$ using \pref{epsilonstarvslambda0}. Higher orders in $\omega \epsilon_\star$ can be included systematically by including higher partial waves and by working to higher order in the small $\omega\epsilon$ expansion of the mode-functions. 

\section{Exact two-point correlator}
\label{sec:exactcorr}

In this section we evaluate the $\phi$ mode functions without perturbing in $\lambda$ and $\tilde g$, and sum these modes to obtain the exact $\langle \phi \phi \rangle$ Wightman function. 

\subsection{Mode functions}
\label{sec:exactmodes}

The first step is to find the mode functions in a way that does not rely on couplings being small. To this end we must solve equations \pref{Seliminated5} and \pref{saeliminated5}, which are repeated here for ease of reference,  for the mode functions $S_\bfp$ and $s^a_\bfp$:
\bea
S_{\bfp}(t,\bfx) & = & - \frac{\lambda \Theta(t - |\bfx|) }{ 4 \pi |\bfx| } \Bigl( e^{-i E_p (t - |\bfx|)} + S_{\bfp}(t - |\bfx|,\mathbf{0}) \Bigr) - \frac{\delta^{ab} g_a g_b }{16 \pi^2 |\bfx| } \, \delta(t - |\bfx|)\\
& \ & \qquad \qquad \qquad \qquad  - \frac{\delta^{ab} g_a g_b }{16 \pi^2 |\bfx|}\, \Theta(t - |\bfx|) \Bigl[ (- i E_p ) e^{-i E_p (t - |\bfx|)} + \partial_t S_{\bfp}(t -|\bfx| ,\mathbf{0}) \Bigr] \nn
\eea
and
\bea
s^{a}_{\bfp}(t,\bfx) &=& - \frac{\delta^{ab} g_b}{4 \pi |\bfx| }\, \Theta(t - |\bfx|) \, e^{-i E_p (t - |\bfx|)} - \frac{\lambda \Theta(t - |\bfx|)}{4 \pi |\bfx| } \,s^{a}_{\bfp}(t - |\bfx|,\mathbf{0}) \nn\\
&&\qquad\qquad\qquad\qquad\qquad - \frac{\delta^{bc} g_b g_c }{16 \pi^2 |\bfx|}\, \Theta(t - |\bfx|)\, \partial_t s^a_{\bfp}(t - |\bfx| ,\mathbf{0})  
\eea
subject to the initial conditions \pref{ICs}.

Clearly, whenever $t<|\bfx|$ the right-hand sides of these equations vanish and so they imply that $S_{\bfp}(t,\bfx)  = s^a_{\bfp}(t,\bfx) =0$, as required by causality. In the opposite case $t > |\bfx|$ they instead become
\bea \label{Seliminated6}
S_{\bfp}(t,\bfx) & = & - \frac{\lambda }{ 4 \pi |\bfx| } \big[ e^{-i E_p (t - |\bfx|)} + S_{\bfp}(t - |\bfx|,\mathbf{0}) \big] \\
& \ & \qquad \qquad \qquad \qquad  - \frac{\delta^{ab} g_a g_b}{16 \pi^2 |\bfx|} \Bigl[ (- i E_p ) e^{-i E_p (t - |\bfx|)} + \partial_t S_{\bfp}(t -|\bfx| ,\mathbf{0}) \Bigr] \nn \qquad \mathrm{when}\ t > |\bfx| \nn
\eea
and
\be \label{saeliminated6}
s^{a}_{\bfp}(t,\bfx) = - \sfrac{\delta^{ab} g_b}{4 \pi  |\bfx| } \, e^{-i E_p (t - |\bfx|)} - \sfrac{\lambda}{4 \pi |\bfx| } s^{a}_{\bfp}(t - |\bfx|,\mathbf{0}) - \sfrac{\delta^{bc} g_b g_c}{16 \pi^2 |\bfx|} \,\partial_t s^a_{\bfp}(t - |\bfx| ,\mathbf{0}) \qquad \mathrm{when}\ t > |\bfx| \ .
\ee
As suggested by the perturbative case, solutions to these equations unsurprisingly divergence at the location of the compact source $\bfx =\mathbf{0}$, which we regulate as before by replacing $\bfx = 0$ with $|\bfx| = \epsilon$: $S_{\bfp}(t-|\bfx|,\mathbf{0}) = S_{\bfp}(t-|\bfx|,\mathbf{y}) \big|_{|\bfy| = \epsilon}$ and $\partial_t S^{(0)}_{\bfp}(t-|\bfx|,\mathbf{0}) = \partial_t S^{(0)}_{\bfp}(t-|\bfx|,\mathbf{y}) \big|_{|\bfy| = \epsilon}$ and so on.

To solve \pref{Seliminated6} and \pref{saeliminated6} (in the special case where $g_a = \tilde{g}/\sqrt{N}$ for all $a$) we make the {\it ans\"atze}
\be
S_{\bfp}(t,\bfx) = F(|\bfx|) e^{ - i E_{p} ( t - |\bfx| ) } \qquad \mathrm{when}\ t > |\bfx| 
\ee
and
\be
s^a_{\bfp}(t,\bfx) = G(|\bfx|) e^{ - i E_{p} ( t - |\bfx| ) }  \quad \hbox{(for all $a$)}\qquad \mathrm{when}\ t > |\bfx| \,,
\ee
and so $S^{(0)}_{\bfp}(t-|\bfx|,\mathbf{0}) := F(\epsilon) \, e^{- i E_{p} (t - |\bfx| - \epsilon)}$ and $\partial_t S^{(0)}_{\bfp}(t-|\bfx|,\mathbf{0}) := F(\epsilon)\,  (- i E_p ) e^{- i E_{p} (t - |\bfx| - \epsilon)}$ and similarly for $s^{a}_{\bfp}$. These {\it ans\"atze} solve \pref{Seliminated6} and \pref{saeliminated6} provided $F$ and $G$ satisfy
\be
F(|\bfx|) = \bigg( - \frac{\lambda}{4 \pi |\bfx| } +  \frac{i \tilde{g}^2 E_p}{16 \pi^2 |\bfx|} \bigg) \big[ 1 + F(\epsilon) \, e^{+ i E_p \epsilon} \big] 
\ee
and
\be
G(|\bfx|) = - \frac{\tilde{g}}{4 \pi \sqrt{N} |\bfx| } + \bigg( - \frac{\lambda}{4 \pi |\bfx| } +  \frac{i \tilde{g}^2 E_p}{16 \pi^2 |\bfx|} \bigg) \,  G(\epsilon) \, e^{+ i E_p \epsilon} 
\ee
whose solutions are
\be
F(|\bfx|) = \frac{ - \dfrac{ \lambda }{4 \pi |\bfx| } +  \dfrac{i \tilde{g}^2 E_p}{16 \pi^2 |\bfx|}}{ 1 - \left( - \dfrac{\lambda}{4 \pi \epsilon } +  \dfrac{i \tilde{g}^2 E_p}{16 \pi^2 \epsilon} \right) e^{+ i E_p \epsilon } } \qquad \mathrm{and} \qquad 
G(|\bfx|) =  \frac{- \dfrac{\tilde{g}}{4 \pi \sqrt{N} |\bfx |} e^{- i E_p (t - |\bfx|)} }{ 1 - \left( - \dfrac{\lambda}{4 \pi \epsilon } +  \dfrac{i \tilde{g}^2 E_p}{16 \pi^2 \epsilon} \right) e^{+ i E_p \epsilon }   } \ .
\ee
Recalling that the derivation of equations \pref{Seliminated5} and \pref{saeliminated5} assume $\epsilon E_p \ll 1$ --- see the discussion below equations \pref{Seliminated3} and \pref{saeliminated3} --- we can take $e^{ i E_p \epsilon} \simeq 1$ without loss, giving the mode functions
\be \label{Ssol}
S_{\bfp}(t,\bfx) \simeq \frac{ - \dfrac{ \lambda }{4 \pi |\bfx| } +  \dfrac{i \tilde{g}^2 E_p}{16 \pi^2 |\bfx|}}{ 1 + \dfrac{\lambda}{4 \pi \epsilon } - \dfrac{i \tilde{g}^2 E_p}{16 \pi^2 \epsilon}  } e^{ - i E_{p} ( t - |\bfx| ) } \qquad \mathrm{when}\ t > |\bfx| 
\ee
and
\be \label{sasol}
s^{a}_{\bfp}(t,\bfx) \simeq \frac{- \dfrac{\tilde{g}}{4 \pi \sqrt{N} |\bfx |} e^{- i E_p (t - |\bfx|)} }{ 1 + \dfrac{\lambda}{4 \pi \epsilon } - \dfrac{i \tilde{g}^2 E_p}{16 \pi^2 \epsilon} } e^{- i E_p (t - |\bfx|) } \qquad \mathrm{when}\ t > |\bfx| \ .
\ee

Comparing \pref{Ssol} and \pref{sasol} to the perturbative solutions \pref{S0order} and \pref{sa0order} in the regime $t > |\bfx|$, it is clear that the perturbative expression are valid only when 
\be
\left| \dfrac{\lambda}{4 \pi \epsilon } - \dfrac{i \tilde{g}^2 E_p}{16 \pi^2 \epsilon} \right| \ll 1 \ .
\ee
For this to be small for all $\bfp$ requires each term to separately be small. If $\Lambda$ is a bulk cutoff, so that $E_{p} < \Lambda$ for all $\bfp$ (which in principle is logically distinct from the UV cutoff $1/\epsilon$ associated with proximity to the hotspot) then at face value the perturbative limit requires
\be
\frac{\lambda}{4 \pi \epsilon} \ll 1 \qquad \mathrm{and} \qquad \frac{\Lambda \tilde{g}^2}{16\pi^2 \epsilon} \ll 1\ .
\ee

\subsection{Performing the mode sum}

Given the mode functions in \pref{Ssol} and \pref{sasol} the Wightman function of \pref{CorrelationStart} can be evaluated as a mode sum. For two points $(t,\bfx)$ and $(t',\bfx')$ satisfying $t > |\bfx|$ and $t' > |\bfx'|$, the result becomes
\bea \label{fullW}
W_{\beta}(t,\bfx ; t', \bfx') & = & \frac{1}{Z_\beta} \mathrm{Tr}\Bigl[ \phi _{\ssH}(t,\bfx) \phi _{\ssH}(t',\bfx') \big( \ket{\mathrm{vac}} \bra{\mathrm{vac}} \otimes  e^{ - \beta \cH_{-}} \big) \Bigr] \nn\\
& =: & \mathscr{S}(t,\bfx ; t, \bfx') + \mathscr{E}_\beta(t,\bfx ; t, \bfx') 
\eea
where the functions $\mathscr{S}$ and $\mathscr{E}_\beta$ are defined by
\bea \label{curlySdef}
\mathscr{S}(t,\bfx ; t', \bfx') & = & \int \frac{\exd^3 \bfp}{(2\pi)^3 2 E_p} \bigg( e^{- i E_p t + i \bfp \cdot \bfx} + S_{\bfp}(t,\bfx) \bigg) \bigg( e^{+ i E_p t' - i \bfp \cdot \bfx'} + S^{\ast}_{\bfp}(t',\bfx')  \bigg)
\eea
and
\bea \label{curlyEdef}
\mathscr{E}_\beta(t,\bfx ; t', \bfx') & = & \frac{1}{Z_{\beta}} \sum_{a,b=1}^{N} \int \frac{\exd^3 \bfp}{\sqrt{(2\pi)^3 2E_p}} \int \frac{\exd^3 \bfk}{\sqrt{(2\pi)^3 2E_k}} \bigg( s^a_{\bfp}(t,\bfx) s^{b\ast}_{\bfk}(t',\bfx') \TrB\left[ \mfb_{\bfp}^{a} \mfb_{\bfk}^{b\ast} e^{- \beta \cH_{-}} \right] \\
& \ & \qquad \qquad \qquad \qquad \qquad \qquad \qquad \qquad \qquad \qquad + s^{a\ast}_{\bfp}(t,\bfx) s^b_{\bfk}(t',\bfx') \TrB\left[ \mfb_{\bfp}^{a\ast} \mfb_{\bfk}^{b} e^{- \beta \cH_{-}} \right] \bigg) \nn \ .
\eea

These mode sums are performed explicitly in Appendix \ref{App:modesumfull}, giving the following result for $\mathscr{S}$
\bea \label{curlySanswer}
\mathscr{S}(t,\bfx ; t', \bfx') & = & \frac{1}{4 \pi^2 \left[ - (t - t' - i \delta)^2 + |\bfx - \bfx'|^2 \right]} \label{curlySanswer} \\
& \ & \qquad + \frac{ 2 \epsilon^2}{\tilde{g}^2 |\bfx| |\bfx'|} \bigg[ I_{-}(t-t'+|\bfx|+|\bfx'|, c) - I_{-}(t-t'-|\bfx|+|\bfx'|, c) \nn \\
& \ & \qquad \qquad \qquad \qquad \qquad \qquad - I_{+}( t - t' - |\bfx| + |\bfx'| , c) + I_{+}( t - t' - |\bfx| - |\bfx'| , c) \bigg] \nn \\
& \ & \qquad + \frac{\epsilon}{8 \pi^2 |\bfx| |\bfx'|} \bigg[ - \frac{1}{t - t' + |\bfx| + |\bfx'| - i \delta} + \frac{1}{t - t' - |\bfx| - |\bfx'| - i \delta } \bigg] \nn \\
& \ & \qquad \quad -  \frac{32 \pi^2 \epsilon^4 ( 1 + \frac{\lambda}{2 \pi \epsilon} )}{ \tilde{g}^4 |\bfx| |\bfx'|} \bigg[ I_{-}( t - t' -|\bfx| + |\bfx'| , c ) + I_{+}( t - t' -|\bfx| + |\bfx'| , c )  \bigg] \nn \\
& \ & \qquad \qquad - \frac{\epsilon^2}{4 \pi^2 |\bfx| |\bfx'| ( t - t' -|\bfx| + |\bfx'| - i \delta )^2} \nn
\eea
where the parameter $c$ is the following combination of couplings and $\epsilon$,
\be \label{cdef}
c := \frac{16 \pi^2 \epsilon}{\tilde{g}^2} \left( 1  + \frac{\lambda}{4 \pi \epsilon} \right) \,.
\ee
The functions $I_{\mp}(\tau)$ are defined by
\be
I_{\mp}(\tau) = e^{\pm c \tau} E_{1}\big( \pm c [\tau - i \delta] \big)
\quad \hbox{where} \quad
 E_{1}(z) := \int_z^\infty \exd u\; \frac{e^{-u}}{u} \,,
\ee
is the $E_{n}$-function with $n=1$ (closely related to the exponential integral function) and the limit $\delta \to 0^{+}$ is (as usual) understood.

The temperature-dependent contribution similarly evaluates to 
\be  \label{curlyEanswer}
\mathscr{E}_\beta(t,\bfx ; t, \bfx') = \frac{2\epsilon^2}{\tilde{g}^2 |\bfx| |\bfx'|} \bigg[ \Phi\bigg( e^{ - \tfrac{2\pi (t - t' - |\bfx| + |\bfx'| - i \delta)}{\beta}}, 1 , \frac{c\beta}{2\pi} \bigg) + \Phi\bigg( e^{ + \tfrac{2\pi (t - t' - |\bfx| + |\bfx'| - i \delta)}{\beta}}, 1 , \frac{c\beta}{2\pi} \bigg) \bigg]  - \frac{2 \pi}{c\beta} \bigg]
\ee
where $\Phi(z,s,a)$ is the Lerch transcendent, defined by the series $\Phi(z,s,a) = \sum_{n=0}^{\infty} \frac{z^n}{(a+n)^s}$ for complex numbers in the unit disc (with $|z|<1$), and by analytic contribution elsewhere in the complex plane. Asymptotic forms for the functions $I_\pm$ and $\Phi$ are given in Appendix \ref{App:curlySpert}. 

These expressions pass all of the smell tests. In particular, the full correlation function $W_{\beta} = \mathscr{S} + \mathscr{E}_\beta$ reduces to the perturbative correlation function quoted in \pref{pertcorr} in the appropriate perturbative limit. The perturbative expression in powers of $\tilde g$ is obtained from the asymptotic form when both $c \tau \gg 1$ in $\mathscr{S}$ (as shown explicitly in \S\ref{App:curlySpert}) and $c\beta \gg 1$ in $\mathscr{E}_\beta$ (see \S\ref{App:curlyEpert}). The expression found in this limit agrees with \pref{WightmanlambdastarResum}, obtained earlier by resumming the perturbative result to all orders in ${\lambda}/{(4 \pi \epsilon)}$. Further taking $\lambda / (4\pi\epsilon) \ll 1$ in this expression then reproduces exactly the perturbative correlation function \pref{pertcorr} found previously.

Finally, this result has a thermal character in the sense that $\mathscr{E}_{\beta}$ satisfies the Kubo-Martin-Schwinger (KMS) condition \cite{Kubo,Martin}, as we show explicitly in Appendix \ref{app:KMScurlyE}.

\section{Conclusions}
\label{sec:Conclusions}

Black hole physics is a puzzle wrapped in an enigma hidden by a horizon, and ongoing studies of information loss show that Hawking radiation is the discovery that keeps on giving. Although Hawking radiation in principle occurs in the weak-field regime where calculation control should be good, reliable and explicit calculations of corrections to Hawking radiation are relatively rare (partly due to the extremely late times involved). Yet resolving issues such as the existence (or not) of firewalls, possible loss of locality and the like are crucial towards gaining an understanding of what a theory of quantum gravity should ultimately look like.

In this paper we construct a Caldeira-Leggett type \cite{CaldeiraLeggett} toy model of a hot compact relativistic object that captures some of the features of black holes, and so can be used as a benchmark against which real calculations can be compared. The model is simple enough to be solved explicitly, but complicated enough to capture some of the open-system effects believed to be important for black holes. Although the toy model cannot in itself resolve the thorniest puzzles associated with horizons, it can show which features are shared with more mundane systems that are hot and relatively small. 

There are several directions in which this model might fruitfully be explored. In a companion work \cite{Companion}, we apply to it several approximate Open-EFT techniques designed to probe late-time evolution, to better identify their domains of validity and whether they can illuminate the extent to which the open nature of the hotspot causes a breakdown of local descriptions of the physics of the field $\phi$ living in $\cR_+$. A second companion \cite{Kaplanek:2021xxx} explores the behaviour of an Unruh detector (or qubit) that couples to the field $\phi$ in the vicinity of the hotspot, to determine the extent to which it thermalizes as a function of its couplings and its distance from the hotspot. 

Other useful directions might explore the regime where the radius $\xi$ of the interaction sphere $\cS_\xi$ is not small, and so parallels the EFT discussion of \cite{Burgess:2018pmm}. By tuning the physics at the interface between the spaces $\cR_{\pm}$ one might hope to mock up the entanglement between modes inside and outside the horizon, and provide a simple analog of the matching calculations often used to extract black hole phenomenology from scattering amplitudes. The model can be further developed to include redshifting and the geometrical effects of gravitational fields in regions $\cR_\pm$. We offer up the hotspot model in the hopes that such comparisons and extensions will prove instructive. 

\section*{Acknowledgements}
We thank Sarah Shandera for making the suggestion that got this project started, as well as Meg Carrington, Walter Goldberger, Gabor Kunstatter and Ira Rothstein for useful conversations. Thanks also to KITP Santa Barbara for hosting the workshop (during a pandemic) that led us to think along these lines. CB's research was partially supported by funds from the Natural Sciences and Engineering Research Council (NSERC) of Canada. Research at Perimeter Institute is supported in part by the Government of Canada through the Department of Innovation, Science and Economic Development Canada and by the Province of Ontario through the Ministry of Colleges and Universities.

\appendix

\section{Thermal correlation functions}

This appendix computes the thermal correlation functions used in the main text. This is partly done as a confidence-building exercise to verify the techniques used elsewhere. 

\subsection{Free thermal correlation function}
\label{sec:freethermaltrace}

The first correlation to compute is the standard two-point function for a free thermal field. Although the main text works with $N$ fields for this correlation function it suffices to work with only one of the $N$ copies and evaluate
\be\label{freethermtrace}
\langle \chi(t,\bfx) \chi(0,\mathbf{0}) \rangle_{\beta} = \frac{1}{Z}\; \mathrm{Tr}\Bigl[ \chi(t,\bfx) \chi(0,\mathbf{0}) \,e^{- \beta \cH }   \Bigr]  \,,
\ee
where translation invariance in $\cR_-$ is used to set one of the fields to $(t',\bfx') = (0,\mathbf{0})$, the relevant free-particle part of the Hamiltonian $\cH_-$ is denoted $\cH$ and $Z$ denotes the partition function
\bea
Z := \mathrm{Tr}[ e^{- \beta \cH  } ] &=& \sum_{N=0}^{\infty} \sum_{\{n_p\}_{\ssN}} \bra{ \{ n_p \}_{\ssN} } e^{- \beta \cH  } \ket{ \{ n_{p} \}_\ssN } \nn\\
 &=&  \prod_{\bfp}  \sum_{n_p = 0}^{\infty}  e^{- \beta E_p n_p } = \prod_{\bfp} \frac{1}{1 - e^{ - \beta E_{p}}} =: \prod_\bfp Z_p\,.
\eea

This last expression temporarily switches for convenience to discretely normalized momenta, as would be appropriate when the spatial volume $\cV$ is finite, and takes the eigenvalues of $\cH$ to be $\sum_p n_p E_p$. The field expansion for this normalization of states is
\be \label{fieldexpFV}
      \chi(t,\bfx) =   \sum_\bfp \frac{1}{\sqrt{2E_p\cV}} \, \Bigl[ b_\bfp \, e^{-iE_pt+i \bfp \cdot \bfx} +  b^*_\bfp \, e^{ iE_pt - i \bfp \cdot \bfx} \Bigr]
\ee
where $b_p$ is the discretely normalized destruction operator, and the conversion between discrete and continuum normalization is given by
\be
    b_\bfp = \left[ \frac{(2\pi)^3}{\cV} \right]^{1/2} \, \mfb_\bfp  \quad \hbox{and} \quad
    \sum_\bfp = \frac{\cV}{(2\pi)^3} \int \exd^3p \,,
\ee
and so on. 

Inserting \pref{fieldexpFV} into \pref{freethermtrace} allows it to be evaluated in the occupation-number basis, leading to
\bea
\langle \chi(t,\bfx) \chi(0,\mathbf{0}) \rangle_{\beta} & = & \frac{1}{2 \cV Z } \sum_{\bfk\bfq} \frac{\mathrm{Tr}\left[ \left( e^{- i E_k t + i \bfk \cdot \bfx} b_{\bfk} + e^{+ i E_k t - i \bfk \cdot \bfx} b_{\bfk}^{\ast} \right) \left( b_{\bfq} + b_{\bfq}^{\ast} \right) e^{- \beta \cH }   \right] }{\sqrt{E_k E_q}} \nn\\
&=&  \frac{1}{2 \cV Z }\sum_{\bfk} \frac{\mathrm{Tr}\left[ \left( e^{- i E_k t + i \bfk \cdot \bfx} b_{\bfk} + e^{+ i E_k t - i \bfk \cdot \bfx} b_{\bfk}^{\ast} \right) \left( b_{\bfk} + b_{\bfk}^{\ast} \right) e^{- \beta \cH }   \right] }{E_k} \\
& = & \frac{1}{2 \cV Z } \sum_{\bfk} \frac{ e^{- i E_k t + i \bfk \cdot \bfx} \mathrm{Tr}\left[b_{\bfk} b_{\bfk}^{\ast} e^{- \beta \cH } \right] + e^{+ i E_k t - i \bfk \cdot \bfx} \mathrm{Tr}\left[ b^{\ast}_{\bfk} b_{\bfk} e^{- \beta \cH } \right]   }{E_k} \,, \nn
\eea
and so, using $b_\bfk b^*_\bfk = b^*_\bfk b_\bfk + 1$, this becomes 
\bea
\langle \chi(t,\bfx) \chi(0,\mathbf{0}) \rangle_{\beta} & = & \frac{1}{2 \cV } \sum_{\bfk} \sfrac{ e^{- i E_k t + i \bfk \cdot \bfx} }{E_k} +  \frac{1}{\cV Z} \sum_{\bfk} \sfrac{ \cos\left( E_k t - \bfk \cdot \bfx \right) }{E_k} \; \mathrm{Tr}\left[b^{\ast}_{\bfk} b_{\bfk} \, e^{- \beta \cH } \right]  \nn \\
& = &  \int \frac{\exd^{3}k}{(2\pi)^3} \frac{ e^{- i E_k t + i \bfk \cdot \bfx} }{2E_k} +  \frac{1}{\cV Z} \sum_{\bfk} \sfrac{ \cos\left( E_k t - \bfk \cdot \bfx \right) }{E_k} \; \mathrm{Tr}\left[b^{\ast}_{\bfk} b_{\bfk} \, e^{- \beta \cH } \right] 
\eea
The required trace is a standard manipulation
\bea \label{navgtherm}
 \frac{1}{Z} \mathrm{Tr}\left[ b^{\ast}_{\bfk} b_{\bfk} e^{- \beta \cH } \right] & = &\frac{1}{Z} \sum_{N=0}^{\infty} \sum_{\{n_p\}_{N}}  \bra{ \{ n_p \}_{\ssN} } b^{\ast}_{\bfk} b_{\bfk} \, e^{- \beta \sum_{\bfp} E_{p} n_p}  \ket{ \{ n_p \}_{\ssN} } \nn\\
&=& \prod_\bfp \frac{1}{Z_p} \sum_{n_p=0}^\infty  n_p \,  e^{- \beta  E_{p} n_p} = \prod_\bfp \frac{1}{e^{\beta E_p} - 1} \,,
\eea
and so
\be \label{thermtoFourier}
\langle \chi(t,\bfx) \chi(0,\mathbf{0}) \rangle_{\beta} =  \int \frac{\exd^{3}k}{(2\pi)^3} \frac{ e^{- i E_k t + i \bfk \cdot \bfx} }{2E_k} + \int \frac{\exd^{3}k}{(2 \pi)^3}\frac{ \cos\left( E_k t - \bfk \cdot \bfx \right) }{E_k \left(e^{\beta E_{k}} - 1\right)} \,.
\ee

It remains to perform the integrals. The first evaluates to the vacuum Wightman function while the second is the thermal correction. That is
\be
\int \frac{\exd^{3}k}{(2\pi)^3} \frac{ e^{- i E_k t + i \bfk \cdot \bfx} }{2E_k} =  \frac{1}{4 \pi^2 \left[ - (t - i \delta)^2 + |\bfx|^2 \right]} \,,
\ee
where the limit $\delta \to 0^{+}$ is understood, and the geometric series 
\be
\frac{1}{e^{\beta k} - 1 } = \sum_{n=1}^{\infty} e^{ -n \beta  k} \ ,
\ee
allows the remaining term to be written
\bea
 \int \frac{\exd^{3}k}{(2 \pi)^3}\frac{ \cos\left( E_k t - \bfk \cdot \bfx \right) }{E_k \left(e^{\beta E_{k}} - 1\right)} & = &  \frac{1}{4 \pi^2 |\bfx|} \int_0^\infty \exd k \; \frac{ e^{- i k t} + e^{+ i k t} }{\left(e^{\beta k} - 1\right)}\; \sin\left(k |\bfx|   \right) \nn\\
 &=& \frac{1}{4 \pi^2 |\bfx|}  \sum_{n=1}^{\infty}  \int_0^\infty \exd k  \; \left( e^{- i k t} + e^{+ i k t} \right) e^{- n \beta k} \sin\left(k |\bfx|  \right) \nn \\
& = & \frac{1}{4 \pi^2 |\bfx|}  \sum_{n=1}^{\infty}  \int_0^\infty \exd k \; \Bigl\{ e^{- n \beta k} \sin\big[k ( t + |\bfx| )   \big]  -   e^{- n \beta k}  \sin\big[k ( t - |\bfx| )   \big] \Bigr\} \notag \\
& = &   \frac{1}{8 \pi \beta |\bfx|} \cdot \frac{2}{\pi}  \sum_{n=1}^{\infty} \bigg[ \frac{({t + |\bfx|})/{\beta} }{n^2 + [({t + |\bfx|})/{\beta}]^2}  - \frac{({t - |\bfx|})/{\beta}}{n^2 + [({t - |\bfx|})/{\beta}]^2} \bigg] \,. 
\eea
This final sum can be performed using identity (1.421.4) from \cite{grad}, which for $z \in \mathbb{R}$ states
\be
\coth\left( \pi z \right) = \frac{1}{\pi z} + \frac{2}{\pi} \sum_{n=1}^{\infty} \frac{z}{n^2 + z^2} ,
\ee
and so
\bea  \label{beforecancel}
\langle \chi(t,\bfx) \chi(0,\mathbf{0}) \rangle_{\beta} & = & \frac{1}{8 \pi^2 |\bfx|} \bigg( \frac{1}{t + |\bfx|- i \delta } - \frac{1}{t - |\bfx| - i \delta }  - \frac{1}{t + |\bfx| } + \frac{1}{ t -  |\bfx| }\bigg) \\
&&\qquad\qquad\qquad\qquad\qquad\qquad+ \frac{\coth\left[ {\pi ( t + | \bfx | ) }/{\beta} \right] - \coth\left[ {\pi ( t -| \bfx | ) }/{\beta} \right] }{8 \pi \beta |\bfx|} \ . \nn
\eea
In the limit $\delta \to 0^{+}$ (for real $t$ and $\bfx$) the real parts of the second line cancel leaving 
\bea
\langle \chi(t,\bfx) \chi(0,\mathbf{0}) \rangle_{\beta} & = &  \frac{\coth\Bigl[  {\pi ( t + | \bfx | - i \delta ) }/{\beta} \Bigr] - \coth\Bigl[ {\pi ( t - | \bfx | - i \delta ) }/{\beta} \Bigr] }{8 \pi \beta |\bfx|} \label{thermalcorrelatorab2}
\eea
which is the result quoted in \pref{thermalcorrelatorab} of the main text (after restoring the arguments $t'$ and $\bfx'$ of the second field using translational invariance).

\subsubsection{The KMS Condition}
\label{app:thermKMS}

A subtlety with the $i \delta$-prescription in the above formula (\ref{thermalcorrelatorab2}) is that it is only correct for {\it real} $t$ and $|\bfx|$. This matters because the thermal correlation function is supposed to obey the Kubo-Martin-Schwinger (KMS) condition given by
\be \label{App:KMSdef}
\langle \chi(t- i \beta,\bfx) \chi(t',\bfx') \rangle_{\beta} = \langle \chi(t',\bfx') \chi(t,\mathbf{x}) \rangle_{\beta} = \langle \chi(t,\bfx) \chi(t',\bfx') \rangle_{\beta}^{\ast} \ ,
\ee
which \pref{thermalcorrelatorab2} apparently does not satisfy because of the $i\delta$-prescription used there. 

Later in Appendix \ref{app:KMScurlyE} we prove that the function $\mathscr{E}_{\beta}$ obeys a KMS-type condition, and so for later use we here flesh out the argument for why the KMS is explicitly obeyed for the free thermal correlator $\langle \chi(t,\bfx) \chi(t',\mathbf{x}') \rangle_{\beta}$. To see how this works, go back to \pref{thermtoFourier} (with arguments $t'$ and $\bfx'$ reinstated) which says
\be
\langle \chi(t,\bfx) \chi(t',\bfx) \rangle_{\beta} = \int \frac{\exd^{3}k}{(2\pi)^3} \frac{ e^{- i E_k (t-t') + i \bfk \cdot (\bfx-\bfx')} }{ 2 E_k  \left(1 - e^{-\beta E_{k}} \right) } + \int \frac{\exd^{3}k}{(2 \pi)^3}\frac{ e^{+ i E_k (t-t') - i \bfk \cdot (\bfx-\bfx')} }{2 E_k \left(e^{\beta E_{k}} - 1\right)}
\ee
after use of the identity $1 + ( e^{\beta E_k} -1 )^{-1} = (1 - e^{-\beta E_k})^{-1}$. Evaluating this with $t \to t - i \beta$ gives\footnote{Notice that replacing $t \to t + i b$ for some imaginary part $b$ in the above makes the integral converge for all $- \beta  < b < 0$, since $(e^{\beta E_{k}} - 1)^{-1} \simeq e^{- \beta E_{k}}$ and $(1 - e^{-\beta E_{k}} )^{-1} \simeq 1 + e^{- \beta E_{k}}$ for large momenta $\beta E_{k} \gg 1$. This means that the correlator is a complex-analytic function of time $t \in \mathbb{C}$ in this strip \cite{Takagi:1986kn} (with $- \beta < \mathrm{Im}[t] < 0$).} 
\bea  \label{KMSanalytic}
\langle \chi(t - i \beta,\bfx) \chi(t',\bfx') \rangle_{\beta} & = & \int \frac{\exd^{3}k}{(2\pi)^3} \frac{ e^{ -  \beta E_k } e^{- i E_k (t-t') + i \bfk \cdot (\bfx-\bfx')} }{ 2 E_k  \left(1 - e^{-\beta E_{k}} \right) } + \int \frac{\exd^{3}k}{(2 \pi)^3}\frac{e^{ +  \beta E_k}  e^{+ i E_k ( t -t') - i \bfk \cdot (\bfx-\bfx')} }{2 E_k \left(e^{\beta E_{k}} - 1\right)}   \nn\\
& = & \int \frac{\exd^{3}k}{(2\pi)^3} \frac{ e^{- i E_k (t-t') + i \bfk \cdot (\bfx-\bfx')} }{ 2 E_k  \left( e^{\beta E_{k}} - 1 \right) } + \int \frac{\exd^{3}k}{(2 \pi)^3}\frac{ e^{+ i E_k (t-t') - i \bfk \cdot (\bfx-\bfx')} }{2 E_k \left(1 - e^{-\beta E_{k}} \right)} \\
& = & \langle \chi(t',\bfx') \chi(t,\mathbf{x}) \rangle_{\beta} \,, \nn
\eea
as required by the KMS condition \pref{App:KMSdef}. 

The validity of the KMS condition can be made manifest in position space if we re-write \pref{thermalcorrelatorab} with an $i\delta$-prescription that is both consistent with the KMS condition and reduces to \pref{thermalcorrelatorab} in the limit of real $t$. To do this notice the identity
\be
\coth(a+b) - \coth(a-b) =  - \frac{\sinh(2b)}{\sinh(a+b)\sinh(a-b)} \qquad \qquad \mathrm{for\ any\ } a,b \in \mathbb{C}\ , 
\ee
which allows the correlation function to be written in the KMS-consistent form ({\it c.f.} formula (\ref{beforecancel}))
\be \label{thermcorrKMS}
\langle \chi(t,\bfx) \chi(t',\bfx') \rangle_{\beta} = - \; \sfrac{\sinh\left( \dfrac{2 \pi | \bfx - \bfx' | }{\beta} \right)}{8 \pi \beta |\bfx-\bfx'| \left[ \sinh\left( \dfrac{\pi ( t - t' + | \bfx - \bfx' | }{\beta} \right) - \dfrac{i \pi \delta}{\beta}  \right] \left[ \sinh\left( \dfrac{\pi ( t-t' - | \bfx - \bfx' | ) }{\beta} \right) - \dfrac{i \pi \delta}{\beta} \right]} \,.
\ee

\section{RG evolution}
\label{AppendixF}

This Appendix summarizes some parts of the renormalization evolution not made explicit in the main text, closely following the discussion in Appendix F of \cite{PPEFTH2}.

\subsection{Universal evolution}
\label{ssec:AppEvoEq}

The boundary conditions of the main text provide examples where the effective couplings are found to satisfy equations of the form
\be \label{appgvseps}
  g(\epsilon) =   \frac{A \rho_\epsilon^{2\zeta} + B}{C \rho_\epsilon^{2\zeta} + D}  \,,
\ee
where $g$ is a representative coupling and $\epsilon$ appears on the right-hand side through the variable $\rho_\epsilon$. In the example of \pref{BClambdavsDC2}, for instance, we have $g = \lambda/4\pi \epsilon$ while $\rho_\epsilon = \frac12 \, \omega \epsilon$, the power is $\zeta = \ell + \frac12$ and the parameters $A,B,C$ and $D$ are given explicitly by
\be \label{App:ABCDform}
   A = \ell  \,, \quad B = ( \ell +1)  \, X_\ell(\omega)  \,, \quad C = 1 \quad \hbox{and} \quad D = -X_\ell(\omega) \,,
\ee
where 
\be  
    X_\ell(\omega) := \frac{1}{\pi} \, \Gamma \left(\ell + \sfrac32 \right) \, \Gamma \left(\ell + \sfrac12 \right) \, \frac{D_\ell(\omega)}{C_\ell(\omega)} \,.
\ee

For later use, eq.~\pref{appgvseps} also inverts to give 
\be \label{useme1}
   \rho_\epsilon^{2\zeta} =  \frac{B-D g }{C g - A} \,.
\ee

The goal is to derive a universal differential version of this evolution (see, for example \cite{PPEFT, PPEFT2, PPEFT3, PPEFTDis} for more details). To start this off directly differentiate \pref{appgvseps} holding $A,B,C,D$ fixed, leading to
\be \label{appgdiffeq}
  \epsilon \, \frac{\exd g}{\exd \epsilon} = 2\zeta \left[ \frac{AD-BC}{(C \rho_\epsilon^{2\zeta} + D)^2} \right] \rho_\epsilon^{2\zeta}=  2\zeta \left[ \frac{(Cg - A)(B-Dg)}{AD-BC} \right] \,,
\ee
where the second equality uses \pref{useme1} to trade $\rho_\epsilon^{2\zeta}$ for $g$. This evolution equation has fixed points at $g = g_*$, where
\be 
   g_* = \frac{A}{C} \quad \hbox{or} \quad g_* = \frac{B}{D} \,,
\ee
which can also be seen as the $\rho_\epsilon \to 0$ and $\rho_\epsilon \to \infty$ limits of \pref{appgvseps}.

Equation \pref{appgdiffeq} can be put into a standard form by redefining $g$ to ensure that $g_* = \pm 1$. To this end write 
\be
  g(\epsilon) = u(\epsilon) + \frac12 \left( \frac{A}{C} + \frac{B}{D} \right) \,,
\ee
in terms of which the fixed points are
\be
  u_* = \pm \frac12 \left(\frac{A}{C} - \frac{B}{D}  \right) = \pm \left( \frac{AD-BC}{2CD} \right) \,,
\ee
and \pref{appgdiffeq} becomes
\be
  \epsilon \, \frac{\exd u}{\exd \epsilon}  
  =  - \frac{2\zeta CD}{AD-BC} \left[ u - \left( \frac{AD-BC}{2CD} \right) \right] \left[ u + \left( \frac{AD-BC}{2CD} \right) \right] \,.
\ee

Finally rescale
\be
   u = \left[ \frac{AD-BC}{2CD} \right] v 
\ee
to see that 
\be \label{appuniversalDE}
  \epsilon \, \frac{\exd v}{\exd \epsilon} =  \zeta (1-v^2) \,
\ee
is an automatic consequence of \pref{appgvseps} once one defines
\be \label{App:gvsv}
  g = u + \frac{AD+BC}{2CD}  
  =\frac12 \left( \frac{A}{C} - \frac{B}{D} \right) v + \frac12 \left( \frac{A}{C} + \frac{B}{D} \right) \,.
\ee

These expressions emphasize that although the positions of the fixed points for $g$ depend on the ratios $A/C$ and $B/D$, the speed of evolution along the RG flow depends only on $\zeta$. Indeed the general solution to \pref{appuniversalDE} is 
\be \label{appRGsolution}
  v(\epsilon) = \frac{ (v_0 + 1)(\epsilon/\epsilon_0)^{2\zeta} + (v_0 - 1)}{(v_0 + 1)(\epsilon/\epsilon_0)^{2\zeta} - (v_0 - 1)} 
\ee
where the integration constant is chosen to ensure $v(\epsilon_0) = v_0$. For $\zeta > 0$ this describes a universal flow that runs from $v = -1$ to $v = +1$ as $\epsilon$ flows from 0 to $\infty$. 

Since the trajectories given in \pref{appRGsolution} cannot cross the lines $v = \pm 1$ for any finite nonzero $\epsilon$ there are two categories of flow, distinguished by the flow-invariant sign of $|v| - 1$ (see Figure \ref{figureFlow}). That is, if $|v_0| - 1$ is negative (positive) for any $0 < \epsilon_0 < \infty$, then $|v(\epsilon)| -1$ is negative (positive) for all $0 < \epsilon < \infty$. Every trajectory is therefore uniquely characterized by a pair of numbers. These can equally well be chosen to be the pair $(\epsilon_0, v_0)$ that specifies an initial condition $v_0 = v(\epsilon_0)$, or it can be taken to be the pair $(\epsilon_\star, y_\star)$ where $y_\star = \hbox{sign}(|v|-1) = \pm 1$ distinguishes the two classes of trajectories, and $\epsilon_\star$ is defined as the value of $\epsilon$ for which $v(\epsilon_\star) = 0$ (if $y_\star = -1$) or the value for which $v(\epsilon_\star) = \infty$ (if $y_\star = + 1$). The parameterization using $(\epsilon_\star, y_\star)$ is useful because physical observables turn out to have particularly transparent expressions in terms of these variables.  

For the specific cases given in \pref{App:ABCDform} the fixed points are
\be
   \frac{A}{C} = \ell  \quad \hbox{and} \quad
   \frac{B}{D} = - \ell - 1 
\ee
and $\lambda$ is related to the universal scaling variable $v$ by \pref{App:gvsv}, which becomes
\be
  \frac{\lambda}{4\pi\epsilon} =\frac12 \Bigl[ \left( 2\ell+1 \right) v - 1 \Bigr]
  \quad\hbox{and so} \quad
  v = \frac{1}{2\ell+1} \left( \frac{\lambda}{2\pi \epsilon} + 1 \right) \,,
\ee
as used in the main text.

\section{Mode properties}

This Appendix evaluates several properties associated with the modes in the presence of a $\frac12 \, \lambda \, \phi^2(\mathbf{0})$ interaction localized at the hotspot. The first subsection computes their normalization constants and the second evaluates the mode sums required for the Wightman function (in an approximate limit).

\subsection{Mode normalization}
\label{sec:norm}

This Appendix computes the $\epsilon_\star$-dependence of the integration constants $C_\ell$ and $D_\ell$, by requiring the mode functions to be properly normalized. As discussed in the main text, we do so using the standard Klein-Gordon inner product \pref{KGinnerproduct}, since the reality of $\lambda$ ensures this remains time-independent even with the $\lambda$-dependent boundary conditions of \S\ref{sec:BC}. 

Our mode functions have the form
\bea
u_{\omega\ell m}(t,\bfx) & = & e^{- i \omega t} \big[ C_{\ell }(\omega) j_{\ell}(\omega r) + D_{\ell }(\omega) y_{\ell}(\omega r) \big] Y_{\ell m} (\theta,\phi) ,
\eea
where we have already seen that the boundary condition implies the ratio $D_\ell/C_\ell$ is given by \pref{BCDoverC2} or  \pref{DCvsEpsStar2}. Inserting this into the Klein-Gordon inner product (\ref{KGinnerproduct}) yields
\bea
\langle u_{\omega\ell m}, u_{\tilde{\omega} \tilde{\ell} \tilde{m} } \rangle & = & i \int \exd^{3}x\; \ \bigg( u^{\ast}_{\omega\ell m}(t,\bfx) \dot{u}_{\tilde{\omega} \tilde{\ell} \tilde{m} }(t,\bfx) - \dot{u}^{\ast}_{\omega\ell m}(t,\bfx) u_{\tilde{\omega} \tilde{\ell} \tilde{m} }(t,\bfx)  \bigg) \nn\\
& = & C_{\ell }^{\ast}(\omega) C_{\tilde{\ell}}(\tilde{\omega})  ( \omega + \tilde{\omega} )  e^{+ i ( \omega - \tilde{\omega} ) t } \int \exd^{3}x\; j_{\ell}(\omega r) Y_{\ell m}(\theta, \phi) j_{\tilde{\ell}}(\tilde{\omega} r) Y_{\ell m}(\theta, \phi) \\
& \ & \quad + C_{\ell }^{\ast}(\omega) D_{\tilde{\ell}}(\tilde{\omega})  ( \omega + \tilde{\omega} )  e^{+ i ( \omega - \tilde{\omega} ) t } \int \exd^{3}x\; j_{\ell}(\omega r) Y_{\ell m}(\theta, \phi) y_{\tilde{\ell}}(\tilde{\omega} r) Y_{\ell m}(\theta, \phi) \nn \\
& \ & \qquad + D_{\ell }^{\ast}(\omega) C_{\tilde{\ell} }(\tilde{\omega})  ( \omega + \tilde{\omega} )  e^{+ i ( \omega - \tilde{\omega} ) t } \int \exd^{3}x\; y_{\ell}(\omega r) Y_{\ell m}(\theta, \phi) j_{\tilde{\ell}}(\tilde{\omega} r) Y_{\ell m}(\theta, \phi) \notag \\
& \ & \qquad \quad + D_{\ell }^{\ast}(\omega) D_{\tilde{\ell} }(\tilde{\omega})  ( \omega + \tilde{\omega} )  e^{+ i ( \omega - \tilde{\omega} ) t } \int \exd^{3}x\; y_{\ell}(\omega r) Y_{\ell m}(\theta, \phi) y_{\tilde{\ell}}(\tilde{\omega} r) Y_{\ell m}(\theta, \phi) \notag \qquad 
\eea

Using orthonormality of the spherical harmonics $Y_{\ell m}(\theta, \phi)$
\be
\int   \exd^2\Omega \;  Y_{\ell m}(\theta, \phi) Y_{\tilde{\ell} \tilde{m}}(\theta,\phi) = \delta_{\ell \tilde{\ell}}\, \delta_{m \tilde{m}} \ ,
\ee
allows the above to be written as
\bea
\langle u_{\omega\ell m}, u_{\tilde{\omega} \tilde{\ell} \tilde{m} } \rangle & = & \delta_{\ell \tilde{\ell}} \, \delta_{m \tilde{m}} \bigg[  C_{\ell }^{\ast}(\omega) C_{\ell }(\tilde{\omega})  ( \omega + \tilde{\omega} )  e^{+ i ( \omega - \tilde{\omega} ) t } \int_0^\infty \exd r\; r^2 j_{\ell}(\omega r) j_{\ell}(\tilde{\omega} r) \\
& \ & \qquad \qquad \quad + C_{\ell }^{\ast}(\omega) D_{\ell  }(\tilde{\omega})  ( \omega + \tilde{\omega} ) e^{+ i ( \omega - \tilde{\omega} ) t } \int_0^\infty \exd r\; r^2 j_{\ell}(\omega r) y_{\ell}(\tilde{\omega} r) \nn  \\
& \ & \qquad \qquad \qquad + D_{\ell }^{\ast}(\omega) C_{\ell }(\tilde{\omega})  ( \omega + \tilde{\omega} )  e^{+ i ( \omega - \tilde{\omega} ) t } \int_0^\infty \exd r\; r^2 y_{\ell}(\omega r) j_{\ell}(\tilde{\omega} r) \notag  \\
& \ & \qquad \qquad \qquad \quad + D_{\ell }^{\ast}(\omega) D_{\ell }(\tilde{\omega})  ( \omega + \tilde{\omega} )  e^{+ i ( \omega - \tilde{\omega} ) t } \int_0^\infty \exd r\; r^2 y_{\ell}(\omega r) y_{\ell}(\tilde{\omega} r) \bigg] \ . \notag 
\eea
The $j$-$j$ and the $y$-$y$ terms can be evaluated using the orthonormality relation for spherical Bessel functions,
\be
\int_0^\infty \exd r\; r^2 j_{\ell}(\omega r) j_{\ell}(\tilde{\omega} r) =
\int_0^\infty \exd r\; r^2 y_{\ell}(\omega r) y_{\ell}(\tilde{\omega} r) = \frac{\pi}{2 \omega^2} \delta(\omega - \tilde{\omega}) \ ,
\ee
while the cross-terms are evaluated in Appendix \ref{App:B}, giving
\bea
\int_0^\infty \exd r\; r^2 j_{\ell}(\omega r) y_{\ell}(\tilde{\omega} r) & = & \frac{\left( {\omega}/{\tilde{\omega}} \right)^{\ell}}{\tilde{\omega}( \tilde{\omega}^2 - \omega^2 )} \ . \label{trickyjy}
\eea
The above manipulations lead to the expression
\bea
\langle u_{\omega\ell m}, u_{\tilde{\omega} \tilde{\ell} \tilde{m} } \rangle & = & \delta_{\ell \tilde{\ell}} \delta_{m \tilde{m}} \bigg[ C_{\ell }^{\ast}(\omega) C_{\ell }(\tilde{\omega})  ( \omega + \tilde{\omega} )  e^{+ i ( \omega - \tilde{\omega} ) t } \frac{\pi}{2 \omega^2} \delta(\omega - \tilde{\omega}) \qquad \qquad  \\
& \ & \qquad \qquad \quad + C_{\ell }^{\ast}(\omega) D_{\ell  }(\tilde{\omega})  ( \omega + \tilde{\omega} ) e^{+ i ( \omega - \tilde{\omega} ) t } \frac{\left( {\omega}/{\tilde{\omega}} \right)^{\ell}}{\tilde{\omega}( \tilde{\omega}^2 - \omega^2 )} \notag \\
& \ & \qquad \qquad \qquad + D_{\ell }^{\ast}(\omega) C_{\ell }(\tilde{\omega})  ( \omega + \tilde{\omega} )  e^{+ i ( \omega - \tilde{\omega} ) t } \frac{\left( {\tilde{\omega}}/{\omega} \right)^{\ell}}{\omega( \omega^2 - \tilde{\omega}^2 )} \notag \\
& \ & \qquad \qquad \qquad \quad+ D_{\ell }^{\ast}(\omega) D_{\ell }(\tilde{\omega})  ( \omega + \tilde{\omega} )  e^{+ i ( \omega - \tilde{\omega} ) t } \frac{\pi}{2 \omega^2} \delta(\omega - \tilde{\omega}) \bigg] \ .\notag 
\eea
Of these, the terms with $\delta$-functions are simplified if we take $\tilde{\omega} \to \omega$, and after some simplification on the cross-terms the above becomes
\bea \label{suspicious}
\langle u_{\omega\ell m}, u_{\tilde{\omega} \tilde{\ell} \tilde{m} } \rangle & = & \delta_{\ell \tilde{\ell}} \; \delta_{m \tilde{m}} \; \bigg( \frac{\pi}{\omega} \big( | C_{\ell  }(\omega) |^2 + | D_{\ell  }(\omega) |^2 \big)  \delta(\omega - \tilde{\omega}) \\
& \ & \qquad \qquad \qquad + \frac{e^{+ i (\omega - \tilde{\omega} ) t }}{\omega - \tilde{\omega}} \bigg[  - C_{\ell  }^{\ast}(\omega) D_{\ell   }(\tilde{\omega}) \frac{\omega^{\ell}}{\tilde{\omega}^{\ell + 1}}  + D_{\ell  }^{\ast}(\omega) C_{\ell  }(\tilde{\omega}) \frac{\tilde{\omega}^{\ell}}{\omega^{\ell + 1}}  \bigg] \bigg) \ .\notag 
\eea
The second line of this last equation seems suspicious because it is time-dependent and the Klein-Gordon inner product should not be when evaluated on a solution to the Klein-Gordon equation.  However, this has not yet accounted for the relation between $C_\ell$ and $D_\ell$ that follows from the boundary condition, which states
\be
\frac{D_{\ell m}(\omega)}{C_{\ell m}(\omega)}  \simeq   \frac{\pi\eta_{\star}}{\Gamma(\ell+\tfrac{1}{2})\Gamma(\ell +\tfrac{3}{2})} \cdot \left( \frac{\omega \epsilon_{\star}}{2} \right)^{2 \ell + 1} \ .
\ee
Using this, the square bracket in \pref{suspicious} becomes 
\bea
  && \bigg[  - C_{\ell  }^{\ast}(\omega) D_{\ell   }(\tilde{\omega}) \frac{\omega^{\ell}}{\tilde{\omega}^{\ell + 1}}  + D_{\ell  }^{\ast}(\omega) C_{\ell  }(\tilde{\omega}) \frac{\tilde{\omega}^{\ell}}{\omega^{\ell + 1}}  \bigg]  =  C_{\ell  }^{\ast}(\omega) C_{\ell  }(\tilde{\omega})  \bigg[  - \frac{D_{\ell   }(\tilde{\omega})}{C_{\ell  }(\tilde{\omega}) } \frac{\omega^{\ell}}{\tilde{\omega}^{\ell + 1}}  + \frac{D_{\ell  }^{\ast}(\omega)}{ C_{\ell  }^{\ast}(\omega) } \frac{\tilde{\omega}^{\ell}}{\omega^{\ell + 1}}  \bigg]  \nn\\
   && \qquad\qquad\qquad\qquad\qquad\qquad\qquad =  \frac{\pi\eta_{\star} C_{\ell m}^{\ast}(\omega) C_{\ell m}(\tilde{\omega})}{\Gamma(\ell+\tfrac{1}{2})\Gamma(\ell +\tfrac{3}{2})} \bigg[  - \left( \frac{\tilde{\omega} \epsilon_{\star}}{2} \right)^{2 \ell + 1} \frac{\omega^{\ell}}{\tilde{\omega}^{\ell + 1}}  + \left( \frac{\omega \epsilon_{\star}}{2} \right)^{2 \ell + 1} \frac{\tilde{\omega}^{\ell}}{\omega^{\ell + 1}}  \bigg] \nn\\
   &&\qquad\qquad\qquad\qquad\qquad\qquad\qquad = 0 \,,
\eea
so, as expected, the boundary conditions ensure the time-independence of the inner product.

The final result then is
\bea
\langle u_{\omega\ell m}, u_{\tilde{\omega} \tilde{\ell} \tilde{m} } \rangle & = & \delta_{\ell \tilde{\ell}} \; \delta_{m \tilde{m}} \; \frac{\pi}{\omega} \big( | C_{\ell m}(\omega) |^2 + | D_{\ell m}(\omega) |^2 \big)  \delta(\omega - \tilde{\omega})  \\
 & = & \delta_{\ell \tilde{\ell}} \; \delta_{m \tilde{m}} \; \frac{\pi}{\omega} \left\{ 1 + \left[ \frac{\pi }{\Gamma(\ell+\tfrac{1}{2})\Gamma(\ell +\tfrac{3}{2})} \right]^2 \bigg( \frac{\omega \epsilon_{\star}}{2} \bigg)^{4 \ell + 2} \right\}  | C_{\ell m}(\omega) |^2  \delta(\omega - \tilde{\omega}) \,, \nn
\eea
which uses $\eta_{\star}^2 = 1$. Proper normalization then implies
\be
C_{\ell m}(\omega) = \sqrt{\frac{\omega}{\pi}} \;  \bigg\{ 1 + \left[ \frac{\pi}{\Gamma(\ell+\tfrac{1}{2})\Gamma(\ell +\tfrac{3}{2})} \bigg( \frac{\omega \epsilon_{\star}}{2} \bigg)^{2 \ell + 1} \right]^2 \bigg\}^{-1/2} \,,
\ee
and
\be
D_{\ell m}(\omega) = \frac{\pi\eta_{\star}}{\Gamma(\ell+\tfrac{1}{2})\Gamma(\ell +\tfrac{3}{2})}   \left( \frac{\omega \epsilon_{\star}}{2} \right)^{2 \ell + 1}  \sqrt{\frac{\omega}{\pi}} \; \bigg\{ 1 + \left[ \frac{\pi}{\Gamma(\ell+\tfrac{1}{2})\Gamma(\ell +\tfrac{3}{2})} \bigg( \frac{\omega \epsilon_{\star}}{2} \bigg)^{2 \ell + 1} \right]^2 \bigg\}^{-1/2}
\ee
as claimed in the main text. It is straightforward to similarly check the other relations $\langle u_{\omega\ell m}, u_{\tilde{\omega} \tilde{\ell} \tilde{m} } \rangle = \delta_{\ell \tilde{\ell}} \; \delta_{m \tilde{m}} \; \delta(\omega - \tilde{\omega})$ and $\langle u_{\omega\ell m}, u^{\ast}_{\tilde{\omega} \tilde{\ell} \tilde{m} } \rangle = 0$.

\subsubsection{Evaluating the $j_{\ell} \cdot y_{\ell}$ product integral}
\label{App:B}

We next compute the integral that appears in (\ref{trickyjy}) above, when calculating the cross terms when normalizing the mode functions. First we use $j_{\ell}(z) = \sqrt{ \frac{\pi}{2z} } J_{\ell + \frac{1}{2}}(z)$ to write
\bea
\int_0^\infty \exd r\; r^2 j_{\ell}(a r) y_{\ell}(b r) & = & (-1)^{\ell + 1} \int_0^\infty \exd r\; r^2 j_{\ell}(a r) j_{-\ell-1}(b r) \nn\\
& = & \frac{ (-1)^{\ell + 1} \pi}{2 \sqrt{ab}} \int_0^\infty \exd r\; r J_{\ell+\frac{1}{2}}( a r) J_{-\ell-\frac{1}{2}}( b r) \\
& = & \frac{ (-1)^{\ell + 1} \pi}{2 \sqrt{ a b } } \lim_{\epsilon \to 0^{+}} \int_0^\infty \exd r\; r^{1 - \epsilon} J_{\ell+\frac{1}{2}}(a r) J_{-\ell-\frac{1}{2}}(b r) \,.\nn
\eea
From here we must use the formula (10.22.56) from \cite{NIST} where
\bea
\int_0^\infty \exd r\; r^{-\lambda} J_{\mu}(ar) J_{\nu}(b r ) = \frac{a^{\mu} \Gamma\left(\frac{\nu + \mu - \lambda + 1}{2} \right) \; _2F_1\left( \sfrac{\mu + \nu - \lambda + 1}{2}, \sfrac{\mu - \nu - \lambda + 1}{2}, \mu + 1; \frac{a^2}{b^2} \right)  }{ 2^{\lambda} b^{\mu - \lambda + 1} \Gamma\left( \frac{\nu - \mu + \lambda + 1}{2} \right) \Gamma(\mu + 1) } \qquad
\eea
which assumes that $0 < a < b$ and $\mathrm{Re}[ \mu + \nu + 1 ]  > \mathrm{Re}[ \lambda ] > -1$. In the case that $0 < a < b$ and picking some tiny $\epsilon > 0$ so that $\lambda = \epsilon - 1$ as well as $\mu = \ell + \frac{1}{2}$ and $\nu = - \ell - \frac{1}{2}$ we get
\bea
\int_0^\infty \exd r\; r^2 j_{\ell}(a r) y_{\ell}(b r) & = & \frac{ (-1)^{\ell + 1} \pi}{2 \sqrt{ a b } } \lim_{\epsilon \to 0^{+}} \bigg\{ \frac{2\cos(\pi \ell)}{\pi} \cdot \frac{ a^{\frac{1}{2} + \ell } b^{ - \frac{1}{2} - \ell }}{ a^2 -  b^2} + \cO(\epsilon) \bigg\} \qquad
\eea
Noting that $\cos(\pi \ell) = (-1)^{\ell}$ and taking the limit $\epsilon \to 0^{+}$ gives
\bea
\int_0^\infty \exd r\; r^2 j_{\ell}(a r) y_{\ell}(b r) & = & \frac{\left( {a}/{b} \right)^{\ell}}{b( b^2 - a^2 )} \qquad \qquad \qquad \mathrm{when\ } 0 < a < b
\eea
which is only true for the case $a < b$. For the other case $a > b$ we switch the positions of the Bessel functions in the formula giving us (we need to simultaneously swap $a$ and $b$, as well as $\ell + \frac{1}{2}$ and $- \ell - \frac{1}{2}$)
\bea
 \int_0^\infty \exd r\; r^2 j_{\ell}(a r) y_{\ell}(b r) & = & (-1)^{\ell + 1} \int_0^\infty \exd r\; r^2 j_{-\ell-1}(b r) j_{\ell}(a r) \nn\\
& = & \frac{ (-1)^{\ell + 1} \pi}{2 \sqrt{ a b } } \lim_{\epsilon \to 0^{+}} \int_0^\infty \exd r\; r^{1 - \epsilon} J_{-\ell-\frac{1}{2}}(b r)  J_{\ell+\frac{1}{2}}(a r)  \\
& = & \frac{\left( {a}/{b} \right)^{\ell}}{b( b^2 - a^2 )} \qquad \qquad \qquad \mathrm{when\ } 0 < b < a \,.\nn
\eea
Combining gives the result for any $a>0$ and $b>0$
\be
\int_0^\infty \exd r\; r^2 j_{\ell}(a r) y_{\ell}(b r) = \frac{\left( {a}/{b} \right)^{\ell}}{b( b^2 - a^2 )}
\ee
as quoted in (\ref{trickyjy}).

\subsection{Mode sum}
\label{App:ModeSum}

This appendix evaluates the mode sum encountered in the main text when computing the Wightman function for $\phi$ in the presence of the localized $\lambda \phi^2(t, \mathbf{0})$ hotspot interaction. As argued in the main text, the Wightman function is given by the mode sum
\bea
\braket{\phi (t,\bfx)\phi(t',\bfx')} & = & \sum_{\ell =0}^{\infty} \sum_{m=-\ell}^{+\ell} \int_0^\infty \exd \omega\; u_{\omega \ell m}(t,\bfx) u^{\ast}_{\omega \ell m}(t',\bfx') \\
& = & \sum_{\ell =0}^{\infty} \sum_{m=-\ell}^{+\ell} \int_0^\infty \exd \omega\; e^{- i \omega (t - t')} |C_{\ell m }(\omega)|^2 \left[ j_{\ell}(\omega |\bfx|) + \sfrac{\pi \eta_{\star}}{\Gamma(\ell+\tfrac{1}{2})\Gamma(\ell +\tfrac{3}{2})} \left( \frac{\omega \epsilon_{\star}}{2} \right)^{2 \ell + 1} y_{\ell}(\omega |\bfx| ) \right] \nn \\
&& \qquad   \times \left[ j_{\ell}(\omega |\bfx'|) + \sfrac{\pi \eta_{\star}}{\Gamma(\ell+\tfrac{1}{2})\Gamma(\ell +\tfrac{3}{2})} \left( \frac{\omega \epsilon_{\star}}{2} \right)^{2 \ell + 1} y_{\ell}(\omega |\bfx'| ) \right] Y_{\ell m} (\theta,\phi) Y^{\ast}_{\ell m} (\theta',\phi' ) \ .\notag
\eea

\subsubsection{Evaluating the sums}

Next exploit spherical symmetry about the origin to rotate our coordinate axes so that the $\bfx'$ direction is the $3$-axis of polar coordinates, in which case we can set $\theta' = 0$ (not specifying $\phi'$). Noting the identity (14.4.30) from \cite{NIST} we can write
\be
Y_{\ell m}(0,\phi') = \delta_{m0} \sqrt{ \frac{2\ell +1}{4\pi} } \ ,
\ee
along with $Y_{\ell 0}(\theta,\phi) = \sqrt{ \frac{2 \ell + 1}{4 \pi} } P_{\ell}(\cos \theta)$ where $P_{\ell}$ is the Legendre polynomial of degree $\ell$. This gives
\bea
\braket{\phi (t,\bfx)\phi(t',\bfx')} & = & \sum_{\ell =0}^{\infty} \frac{2\ell + 1}{4\pi} \int_0^\infty \exd \omega\; e^{- i \omega (t - t')} |C_{\ell 0 }(\omega)|^2 \big[ j_{\ell}(\omega |\bfx|) + \sfrac{\pi \eta_{\star}}{\Gamma(\ell+\tfrac{1}{2})\Gamma(\ell +\tfrac{3}{2})} \left( \frac{\omega \epsilon_{\star}}{2} \right)^{2 \ell + 1} y_{\ell}( \omega |\bfx| ) \big] \notag \\
& \ & \qquad \qquad \qquad \times \big[ j_{\ell}(\omega |\bfx'|) + \sfrac{\pi \eta_{\star}}{\Gamma(\ell+\tfrac{1}{2})\Gamma(\ell +\tfrac{3}{2})} \left( \frac{\omega \epsilon_{\star}}{2} \right)^{2 \ell + 1} y_{\ell}(\omega |\bfx'| ) \big] P_{\ell}( \cos \theta )\ . 
\eea

Notice also in passing that the Gamma-matrix identities $\Gamma(n+1) = n \Gamma(n)$ and $\Gamma(n) = (n-1)!$ and $\Gamma(n+\tfrac{1}{2}) = 2^{1 - 2n} \sqrt{\pi} \Gamma(2n) / \Gamma(n)$ imply
\be
\frac{\pi}{\Gamma(\ell+\tfrac{1}{2})\Gamma(\ell +\tfrac{3}{2})} = \frac{2\pi}{(2\ell + 1)\Gamma(\ell + \frac{1}{2})^2} = \frac{2^{4 \ell + 2} \Gamma(\ell)^2}{8 (2\ell + 1)\Gamma(2\ell)^2} = \frac{2^{4 \ell + 2} [ \ell !]^2 }{2 (2\ell + 1) [ (2\ell)! ]^2} \,.
\ee

\subsubsection*{Perturbative limit}

For further progress assume $\omega\epsilon_\star \ll 1$ and seek only the leading $\epsilon_\star$-dependent contribution. Because of the factor $(\omega\epsilon_\star)^{2\ell + 1}$ this allows restricting to $\ell = 0$ in the $\epsilon_\star$-dependent term. In this regime we can approximate the mode sum
\bea
\braket{\phi (t,\bfx)\phi(t',\bfx')} & \simeq & \sum_{\ell =0}^{\infty} \frac{2\ell + 1}{4\pi} \int_0^\infty \exd \omega\; e^{- i \omega (t - t')} \left( \frac{\omega}{\pi} \right) j_{\ell}(\omega |\bfx|) j_{\ell}(\omega |\bfx'|) P_{\ell}( \cos \theta ) \\
&& \qquad \quad + \frac{1}{4\pi} \int_0^\infty \exd \omega\; e^{- i \omega (t - t')} \left( \frac{\omega}{\pi} \right) j_{0}(\omega |\bfx|) \sfrac{\pi \eta_{\star}}{\Gamma( \tfrac{1}{2})\Gamma( \tfrac{3}{2})} \left( \frac{\omega \epsilon_{\star}}{2} \right)  y_{0}(\omega |\bfx'| ) P_{0}( \cos \theta ) \notag \\
& \ & \qquad \qquad +\frac{  1}{4\pi} \int_0^\infty \exd \omega\; e^{- i \omega (t - t')} \left( \frac{\omega}{\pi} \right) \sfrac{\pi \eta_{\star}}{\Gamma( \tfrac{1}{2})\Gamma( \tfrac{3}{2})} \left( \frac{\omega \epsilon_{\star}}{2} \right)   y_{0}(\omega |\bfx|) j_{0}(\omega |\bfx'| ) P_{0}( \cos \theta ) \notag \\
& \ & \qquad \qquad \qquad + \cO(\omega^2 \epsilon_{\star}^2) \,.  \notag
\eea
where the $s$-wave normalization simplifies to $|C_{\ell m}(\omega)|^2 \simeq \frac{\omega}{\pi} [ 1 + \cO(\omega^2 \epsilon_{\star}^2) ]$. Next use the explicit form for the low-order spherical Bessel functions, 
\be
j_{0}( x ) = \frac{\sin(x)}{x} \qquad  \hbox{and} \qquad y_{0}(x) = - \frac{\cos(x)}{x} \ ,
\ee
and along with $\Gamma(\tfrac{1}{2}) \Gamma(\tfrac{3}{2}) = \frac{\pi}{2}$ and $P_0(x) = 1$ to get
\bea
\braket{\phi (t,\bfx)\phi(t',\bfx')} & \simeq &\frac{1}{4\pi^2} \int_0^\infty \exd \omega\; e^{- i \omega (t - t')} \left\{ \omega \sum_{\ell =0}^{\infty}(2\ell + 1)  j_{\ell}(\omega |\bfx|) j_{\ell}(\omega |\bfx'|) P_{\ell}( \cos \theta ) \right.\\
&& \; \left. - \frac{ (4 \pi \eta_{\star} \epsilon_{\star} )}{16\pi^3|\bfx| |\bfx'|} \int_0^\infty \exd \omega\; e^{- i \omega (t-t')} \bigg[ \sin(\omega |\bfx|) \cos(\omega |\bfx'|) + \cos(\omega |\bfx|) \sin(\omega |\bfx'|) \bigg] \right\} \, \notag
\eea
which drops $(\omega\epsilon_\star)^2$ terms. The $\ell$ sum is performed using (10.60.2) from \cite{NIST} which says
\bea
\sum_{\ell=0}^{\infty} (2\ell + 1)  j_{\ell}( u ) j_{\ell}( v ) P_{\ell}(\cos \alpha)  \ = \ \frac{\sin \sqrt{ u^2 + v^2 - 2 u v \cos ( \alpha ) } }{ \sqrt{ u^2 + v^2 - 2 u v \cos ( \alpha ) }  }
\eea
and so 
\bea
\braket{\phi (t,\bfx)\phi(t',\bfx')} & \simeq & \frac{1}{4\pi^2} \int_0^\infty \exd \omega\; e^{- i \omega (t - t')} \omega \cdot \frac{\sin \left(  \omega \sqrt{ |\bfx|^2 + |\bfx'|^2 - 2 |\bfx| |\bfx'| \cos ( \theta ) } \right) }{ \omega \; \sqrt{ |\bfx|^2 + |\bfx'|^2 - 2 |\bfx| |\bfx'| \cos ( \theta ) }  } \nn\\
& \ & \qquad \quad - \frac{ (4 \pi \eta_{\star} \epsilon_{\star} )}{16\pi^3|\bfx| |\bfx'|} \int_0^\infty \exd \omega\; e^{- i \omega (t-t')} \bigg[ \sin(\omega |\bfx|) \cos(\omega |\bfx'|) + \cos(\omega |\bfx|) \sin(\omega |\bfx'|) \bigg]  \nn \\
& = & \frac{1}{4\pi^2|\bfx - \bfx'|} \int_0^\infty \exd \omega\; e^{- i \omega (t - t')} \sin \left( \omega|\bfx - \bfx'|  \right) \\
& \ & \qquad \qquad \qquad \qquad - \frac{ (4 \pi \eta_{\star} \epsilon_{\star} )}{16\pi^3|\bfx| |\bfx'|} \int_0^\infty \exd \omega\; e^{- i \omega (t-t')} \sin\Bigl[ \omega \big( |\bfx| + |\bfx'| \big)  \Bigr]   \ . \notag
\eea
The above result uses the alignment of the coordinates so that $\bfx'$ points along the 3-axis to write
\be
|\bfx|^2 + |\bfx'|^2 - 2 |\bfx| |\bfx'| \cos ( \theta ) = |\bfx - \bfx'|^2 \,.
\ee
The frequency integral finally is
\bea
\int_0^\infty \exd \omega\; e^{- i T \omega} \sin(X \omega) & = & \frac{i}{2} \int_{-\infty}^\infty \exd \omega\; \Theta(\omega) \bigg[ e^{- i (T + X) \omega} + e^{- i (T - X) \omega} \bigg] \\
& = &  \frac{1}{2} \bigg( \frac{1}{T + X - i \delta} - \frac{1}{T - X - i \delta} \bigg) \nn
\eea
where $\delta$ is, as usual, the positive infinitesimal that arises in the Fourier transform of the Heaviside step function. In this way the mode sum evaluates to the result quoted in the main text:
\bea
\braket{\phi (t,\bfx)\phi(t',\bfx')} & \simeq &\frac{1}{4\pi^2} \cdot \frac{1}{- (t-t' - i\delta)^2 + |\bfx - \bfx' |^2 } \\
& \ & \qquad \qquad \qquad - \frac{ (4 \pi \eta_{\star} \epsilon_{\star} )}{32\pi^3|\bfx| |\bfx'|} \bigg[ \frac{1}{t - t' + |\bfx| + |\bfx'| - i \delta} - \frac{1}{t - t' - |\bfx| - |\bfx'| - i \delta} \bigg] \ . \notag
\eea

\section{Mode sum for the exact two-point correlator}
\label{App:modesumfull}

In this appendix we explicitly evaluate the mode sums for the functions $\mathscr{S}(t,\bfx ; t', \bfx')$ and $\mathscr{E}_{\beta}(t,\bfx ; t', \bfx')$ defined in \pref{curlySdef} and \pref{curlyEdef}, giving us a non-perturbative expression for the Wightman function $W_{\beta}(t,\bfx ; t', \bfx') = \mathscr{S}(t,\bfx ; t', \bfx') + \mathscr{E}_{\beta}(t,\bfx ; t', \bfx')$.

\subsection{The temperature-independent contribution, $\mathscr{S}(t,\bfx ; t', \bfx')$}

Using the explicit form for the mode function $S_{\bfp}(t,\bfx)$ given in \pref{Ssol}, the function $\mathscr{S}(t,\bfx ; t, \bfx')$ defined in \pref{curlySdef} simplifies to
\be \label{App:curlySsum}
\mathscr{S}(t,\bfx ; t', \bfx') = \sfrac{1}{4 \pi^2} \cdot \sfrac{1}{ - (t - t' - i \delta)^2 + |\bfx - \bfx'|^2 } + \mathscr{S}_1(t,\bfx ; t, \bfx')  +  \mathscr{S}_2(t,\bfx ; t, \bfx') + \mathscr{S}_3(t,\bfx ; t, \bfx') 
\ee
with the definitions 
\be
\mathscr{S}_1(t,\bfx ; t', \bfx') := \frac{1}{|\bfx'|} \int \frac{\exd^3 \bfp}{(2\pi)^3 2 E_p} \; e^{- i E_{p}(t - t' + |\bfx'|) + i \bfp \cdot \bfx} \frac{ - \frac{\lambda}{4 \pi} - \frac{i \tilde{g}^2 E_p}{16 \pi^2}}{ 1 + \frac{\lambda}{4 \pi \epsilon} + \frac{ i \tilde{g}^2 E_p}{ 16 \pi^2  \epsilon} }
\ee
and
\be
\mathscr{S}_2(t,\bfx ; t', \bfx') := \frac{1}{|\bfx|} \int \frac{\exd^3 \bfp}{(2\pi)^3 2 E_p} \; e^{- i E_{p}(t - t' - |\bfx|) - i \bfp \cdot \bfx'} \frac{ - \frac{\lambda}{4 \pi} + \frac{i \tilde{g}^2 E_p}{16 \pi^2 }}{ 1 + \frac{\lambda}{4 \pi \epsilon} - \frac{ i \tilde{g}^2 E_p}{ 16 \pi^2  \epsilon} }
\ee
as well as
\be
\mathscr{S}_3(t,\bfx ; t', \bfx') := \frac{1}{|\bfx| |\bfx'|} \int \frac{\exd^3 \bfp}{(2\pi)^3 2 E_p} e^{- i E_{p}(t - t' -|\bfx| + |\bfx'|)} \frac{ \left(\frac{\lambda}{4 \pi} \right)^2 + \left( \frac{\tilde{g}^2 E_p}{16 \pi^2 } \right)^2 }{ \left( 1 + \frac{\lambda}{4 \pi \epsilon} \right)^2 + \left( \frac{ \tilde{g}^2 E_p}{ 16 \pi^2  \epsilon} \right)^2 } \ .
\ee
Integrating the momentum angles away in spherical coordinates and simplifying turns the above into 
\be
\mathscr{S}_1(t,\bfx ; t', \bfx') = \frac{\epsilon}{4 \pi^2 |\bfx| |\bfx'|} \int_0^\infty \exd p \; e^{- i p(t - t' + |\bfx'|)} \sin( |\bfx| p ) \frac{ - {4 \pi \lambda}/{\tilde{g}^2} - i p}{ c + i p } \label{S1mom}
\ee
and
\be
\mathscr{S}_2(t,\bfx ; t', \bfx') = \frac{\epsilon}{4 \pi^2 |\bfx| |\bfx'|} \int_0^\infty \exd p \; e^{- i p(t - t' - |\bfx|)} \sin(|\bfx'| p)  \frac{ - {4 \pi \lambda}/{\tilde{g}^2} + i p}{ c - i p } \label{S2mom}
\ee
as well as
\be
\mathscr{S}_3(t,\bfx ; t', \bfx') = \frac{\epsilon^2}{4 \pi^2 |\bfx| |\bfx'|} \int_0^\infty \exd p\; p e^{- i p(t - t' -|\bfx| + |\bfx'|)} \frac{ ({4 \pi \lambda}/{\tilde{g}^2})^2 + p^2 }{ c^2 + p^2 } \label{S3mom}
\ee
where we define the constant
\be
c := \frac{16 \pi^2 \epsilon}{\tilde{g}^2} +  \frac{4 \pi \lambda}{\tilde{g}^2}\ .
\ee
It is more convenient to arrange the integrals as
\bea
\mathscr{S}_1(t,\bfx ; t', \bfx') & = & \frac{i \epsilon}{8 \pi^2 |\bfx| |\bfx'|} \int_0^\infty \exd p \; \bigg( e^{- i p(t - t' + |\bfx| + |\bfx'|)} - e^{- i p(t - t' - |\bfx| + |\bfx'|)} \bigg) \bigg[ \frac{ i ( {4 \pi \lambda}/{\tilde{g}^2} - c ) }{ p - i c } - 1 \bigg] \qquad \quad  \\
& = & \frac{ 2 \epsilon^2}{\tilde{g}^2 |\bfx| |\bfx'|} \int_0^\infty \exd p \; \bigg( e^{- i p(t - t' + |\bfx| + |\bfx'|)} - e^{- i p(t - t' - |\bfx| + |\bfx'|)} \bigg) \frac{ 1 }{ p - i c } \nn \\
& \ & \qquad \qquad - \frac{i \epsilon}{8 \pi^2 |\bfx| |\bfx'|} \int_0^\infty \exd p \; \bigg( e^{- i p(t - t' + |\bfx| + |\bfx'|)} - e^{- i p(t - t' - |\bfx| + |\bfx'|)} \bigg) \nn
\eea
and
\bea
\mathscr{S}_2(t,\bfx ; t', \bfx') & = & - \frac{ 2 \epsilon^2}{\tilde{g}^2 |\bfx| |\bfx'|} \int_0^\infty \exd p \; \bigg( e^{- i p(t - t' - |\bfx| + |\bfx'|)} - e^{- i p(t - t' - |\bfx| - |\bfx'|)} \bigg)\frac{1}{ p + i c } \\
& \ & \qquad \qquad - \frac{i\epsilon}{8 \pi^2 |\bfx| |\bfx'|} \int_0^\infty \exd p \; \bigg( e^{- i p(t - t' - |\bfx| + |\bfx'|)} - e^{- i p(t - t' - |\bfx| - |\bfx'|)} \bigg) \nn
\eea
as well as
\bea
\mathscr{S}_3(t,\bfx ; t', \bfx') & = & \frac{\epsilon^2}{4 \pi^2 |\bfx| |\bfx'|} \int_0^\infty \exd p\; e^{- i p(t - t' -|\bfx| + |\bfx'|)} \bigg[ p - \frac{ \big[ c^2 - ({4 \pi \lambda}/{\tilde{g}^2})^2 \big]p}{ c^2 + p^2 } \bigg] \\
& = & \frac{\epsilon^2}{4 \pi^2 |\bfx| |\bfx'|} \int_0^\infty \exd p\; p e^{- i p(t - t' -|\bfx| + |\bfx'|)}  \nn \\
& \  & \qquad \qquad -  \frac{\epsilon^2 \big[ c^2 - ({4 \pi \lambda}/{\tilde{g}^2})^2 \big]}{8 \pi^2 |\bfx| |\bfx'|} \int_0^\infty \exd p\; e^{- i p(t - t' -|\bfx| + |\bfx'|)} \bigg[ \frac{1}{ p - i c} + \frac{1}{ p + i c}  \bigg] \ . \nn
\eea
From here we note the elementary integrals (where the limit $\delta \to 0^{+}$ is understood)
\bea
\int_{-\infty}^{\infty} \exd p \; \Theta(p) e^{- i \tau p } \ = \ \frac{- i }{\tau - i \delta} \qquad \qquad \mathrm{and} \qquad \qquad \int_{-\infty}^{\infty} \exd p \; \Theta(p) p e^{- i \tau p } \ = \ \frac{-1}{(\tau - i \delta)^2} \ ,
\eea
as well as the integrals 
\be
I_{\mp}(\tau, c) = \int_0^\infty \exd p\; \frac{e^{- i p \tau}}{p \mp i c } \ . \label{Iintpmdef}
\ee
We defer the calculation of the integrals to \S\ref{App:Imp}, where the result \pref{Impfinal} is given by
\bea
I_{\mp}(\tau, c) & = & e^{\pm c \tau} E_{1}\big(\pm c(\tau - i \delta)\big)  \ ,
\eea
where $E_{1}(z) := \int_z^\infty \exd u\; \frac{e^{-u}}{u}$ is the so-called exponential $E_{n}$-function with $n=1$, and where the limit $\delta \to 0^{+}$ is understood (note that we have $E_{1}^{\ast}(z) = E_{1}(z^{\ast})$ is satisfied, and so $I_{-}(\tau) = I_{+}(-\tau)$ remains true). 

With the above we find 
\bea
\mathscr{S}_1(t,\bfx ; t', \bfx') & = & \frac{ 2 \epsilon^2}{\tilde{g}^2 |\bfx| |\bfx'|} \bigg[ I_{-}(t-t'+|\bfx|+|\bfx'|, c) - I_{-}(t-t'-|\bfx|+|\bfx'|, c) \bigg] \\
& \ & \qquad \qquad - \frac{\epsilon}{8 \pi^2 |\bfx| |\bfx'|} \bigg[ \frac{1}{t - t' + |\bfx| + |\bfx'| - i \delta} - \frac{1}{t - t' - |\bfx| + |\bfx'| - i \delta } \bigg] \nn
\eea
and
\bea
\mathscr{S}_2(t,\bfx ; t', \bfx') & = & - \frac{ 2 \epsilon^2}{\tilde{g}^2 |\bfx| |\bfx'|} \bigg[ I_{+}( t - t' - |\bfx| + |\bfx'| , c)  - I_{+}( t - t' - |\bfx| - |\bfx'| , c) \bigg] \\
& \ & \qquad \qquad - \frac{\epsilon}{8 \pi^2 |\bfx| |\bfx'|} \bigg[ \frac{1}{t - t' - |\bfx| + |\bfx'| - i \delta} - \frac{1}{t - t' - |\bfx| - |\bfx'| - i \delta} \bigg] \nn
\eea
as well as
\bea
\mathscr{S}_3(t,\bfx ; t', \bfx') & = & - \frac{\epsilon^2}{4 \pi^2 |\bfx| |\bfx'|} \cdot \frac{1}{( t - t' -|\bfx| + |\bfx'| - i \delta )^2} \\
& \  & \qquad \qquad -  \frac{32 \pi^2 \epsilon^4 ( 1 + \frac{\lambda}{2 \pi \epsilon} )}{ \tilde{g}^4 |\bfx| |\bfx'|} \bigg[ I_{-}( t - t' -|\bfx| + |\bfx'| , c ) + I_{+}( t - t' -|\bfx| + |\bfx'| , c )  \bigg] \ . \nn 
\eea
Finally, putting the above all together into the sum \pref{App:curlySsum} gives
\bea
\mathscr{S}(t,\bfx ; t', \bfx') & = & \frac{1}{4 \pi^2 \left[ - (t - t' - i \delta)^2 + |\bfx - \bfx'|^2 \right]} \\
& \ & \qquad + \frac{ 2 \epsilon^2}{\tilde{g}^2 |\bfx| |\bfx'|} \bigg[ I_{-}(t-t'+|\bfx|+|\bfx'|, c) - I_{-}(t-t'-|\bfx|+|\bfx'|, c) \nn \\
& \ & \qquad \qquad \qquad \qquad \qquad \qquad - I_{+}( t - t' - |\bfx| + |\bfx'| , c) + I_{+}( t - t' - |\bfx| - |\bfx'| , c) \bigg] \nn \\
& \ & \qquad \quad  + \frac{\epsilon}{8 \pi^2 |\bfx| |\bfx'|} \bigg[ - \frac{1}{t - t' + |\bfx| + |\bfx'| - i \delta} + \frac{1}{t - t' - |\bfx| - |\bfx'| - i \delta } \bigg] \nn \\
& \  & \qquad \qquad \quad -  \frac{32 \pi^2 \epsilon^4 ( 1 + \frac{\lambda}{2 \pi \epsilon} )}{ \tilde{g}^4 |\bfx| |\bfx'|} \bigg[ I_{-}( t - t' -|\bfx| + |\bfx'| , c ) + I_{+}( t - t' -|\bfx| + |\bfx'| , c )  \bigg] \nn \\
& \ & \qquad \qquad \qquad \quad - \frac{\epsilon^2}{4 \pi^2 |\bfx| |\bfx'| ( t - t' -|\bfx| + |\bfx'| - i \delta )^2} \nn
\eea
which is the result quoted in \pref{curlySanswer}.

\subsubsection{Perturbative Limit of $\mathscr{S}$}
\label{App:curlySpert}

From here we wish to consider the perturbative limit of the above, which is taken by assuming that
\be
c \tau \ = \ \left( \frac{16 \pi^2 \epsilon}{\tilde{g}^2} + \frac{4 \pi \lambda}{\tilde{g}^2} \right) \tau \gg 1 \ .
\ee
Note that the function $E_{1}(z)$ has the following asymptotic series 
\be
E_{1}(z)\ \simeq \ e^{ - z} \bigg[ \frac{1}{z} - \frac{1}{z^2} + \cO(z^{-3})  \bigg] \qquad \qquad \mathrm{for}\ |z| \gg 1
\ee
which implies that the functions $I_{\mp}(\tau, c )$ have the following asymptotic series for $| c \tau | \gg 1$
\be
I_{\mp}(\tau, c) \ \simeq \ \pm \frac{1}{c(\tau - i \delta)} - \frac{1}{c^2(\tau - i \delta)^2} + \cO(|c\tau|^{-3}) \qquad \qquad \mathrm{for}\  |c \tau| \gg 1 \ .
\ee
We also note in passing that for any $z \in \mathbb{C}$ (with $|\mathrm{Arg}(z)| < \pi$ so not directly on the branch cut) has the series expansion
\be
E_{1}(z) \simeq - \gamma - \log(z) - \sum_{k=1}^{\infty} \frac{(-z)^k}{k \cdot k!}
\ee
which is a convergent sum for any $z \in \mathbb{C}$ but is particularly useful when $|z| \ll 1$. This means that for $|c \tau | \ll 1$ we have 
\be
I_{\mp}(\tau, c) \ \simeq \ - \gamma - \log\big( c (\tau - i \delta) \big) + \cO(c \tau) \qquad \qquad |c \tau| \ll 1 \ ,
\ee
where this limit will clearly suffer from secular growth problems once $c \tau$ is no longer small.

Taking the $|c\tau| \gg 1$ limit of the expression $\mathscr{S}$ here (dropping $\cO(|c\tau|^{-3})$ contributions), and simplifying after using $c = ( 16 \pi^2 \epsilon + 4 \pi \lambda ) / \tilde{g}^2$ yields 
\bea
\mathscr{S}(t,\bfx ; t', \bfx') & \simeq & \frac{1}{4 \pi^2 \left[ - (t - t' - i \delta)^2 + |\bfx - \bfx'|^2 \right]} \\
& \ & \qquad + \frac{ 1}{16 \pi^3 |\bfx| |\bfx'|} \cdot  \frac{\lambda}{1 + \frac{\lambda}{4 \pi \epsilon} } \cdot \frac{|\bfx| + |\bfx'|}{(t-t' - i \delta)^2 - (|\bfx|+|\bfx'|)^2} \nn \\
& \ & \qquad + \frac{\tilde{g}^2}{32 \pi^4 \big(1 + \frac{\lambda}{4 \pi \epsilon}\big)^2} \bigg[   - \frac{1}{|\bfx|} \frac{t-t'-|\bfx|}{\big[ (t - t' - |\bfx| - i \delta)^2 - |\bfx'|^2 \big]^2}  + \frac{1}{|\bfx'|} \frac{t-t'+|\bfx'|}{\big[ (t - t' + |\bfx'| - i \delta)^2 - |\bfx|^2 \big]^2}\bigg] \nn \\
& \  & \qquad - \frac{1}{ 64 \pi^4 |\bfx| |\bfx'|} \cdot \frac{\lambda^2}{\big( 1 + \frac{\lambda}{4 \pi \epsilon} \big)^2} \cdot \frac{1}{( t - t' -|\bfx| + |\bfx'|  - i \delta )^2 }   \ . \nn
\eea
For perturbatively small $\lambda$ (meaning $\lambda/(4\pi\epsilon) \ll 1$) the above turns into 
\bea
\mathscr{S}(t,\bfx ; t', \bfx') & \simeq & \frac{1}{4 \pi^2 \left[ - (t - t' - i \delta)^2 + |\bfx - \bfx'|^2 \right]} \\
& \ & \qquad + \frac{\lambda}{16 \pi^3 |\bfx| |\bfx'|} \cdot \frac{|\bfx| + |\bfx'|}{(t-t' - i \delta)^2 - (|\bfx|+|\bfx'|)^2} \nn \\
& \ & \qquad + \frac{\tilde{g}^2}{32 \pi^4} \bigg[   - \frac{1}{|\bfx|} \frac{t-t'-|\bfx|}{\big[ (t - t' - |\bfx| - i \delta)^2 - |\bfx'|^2 \big]^2}  + \frac{1}{|\bfx'|} \frac{t-t'+|\bfx'|}{\big[ (t - t' + |\bfx'| - i \delta)^2 - |\bfx|^2 \big]^2}\bigg] \nn 
\eea
where we neglect $\cO(\lambda^2)$ contributions. Notice that this exactly matches the temperature-independent contribution to the correlator in the perturbative limit (see \pref{pertcorr}).

\subsubsection{Evaluating the integrals $I_{\mp}(\tau,c)$}
\label{App:Imp}

Here we evaluate the integrals $I_{\mp}(\tau,c)$ defined in \pref{Iintpmdef}. Since $I_{-}(\tau,c) = I_{+}^{\ast}(-\tau, c)$, it suffices to compute $I_{-}(\tau,c)$ here. Splitting apart $I_{-}(\tau,c)$ into real and imaginary parts gives
\be
I_{-}(\tau) = \int_0^\infty \exd p\; \frac{ p \cos(p \tau) + c \sin(p\tau) }{p^2 + c^2} + i  \int_0^\infty \exd p\; \frac{ c \cos(p \tau) + p \sin(p\tau) }{p^2 + c^2} \ .
\ee
Assuming that $c >0$ and $\tau \in \mathbb{R}$ and using formulas (3.723.1)-(3.723.4) from \cite{grad} the above can be easily evaluated to give
\be
I_{-}(\tau, c)  =  - e^{ c \tau } \mathrm{Ei}( - c \tau ) + i \pi e^{ c \tau } \Theta( - \tau) 
\ee
where $\mathrm{Ei}(x) := - \int_{-x}^{\infty} \exd u\; \frac{e^{- u }}{u}$ is the exponential integral function. Since $I_{-}(\tau,c) = I_{+}^{\ast}(-\tau, c)$ this immediately implies that 
\be
I_{+}(\tau, c) =  - e^{ - c \tau } \mathrm{Ei}( c \tau ) - i \pi e^{ - c \tau } \Theta(\tau)
\ee
The above formulae can be simplified into a more useful form by relating it to the function $E_{1}(z) := \int_{z}^{\infty} \exd u\; \frac{e^{- u}}{u}$ (the so-called exponential $E_{n}$-function with $n=1$, related to the exponential integral function for $x>0$ by $E_{1}(x) = - \mathrm{Ei}(-x)$).  Obviously $E_{1}$ closely related to $\mathrm{Ei}$ --- however for complex arguments, the definition of $\mathrm{Ei}$ becomes somewhat ambiguous due to branch points at $0$ and $\infty$, and so $E_{1}$ is better defined for this reason.

Noting the behaviour of the function $E_{1}(z)$ nearby its branch cut (along the negative real axis) where
\be
\lim_{\eta \to 0^{+}} E_{1}( - x \pm i \delta ) \ = \ -  \mathrm{Ei}( x ) \mp i \pi  \qquad \qquad \mathrm{for}\ x > 0 \ ,
\ee
the functions $I_{\mp}(\tau, c) $ can be written in the more useful form 
\bea
I_{-}(\tau, c) & = & e^{c \tau} E_{1}\big(c(\tau - i \delta)\big) \label{Impfinal} \\
I_{+}(\tau, c) & = & e^{-c \tau} E_{1}\big(-c(\tau - i \delta)\big) \nn
\eea
where the limit $\delta \to 0^{+}$ is understood as usual (note that we have $E_{1}^{\ast}(z) = E_{1}(z^{\ast})$ is satisfied, and so $I_{-}(\tau) = I_{+}(-\tau)$ remains true).

\subsection{The temperature-dependent contribution, $\mathscr{E}_{\beta}(t,\bfx ; t', \bfx')$}

In order to evaluate the trace, we put the system in a box (as done in Appendix \ref{sec:freethermaltrace} for the free thermal correlation function). Performing the trace, and then reverting back to the continuum limit, yields the mode sum
\bea
\mathscr{E}_{\beta}(t,\bfx ; t', \bfx') & = & \sum_{a=1}^{N} \int \frac{\exd^3 \bfp}{(2\pi)^3 2 E_p} \bigg[ s^a_{\bfp}(t,\bfx) s^{a\ast}_{\bfp}(t',\bfx') + \frac{ 2 \mathrm{Re}\left[ s^a_{\bfp}(t,\bfx) s^{a\ast}_{\bfp}(t',\bfx') \right]}{e^{\beta E_{p} } - 1} \bigg] \ .
\eea
Using the explicit form of the mode function $s^a_{\bfp}$ given in \pref{sasol} the function $\mathscr{E}_{\beta}(t,\bfx ; t', \bfx')$ turns into the integral
\be
\mathscr{E}_{\beta}(t,\bfx ; t', \bfx') = \sfrac{\tilde{g}^2}{16 \pi^2 |\bfx| |\bfx'|} \int \frac{\exd^3 \bfp}{(2\pi)^3 2 E_p} \bigg[ \; \sfrac{e^{- i E_p (t - |\bfx| - t' + |\bfx'|)}}{\left( \frac{\tilde{g}^2 E_p}{16\pi^2 \epsilon} \right)^2 + \left( 1 +  \frac{\lambda}{4 \pi \epsilon} \right)^2} + \sfrac{2 \cos \big[ E_p (t - |\bfx| - t' + |\bfx'|)\big]}{\left[ \left( \frac{\tilde{g}^2 E_p}{16\pi^2 \epsilon} \right)^2 + \left( 1 +  \frac{\lambda}{4 \pi \epsilon} \right)^2 \right]\left[ e^{ \beta E_{p} }  - 1 \right]} \; \bigg] \ .
\ee
Integrating the angles away and simplifying the above yields
\bea
\mathscr{E}_{\beta}(t,\bfx ; t', \bfx') & = & \frac{4\epsilon^2}{\tilde{g}^2 |\bfx| |\bfx'|} \int_0^\infty \exd p\;  \frac{p}{p^2 + c^2} \bigg[ \; e^{- i p (t - |\bfx| - t' + |\bfx'|)} + \frac{2 \cos \big[ p (t - |\bfx| - t' + |\bfx'|)\big]}{e^{ \beta p }  - 1 } \; \bigg] \qquad \label{curlyE2} \\
& = & \frac{4\epsilon^2}{\tilde{g}^2 |\bfx| |\bfx'|} I\big( t - |\bfx| - t' + |\bfx'|, c, \beta \big)\nn
\eea
where $c$ is the constant \pref{cdef} consisting of the couplings and $\epsilon$ defined by
\be
c  \ := \ \frac{1 + \frac{\lambda}{4 \pi \epsilon} }{ \frac{\tilde{g}^2}{16\pi^2 \epsilon} } \ = \ \frac{16 \pi^2 \epsilon}{\tilde{g}^2} + \frac{4 \pi \lambda}{\tilde{g}^2} \ ,
\ee
and we define the integral 
\be 
I(\tau, c, \beta) = \int_0^\infty \exd p\;  \frac{p}{p^2 + c^2} \bigg[ \; e^{- i \tau p} + \frac{2 \cos ( \tau p )}{e^{ \beta p }  - 1 } \; \bigg] \ , \label{Iintdef}
\ee
which we evaluate here for $c, \beta > 0$ and $\tau \in \mathbb{R}$. To compute $I$, use $\int_0^\infty \exd q\; e^{ - c q} \sin( p q ) = \frac{p}{p^2 + c^2}$ giving
\be
I(\tau, c, \beta) = \int_0^\infty \exd q \int_0^\infty \exd p\; e^{ - c q } \sin( p q ) \bigg[ \; e^{- i \tau p} + \frac{2 \cos ( \tau p )}{e^{ \beta p }  - 1 } \; \bigg] \ .
\ee
Next we rearrange the above into the form
\be  \label{Iint1}
I(\tau, c, \beta) = \int_0^\infty \exd q \; e^{-c q} \bigg[ \int_0^\infty \exd p\; \sfrac{i \big( e^{- i ( \tau + q ) p} - e^{- i ( \tau - q ) p} \big) }{2}  + \int_0^\infty \exd p\; \sfrac{\sin\big( ( \tau + q ) p \big) - \sin\big( ( \tau - q ) p \big)}{e^{ \beta p }  - 1 } \bigg] \ .
\ee
We note the elementary result (with the limit $\delta \to 0^{+}$ understood)
\be
\int_{-\infty}^{\infty} \Theta(x) e^{- i y x} \ = \ \int_0^\infty \exd x \; e^{- i y x} \ = \ \frac{ - i }{y - i \delta }
\ee
as well as formula (3.911.2) from \cite{grad} (valid for $a \neq 0$ and $\mathrm{Re}[\beta] > 0$)
\be
\int_0^\infty \exd x \; \frac{\sin(a x)}{e^{\beta x } - 1} \ = \ \frac{\pi}{2\beta} \coth\left( \frac{\pi a}{\beta} \right) - \frac{1}{2a} \ .
\ee
With these formulae the integral \pref{Iint1} becomes
\bea  \label{Iint2}
I(\tau, c, \beta) & = & \frac{1}{2}\int_0^\infty \exd q \; e^{-c q} \bigg[ \frac{1}{\tau + q - i \delta} - \frac{1}{\tau - q - i \delta}  \\
& \ & \qquad \qquad \qquad \qquad + \frac{\pi}{\beta} \coth\left( \frac{\pi (\tau + q)}{\beta} \right) - \frac{1}{\tau + q} - \frac{\pi}{\beta} \coth\left( \frac{\pi (\tau - q)}{\beta} \right) + \frac{1}{\tau - q} \bigg] \ . \nn
\eea
Using the result of the Sochocki-Plemelj theorem
\be
\frac{1}{x \pm i \delta} \ = \ \frac{1}{x} \mp i \pi \delta(x) \ , \label{soch}
\ee
the integral \pref{Iint2} simplifies to 
\be
I(\tau, c, \beta) = \frac{1}{2}\int_0^\infty \exd q \; e^{- c q} \bigg[ i \pi \delta(\tau + q) - i \pi \delta(\tau - q) + \frac{\pi}{\beta} \coth\left( \sfrac{\pi (\tau + q)}{\beta} \right) - \frac{\pi}{\beta} \coth\left( \sfrac{\pi (\tau - q)}{\beta} \right) \bigg] \ . \label{Iint3}
\ee
Applying the Sochocki-Plemelj theorem yet again to the above yields 
\bea 
I(\tau, c, \beta) = \frac{\pi}{2\beta} \int_0^\infty \exd q \; e^{- c q} \bigg[ \coth\left( \frac{\pi (q + [ \tau - i \delta ] )}{\beta} \right) + \coth\left( \frac{\pi (q - [\tau - i\delta])}{\beta} \right) \bigg]
\eea
where the behaviour $\coth( z ) \simeq {1}/{z}$ near $z \simeq 0$ allows use of formula \pref{soch} (and so justifying the $i \delta$-prescription in the arguments of the $\coth(\cdot)$ functions). From here we note the identity 
\be
\coth\left( \frac{z}{2} \right)  = \frac{2}{1 - e^{-z}} - 1 \ ,
\ee
and make the change of integration variable to $Q = {2\pi q}/{\beta}$ giving
\bea 
I(\tau, c, \beta) & = & \frac{1}{4} \int_0^\infty \exd Q \; e^{- \tfrac{c\beta}{2\pi} Q} \bigg[ \coth\left( \frac{Q + {2\pi [ \tau - i \delta ]}/{\beta} )}{2} \right) + \coth\left( \frac{Q - {2\pi [ \tau - i \delta ]}/{\beta} )}{2} \right) \bigg] \qquad \qquad \nn \\
& = & \frac{1}{2} \int_0^\infty \exd Q \; e^{- \tfrac{c\beta}{2\pi} Q} \left[ \frac{1}{1- e^{ - {2\pi (\tau - i \delta)}/{\beta} } e^{-Q} } + \frac{1}{1- e^{ + {2\pi (\tau - i \delta)}/{\beta} } e^{-Q} } - 1 \right] \nn \\
& = & \frac{1}{2} \Phi\bigg( e^{ - \tfrac{2\pi (\tau - i \delta)}{\beta}}, 1 , \frac{c\beta}{2\pi} \bigg) + \frac{1}{2} \Phi\bigg( e^{ + \tfrac{2\pi (\tau - i \delta)}{\beta}}, 1 , \frac{c\beta}{2\pi} \bigg) - \frac{\pi}{c\beta} \label{Iint4} \ ,
\eea
where we use the integral representation (see formula (25.14.5) in \cite{NIST})
\be
\Phi(z,s,a) \ = \ \frac{1}{\Gamma(s)} \int_0^\infty \exd x \; \frac{x^{s-1} e^{-ax} }{ 1 - z e^{-x} } \qquad \mathrm{valid\ for\ }\mathrm{Re}[s] > 0,\ \mathrm{Re}[a]>0\ \& \ z\in \mathbb{C} \setminus [1,\infty)
\ee
where $\Phi(z,s,a)$ is the Lerch Transcendent, usually defined by the series (see formula (25.14.1) in \cite{NIST})
\be
\Phi(z,s,a) \ = \ \sum_{n=0}^{\infty} \frac{z^n}{(a+n)^s} \qquad \mathrm{valid\ for\ }|z| < 1 \ .
\ee
For other values of $z \in \mathbb{C}$ ({\it ie.} not inside the unit disc) the function $\Phi$ is defined via analytic continuation in the complex plane. At the end of the day, using the above formula \pref{Iint4} in \pref{curlyE2} yields
\be
\mathscr{E}_{\beta}(t,\bfx ; t', \bfx') = \frac{2\epsilon^2}{\tilde{g}^2 |\bfx| |\bfx'|} \bigg[ \Phi\bigg( e^{ - \tfrac{2\pi (t - t' - |\bfx| + |\bfx'| - i \delta)}{\beta}}, 1 , \frac{c\beta}{2\pi} \bigg) + \Phi\bigg( e^{ + \tfrac{2\pi (t - t' - |\bfx| + |\bfx'| - i \delta)}{\beta}}, 1 , \frac{c\beta}{2\pi} \bigg)  - \frac{2\pi}{c\beta} \bigg]
\ee
which is the result quoted in \pref{curlyEanswer} in the main text.

\subsubsection{Perturbative Limit of $\mathscr{E}_{\beta}$}
\label{App:curlyEpert}

Here we take the perturbative limit of $\mathscr{E}_{\beta}$ which turns out to be the limit in which
\be
\frac{c\beta}{2\pi} \gg 1\ .
\ee
We first note the asymptotic series of the Lerch transcendent for any large $a$
\be
\Phi(z,s,a) \simeq \frac{1}{1-z} \frac{1}{a^s} + \sum_{n=1}^{N-1} \frac{(-1)^n \Gamma(s+n)}{n!\; \Gamma(s)} \cdot  \frac{\mathrm{Li}_{-n}(z)}{a^{s+n}} + \cO(a^{-s-N}) \qquad \quad \mathrm{for\ } a \gg 1
\ee
for fixed $s \in \mathbb{C}$ and fixed $z \in \mathbb{C}  \setminus [ 1, \infty )$, where $\mathrm{Li}_{-n}(z) = \left( z \partial_{z} \right)^n \frac{z}{1-z}$ are polylogarithm functions of negative integer order. Using $\mathrm{Li}_{-1}(z) = \frac{z}{(1-z)^2}$ and $\mathrm{Li}_{-2}(z) = \frac{z+ z^2}{(1-z)^3}$ as well as $\mathrm{Li}_{-3}(z) = \frac{z+4z^2+z^3}{(1-z)^4}$ we find that for $c \beta /(2\pi) \gg 1$ we have 
\bea
\Phi\left( z, 1 , \sfrac{c\beta}{2\pi} \right) & \simeq &  \sfrac{1}{1 - z} \sfrac{2\pi}{c\beta} - \sfrac{z}{(1-z)^2} \left( \sfrac{2\pi}{c\beta} \right)^2 + \sfrac{z+ z^2}{(1-z)^3}  \left( \sfrac{2\pi}{c\beta} \right)^3 + \cO\big( (c\beta)^{-4} \big) \\
 \Phi\left( \sfrac{1}{z}, 1 , \sfrac{c\beta}{2\pi} \right) & \simeq &  \left[ 1 - \sfrac{1}{1 - z} \right] \sfrac{2\pi}{c\beta} - \sfrac{z}{(1-z)^2} \left( \sfrac{2\pi}{c\beta} \right)^2 - \sfrac{z+ z^2}{(1-z)^3} \left( \sfrac{2\pi}{c\beta} \right)^3 + \cO\big( (c\beta)^{-4} \big)\nn
\eea
Which means that (half of) the sum of these two functions has the asymptotics 
\bea
\frac{1}{2}\Phi\left( z, 1 , \sfrac{c\beta}{2\pi} \right) +\frac{1}{2}\Phi\left( \sfrac{1}{z}, 1 , \sfrac{c\beta}{2\pi} \right)  & \simeq & \frac{\pi}{c\beta} - \frac{ z}{(1-z)^2} \bigg( \frac{2\pi}{c\beta} \bigg)^2 + \cO\big( (c\beta)^{-4} \big) \ , \qquad \qquad
\eea
which when used for the function $I(\tau,c,\beta)$ given in \pref{Iint4} yields
\bea
I(\tau, c, \beta) \simeq - \sfrac{e^{ - \tfrac{2\pi (\tau - i \delta)}{\beta}}}{\bigg[ 1-e^{ - \tfrac{2\pi (\tau - i \delta)}{\beta}} \bigg]^2} \bigg( \frac{2\pi}{c\beta} \bigg)^2 + \cO\big( (c\beta)^{-4} \big) = - \frac{\pi^2}{c^2 \beta^2} \mathrm{csch}^2\left( \sfrac{\pi [\tau - i \delta]}{\beta} \right) + \cO\big( (c\beta)^{-4} \big) \qquad \qquad 
\eea
Using this and $c = \frac{16\pi^2 \epsilon}{\tilde{g}^2} \left( 1  + \frac{\lambda}{ 4\pi \epsilon} \right)$, at the end of the day we find that $\mathscr{E}_{\beta}$ in the perturbative limit is
\bea
\mathscr{E}_{\beta}(t,\bfx ; t', \bfx') &\simeq & -  \frac{\tilde{g}^2}{64 \pi^2 \beta^2 |\bfx| |\bfx'|\left( 1 + \frac{\lambda}{4 \pi \epsilon} \right)^2} \mathrm{csch}^2\left( \frac{\pi [t - t' - |\bfx| + |\bfx'| - i \delta]}{\beta} \right) + \ldots \nn
\eea
which (at leading-order) is exactly the expected temperature-dependent contribution to the perturbative result when $\lambda/(4\pi\epsilon) \ll 1$ (see formula \pref{pertcorr}, which neglects $\cO(\lambda^2)$ contributions).

\subsubsection{$I(\tau,c,\beta)$ in the limit $\delta \to 0^{+}$}

Because $\Phi(z,s,a)$ has a branch cut along $z \in [1,\infty)$, the limit $\delta \to 0^{+}$ of the expression \pref{Iint4} is somewhat tricky to take. For completeness we take this limit here: to this end, revert back to the integral form \pref{Iint3} and integrate the $\delta$-functions explicitly to get
\be
I(\tau, c, \beta) = \frac{\pi}{2\beta} \int_0^\infty \exd q \; e^{- c q} \bigg[ \coth\left( \sfrac{\pi (q+\tau)}{\beta} \right) + \coth\left( \sfrac{\pi (q - \tau )}{\beta} \right) \bigg] + \frac{i \pi}{2} \bigg[ e^{+ c \tau} \Theta(- \tau) - e^{- c \tau} \Theta( \tau)  \bigg]
\ee
The remaining integrals over $q$ are now ill-defined for general $\tau$ (the reason for this being that for any $\tau \in \mathbb{R}$ one of the two $\coth(\cdot)$ functions gets integrated over a singularity at $q = \tau$). Since the remaining integral is clearly symmetric under $\tau \to - \tau$, we assume for now that $\tau > 0$. For $\tau > 0$ the first integral is easier to compute, where we change the integration variable to $u = \exp\big(- 2\pi ( q + \tau ) / \beta \big)$ where 
\bea 
\frac{\pi}{2\beta} \int_0^\infty \exd q \; e^{- c q} \coth\left( \frac{\pi (q + \tau )}{\beta} \right) & = & \frac{1}{4} e^{c \tau} \int_0^{\exp\left( - 2 \pi \tau / \beta \right)} \exd u\; u^{\tfrac{c\beta}{2\pi} - 1} \bigg[ \frac{2}{1 - u} - 1 \bigg] \\
& = & \frac{1}{2} e^{c \tau} \mathrm{B}\bigg( e^{ - \tfrac{2 \pi \tau}{\beta} } ; \sfrac{c\beta}{2\pi}, 0 \bigg) - \frac{\pi}{2 c \beta } \nn
\eea
where $\mathrm{B}(z;a,b) = \int_0^z \exd u \; u^{a-1} (1-u)^{b-1}$ is the incomplete Beta function. For the second integral, the procedure is similar with the added complication that the integrand gets integrated over the root at $q = - \tau$ (since we assume here that $\tau > 0$). We find in much the same way that 
\be
\frac{\pi}{2\beta} \int_0^\infty \exd q \; e^{- c q} \coth\left( \sfrac{\pi (q - \tau )}{\beta} \right) = \frac{e^{-c \tau} }{2} \int_0^{\exp\left( + 2 \pi \tau / \beta \right)} \exd u\; u^{\tfrac{c\beta}{2\pi} - 1} (1 - u)^{-1}  - \frac{\pi}{2 c \beta } \ ,
\ee
however since the upper limit on the integral is greater than 1 (since $\tau>0$ is assumed), the above integrand gets integrated over a singularity at $u=1$. Interpreting the above integral as a Cauchy Principal value turns the above into
\be
\sfrac{\pi}{2\beta} \int_0^\infty \exd q \; e^{- c q} \coth\left( \sfrac{\pi (q - \tau )}{\beta} \right) = \sfrac{e^{-c \tau}}{2}  \lim_{\eta \to 0^{+}} \bigg[ \int_0^{1 - \eta} + \int_{1 + \eta}^{\exp\left( + 2 \pi \tau / \beta \right)} \bigg] \exd u\; u^{\tfrac{c\beta}{2\pi} - 1} (1 - u)^{-1}  - \sfrac{\pi}{2 c \beta } \ .
\ee
The first integral is easily seen to evaluate to $\mathrm{B}\big(1-\delta;\frac{c\beta}{2\pi},0\big)$, while the second integral requires a variable change $u = 1/v$ giving
\bea
\int_{1 + \eta}^{\exp\left( + 2 \pi \tau / \beta \right)} \exd u\; u^{\tfrac{c\beta}{2\pi} - 1} (1 - u)^{-1} & = & \int_{1/(1 + \eta)}^{\exp\left( - 2 \pi \tau / \beta \right)} \exd v\; v^{-\tfrac{c\beta}{2\pi} } (1 - v)^{-1}  \\
& = & \bigg[ \int_{0}^{\exp\left( - 2 \pi \tau / \beta \right)} -  \int_0^{1/(1 + \eta)} \bigg] \exd v\; v^{-\tfrac{c\beta}{2\pi} } (1 - v)^{-1} \nn \\
& = & \mathrm{B}\bigg( e^{ - \tfrac{2 \pi \tau}{\beta} } ; 1 - \sfrac{c\beta}{2\pi}, 0 \bigg) - \mathrm{B}\bigg( \sfrac{1}{1+\eta} ; 1 - \sfrac{c\beta}{2\pi}, 0 \bigg) \nn
\eea
which then implies that
\bea 
\frac{\pi}{2\beta} \int_0^\infty \exd q \; e^{- c q} \coth\left( \sfrac{\pi (q - \tau )}{\beta} \right) & = & \frac{e^{-c \tau}}{2} \bigg[ \mathrm{B}\bigg( e^{ - \tfrac{2 \pi \tau}{\beta} } ; 1 - \sfrac{c\beta}{2\pi}, 0 \bigg) \\
& \ & \qquad \quad  + \lim_{\eta \to 0^{+}} \bigg\{ \mathrm{B}\bigg(1-\eta;\frac{c\beta}{2\pi},0\bigg) - \mathrm{B}\bigg( \sfrac{1}{1+\eta} ; 1 - \sfrac{c\beta}{2\pi}, 0 \bigg) \bigg\}  \bigg]  - \frac{\pi}{2 c \beta } \nn \ .
\eea
The limit can be taken noting $\mathrm{B}(z;a,0) \simeq  - \log(1-z) - \psi^{(0)}(a)-\gamma + \cO(z)$ for $z \to 1^{-}$ (with $\psi^{(0)}(z) := \Gamma'(z) / \Gamma(z)$ the digamma function) giving 
\bea 
\frac{\pi}{2\beta} \int_0^\infty \exd q \; e^{- c q} \coth\left( \frac{\pi (q - \tau )}{\beta} \right) & = & \frac{e^{-c \tau}}{2} \bigg[ \mathrm{B}\bigg( e^{ - \tfrac{2 \pi \tau}{\beta} } ; 1 - \sfrac{c\beta}{2\pi}, 0 \bigg) + \pi \cot\left(\frac{\beta c}{2}\right) \bigg]  - \frac{\pi}{2 c \beta } \nn \ .
\eea
where the identity $\psi^{(0)}(1-a)- \psi^{(0)}(a) = \pi \cot(\pi a)$ has been used. Putting the above all together (extending the domain to $\tau>0$ for the above integrals) leaves
\be 
I(\tau, c, \beta) = \frac{ e^{c |\tau |}}{2} \mathrm{B}\bigg( e^{ - \tfrac{2 \pi |\tau|}{\beta} } ; \sfrac{c\beta}{2\pi}, 0 \bigg) + \frac{ e^{- c |\tau|}}{2} \bigg[ \mathrm{B}\bigg( e^{ - \tfrac{2 \pi |\tau|}{\beta} } ; 1 - \sfrac{c\beta}{2\pi}, 0 \bigg) + \pi \cot\left(\frac{c \beta}{2}\right) \bigg] - \frac{\pi}{c\beta} - \frac{i \pi}{2} \mathrm{sign}(\tau) e^{- c |\tau|} \ .
\ee
To write the above formula in a slightly more convenient manner, we note formula (8.17.20) from \cite{NIST} which implies that
\be
B(z;1-a,0) = B(z;-a,0) + \frac{z^{-a}}{a} \ ,
\ee
and so allows us to write the above formula as 
\be 
I(\tau, c, \beta) = \frac{ e^{c |\tau |}}{2} \mathrm{B}\bigg( e^{ - \tfrac{2 \pi |\tau|}{\beta} } ; \sfrac{c\beta}{2\pi}, 0 \bigg) + \frac{ e^{- c |\tau|}}{2} \bigg[ \mathrm{B}\bigg( e^{ - \tfrac{2 \pi |\tau|}{\beta} } ; - \sfrac{c\beta}{2\pi}, 0 \bigg) + \pi \cot\left(\frac{c \beta}{2}\right) \bigg] - \frac{i \pi}{2} \mathrm{sign}(\tau) e^{- c |\tau|} \ .
\ee
The beta function can only be related to the Lerch transcendent for arguments $|z| < 1$, which explains why the limit $\delta \to 0^{+}$ is not straightforward from the representation \pref{Iint4}.

\subsubsection{KMS-like Condition for $\mathscr{E}_{\beta}$}
\label{app:KMScurlyE}

Here we show that the function $\mathscr{E}_{\beta}$ is thermal, in the sense that it obeys a KMS-like condition (as does the free thermal correlator of Appendix \ref{app:thermKMS}) where
\be \label{curlyKMSApp}
\mathscr{E}_{\beta}(t- i \beta,\bfx ; t', \bfx') = \mathscr{E}_{\beta}(t',\bfx' ; t, \bfx)  
\ee
{\it c.f.} equation \pref{App:KMSdef}. The proof for this follows almost identically as the proof given in Appendix \ref{app:thermKMS}, save for the fact that $\mathscr{E}_{\beta}$ enjoys a time-translation invariance only when $t>|\bfx|$ and $t '>|\bfx'|$ --- in this limit, we have the representation \pref{curlyE2}
\bea
\mathscr{E}_{\beta}(t,\bfx ; t', \bfx') & = & \frac{4\epsilon^2}{\tilde{g}^2 |\bfx| |\bfx'|} \int_0^\infty \exd p\;  \frac{p}{p^2 + c^2} \bigg[ \; \frac{e^{- i p (t - |\bfx| - t' + |\bfx'|)}}{1 - e^{- \beta p}} + \frac{e^{+ i p (t - |\bfx| - t' + |\bfx'|)}}{e^{ \beta p }  - 1 } \; \bigg] \ ,
\eea
after using the identity $1 +(e^{\beta p} - 1)^{-1} = (1 - e^{-\beta p})^{-1}$. The above function is a complex analytic function of time for in the strip where $- \beta < \mathrm{Im}[t] < 0$ --- we then clearly have 
\bea
\mathscr{E}_{\beta}(t - i \beta,\bfx ; t', \bfx') & = & \frac{4\epsilon^2}{\tilde{g}^2 |\bfx| |\bfx'|} \int_0^\infty \exd p\;  \frac{p}{p^2 + c^2} \bigg[ \; \frac{e^{-\beta p} e^{- i p (t - |\bfx| - t' + |\bfx'|)}}{1 - e^{- \beta p}} + \frac{ e^{+ \beta p} e^{+ i p (t - |\bfx| - t' + |\bfx'|)}}{e^{ \beta p }  - 1 } \; \bigg] \\
& = & \frac{4\epsilon^2}{\tilde{g}^2 |\bfx| |\bfx'|} \int_0^\infty \exd p\;  \frac{p}{p^2 + c^2} \bigg[ \; \frac{ e^{- i p (t - |\bfx| - t' + |\bfx'|)}}{e^{\beta p}-1} + \frac{ e^{+ i p (t - |\bfx| - t' + |\bfx'|)}}{1 - e^{ -\beta p }  } \; \bigg] \\
&=& \mathscr{E}_{\beta}(t',\bfx' ; t, \bfx)  
\eea
which shows that \pref{curlyKMSApp} holds true. Note however that the full correlation function $W_{\beta} = \mathscr{S} + \mathscr{E}_{\beta}$ is not explicitly thermal since the temperature-independent contribution $\mathscr{S}$ obviously fails to satisfy a KMS-like condition analogous to \pref{curlyKMSApp}. However, in a limit where $\mathscr{E}_{\beta}$ dominates over $\mathscr{S}$ (provided $t>|\bfx|$ and $t '>|\bfx'|$) one should expect to see thermality manifest itself more directly (this is explored using an Unruh-Dewitt detector model in \cite{Companion}, where it is shown that a stationary qubit outside the hotspot thermalizes whilst only interacting with $\phi$).

\end{document}